\documentclass[usenatbib]{mn2e}

\usepackage{times}
\usepackage{graphicx}
\usepackage{amsmath}
\usepackage{amssymb}
\usepackage{natbib}
\usepackage{color}
\usepackage{epsfig}
\usepackage{psfig}
\usepackage{subfig}
\usepackage{journal_names}
\usepackage{paralist}

\newcommand{\kms}{\mbox{\,km~s$^{-1}$}}
\newcommand{\Reff}{\mbox{$R_{\rm eff}$}}
\newcommand{\comment}[1]{}
\newcommand{\Vrms}{\mbox{$V_{\rm rms}$}}

\title[Globular Cluster Kinematics in Twelve Early-type Galaxies]
{The SLUGGS Survey: Kinematics for over 2500 Globular Clusters in Twelve Early-type Galaxies}
\author[
Pota~et~al.~ ]
{Vincenzo Pota$^{1}$, Duncan A. Forbes$^{1}$, Aaron J. Romanowsky$^{2}$, Jean P. Brodie$^{3}$,  \\
\\
\normalfont{\LARGE Lee R. Spitler$^{1}$, Jay Strader$^{4}$,  Caroline Foster$^{5}$, Jacob A. Arnold$^{3}$, Andrew Benson$^{6}$,}  \\ 
\\
\normalfont{\LARGE Christina Blom$^{1}$, Jonathan R. Hargis$^{7}$, Katherine L. Rhode$^{7}$, Christopher Usher$^{1}$} \\
\\
$^1$ Centre for Astrophysics \& Supercomputing, Swinburne University, Hawthorn VIC 3122, Australia\\
$^2$ Department of Physics and Astronomy, San Jos\'e State University, San Jose, CA 95192, USA\\
$^3$ University of California Observatories, 1156 High Street, Santa Cruz, CA 95064, USA\\
$^4$ Department of Physics and Astronomy, Michigan State University, East Lansing, Michigan 48824, USA\\
$^5$ European Southern Observatory, Alonso de Cordova 3107, Vitacura, Santiago, Chile\\
$^6$ The Observatories of the Carnegie Institution for Science, 813 Santa Barbara Street, Pasadena, CA 91101, USA\\
$^7$ Indiana University, 727 East 3rd Street, Swain West 319, Blomington, IN 47405, USA\\
Email: vpota@astro.swin.edu.au, dforbes@astro.swin.edu.au}
 
\date{Released 2012 Xxxxx XX}

\pagerange{\pageref{firstpage}--\pageref{lastpage}} \pubyear{2012}

\begin{document}

\label{firstpage}

\maketitle

\begin{abstract}
We present a spectro-photometric survey of $2522$ extragalactic globular clusters (GCs) around twelve early-type galaxies, nine of which
have not been published previously. Combining space--based and multi--colour wide field ground--based imaging, with spectra from the 
Keck DEIMOS instrument, we obtain an average of $160$ GC radial velocities per galaxy, with a high velocity precision of $\sim 15\kms$ per GC.
After studying the photometric properties of the GC systems, such as their spatial and colour distributions, we focus on the kinematics of metal-poor (blue) 
and metal-rich (red) GC subpopulations to an average distance of $\sim~8$~effective radii from the galaxy centre. 

Our results show that for some systems the bimodality in GC colour is also present in GC kinematics. 
The kinematics of the red GC subpopulations are strongly coupled with the host galaxy stellar kinematics. 
The blue GC subpopulations are more dominated by random motions, especially in the outer regions, and decoupled from the red GCs. 
Peculiar GC kinematic profiles are seen in some galaxies: the blue GCs in NGC~$821$ rotate along the galaxy minor axis, whereas
the GC system of the lenticular galaxy NGC~$7457$ appears to be strongly rotation supported in the outer region. 

We supplement our galaxy sample with data from the literature and carry out a number of tests to study the kinematic differences between the two GC subpopulations. 
We confirm that the GC kinematics are coupled with the host galaxy properties and find that the velocity kurtosis and the slope of their velocity dispersion profiles 
is different between the two GC subpopulations in more massive galaxies. 
\end{abstract}

\begin{keywords}
galaxies:star clusters -- galaxies:evolution -- galaxies: kinematics and dynamics -- GCs
\end{keywords}

\section{Introduction}

The reconstruction of the evolutionary history of galaxies in the local universe requires a comprehensive knowledge of
their chemo-dynamic properties at all scales. Along these lines, the ATLAS$^{\rm 3D}$ project \citep{Cappellari11}
has carried out a rich survey of galaxies employing an integral field unit (IFU) technique to map out the chemo-dynamics of galaxies
within $1$ effective radius (\Reff). However, such a radius encloses $<10$\% of the total galaxy mass (baryonic $+$ dark)
and hence it may not be representative of the overall galaxy \citep[e.g.,][]{Proctor}. In contrast, beyond this radius 
($\Reff>1$), the halo of the galaxy contains precious dynamical signatures of early merging or early collapse events
that might have eventually built up the galaxy \citep{HopkinsA,HopkinsB,Oser,Hoffman}. 
Some recent examples of the dynamical studies of the outskirts of galaxies beyond the Local Group has mainly involved a few deep long-slit observations (e.g., \citealt{Coccato10}), 
the pioneering technique of \citet{Proctor} and \citet{Norris08} that probed the two-dimensional stellar kinematics to $3 \Reff$, 
and extragalactic planetary nebulae (PNe, \citealt{Coccato}) that can probe the galaxy potential up to $\sim 10 \Reff$.

Often labeled the ``fossil record'' of galaxy formation, GCs have been used to study galaxy haloes.
GC systems have been found in galaxies of all morphological type and they extend beyond the detectable galaxy diffuse light, 
probing galactocentric distances $\ge10\Reff$ where the are only few other gravitational constrains \citep{Rhode,Rhode04,Dirsch03,Tamura06,Forbes11}. 
Moreover, their old ages \citep[$>10$ Gyr,][]{Kissler-Patig98,Cohen,Beasley00,Forbes01,Schroder,Brodie02,Beasley04B,Strader05} 
suggest  they have survived violent merging events, preserving the chemo-dynamical record of their parent galaxies.

Moreover, the well-studied dichotomy observed in the colour distribution of most GC systems \citep{Zepf93,Ostrov,Whitmore,Gebhardt,Larsen01,Kundu,Peng06,Sinnott}
is thought to stem from different formation mechanisms that shaped the underlying host galaxy  \citep{Ashman92,Forbes97,Cote98}.
Although the reality of the GC metallicity bimodality has been recently called into question as the result of a strongly non-linear color-metallicity relation 
\citep{Yoon06,YoonA,Blakeslee10},  observations have shown that physical dissimilarities exist among these two subpopulations 
\citep{Cote99,Brodie,Peng06,Chies-SantosAGE,Forbes11}. For instance, the two GC subpopulations are found to have different 
physical sizes \citep{Kundu,Jordan05,Masters} and diverse spatial distributions around the host galaxy, with the
metal-rich (red) GCs more centrally concentrated than the metal-poor (blue) GCs  \citep{Geisler,Ashman98,Brodie,Bassino,Faifer11,Strader11,Forbes12}.

Current spectroscopic studies of GC systems have shown that the kinematics (e.g. rotation directions and rotation amplitudes) of the two GC 
subpopulations are somewhat diverse. The kinematics of the red GCs is usually akin to that of the host galaxy stars \citep{Schuberth,Strader11}, perhaps 
due to a similar formation history \citep{Shapiro}. Whereas, the velocity dispersion of the blue GCs is typically larger than that of the red GCs \citep[e.g.,][]{Lee08}. 
Also intriguing is the fact that rotation has been detected for both the blue and the red GCs, regardless the mass or morphology of the host galaxy \citep{Foster11,Arnold}.  

Interpreting this variety of GC kinematics in the context of galaxy and GC formation has been limited due to the low number of galaxies with 
large GC radial velocity datasets. To date, this set includes only a dozen GC systems, most of which are nearby very massive ellipticals (see \citealt{Lee}, for a summary).
On the other side, numerical simulations on this front have mainly focused on the origin of GC metallicity and colour bimodality 
\citep[e.g.,][]{Weil,Kravtsov,Yoon06,YoonA}, rather than on GC kinematic properties \citep[e.g.,][]{Bekki2005,Bekki2008,Prieto}.

This scenario has left open several questions regarding the kinematics of GC systems: does the colour bimodality also imply kinematic bimodality? If so, do the kinematical differences
between the blue and red GC subpopulations found in the most massive ellipticals also hold for $\sim L^*$ galaxies over the whole early-type sequence of the Hubble diagram? 
Furthermore, do the blue and red GC subpopulations rotate faster in the outer regions, as predicted in a formation in a disk-disk 
merging scenario \citep{Bekki2005}? Can GC kinematics contribute to our understanding of the formation of lenticular galaxies \citep[e.g.,][]{Barr}?

We have been carrying out a project named SLUGGS\footnote{http://sluggs.swin.edu.au/} to investigate the GC systems in external galaxies (Brodie et al. 2012, in preparation). 
SLUGGS is the SAGES Legacy Unifying Globulars and Galaxies Survey, where SAGES is the Study of the Astrophysics of Globular Clusters in Extragalactic Systems. This survey exploits  
the combination of Subaru/Suprime-Cam wide-field imaging with spectra from the Keck/DEIMOS multi-object spectrograph. The results released so far have shown
that the wide-field imaging can give clues about assembly history of the host galaxy \citep{Blom,Forbes11}. If combined with the 
high velocity resolution of DEIMOS, this dataset can unravel, at the same time, the dynamics and the metallicity of the field stars 
\citep{Foster09,Proctor} and of the GCs \citep{Foster11,Arnold} deep into the galaxy halo, as well as a giving compelling view of galaxy dynamics
\citep{Strader11,Romanowsky09,Romanowsky11}.

In this work, we aim to study the GC kinematics for an unprecedented large sample of early-type galaxies (ETGs) with high-quality data. We 
investigate the global kinematics of blue and red subpopulations to study how they, and the underlying galaxy itself, formed. We also supplement our galaxy 
sample with literature data, and we compare the properties of this large sample with the existing numerical predictions.

The plan of this paper is as follows. From Section 1 to Section 5 we describe the reduction and the analysis of  
both the photometric and the spectroscopic data. In Section \ref{sec:individualgalaxies}, we briefly discuss the significant findings for 
each galaxy. In Section \ref{sec:GC formation models}, we
give an overview on the current state of the GC formation models in order to compare their predictions to our generic results 
discussed in Section \ref{sec:Summary of generic results}. In Section \ref{sec:Merging} we supplement our galaxy sample 
with data from the literature. In Section \ref{sec:LiteratureResults} we analyse the enlarged GC system dataset (our data plus literature) and discuss
the results in Section \ref{sec:discussion}. The summary of the paper is given in Section \ref{sec:Summary}.
 
\section{The sample}
In this paper we discuss a subset sample from our survey. This includes nine new galaxies, in addition to other three galaxies already published: 
NGC~$4494$ \citep{Foster11}, NGC~$3115$ \citep{Arnold} and NGC~$4486$ \citep{Strader11}. The analysis and the specific results for these 
three galaxies have been extensively discussed in the respective papers. Therefore, their overall results will be discussed together with the other 
nine starting from Section \ref{sec:Summary of generic results}. 

\begin{figure*}
\centering
\vspace{-0.2cm}
\includegraphics[trim = 0mm 0mm 0mm 0mm, clip, scale=.45]{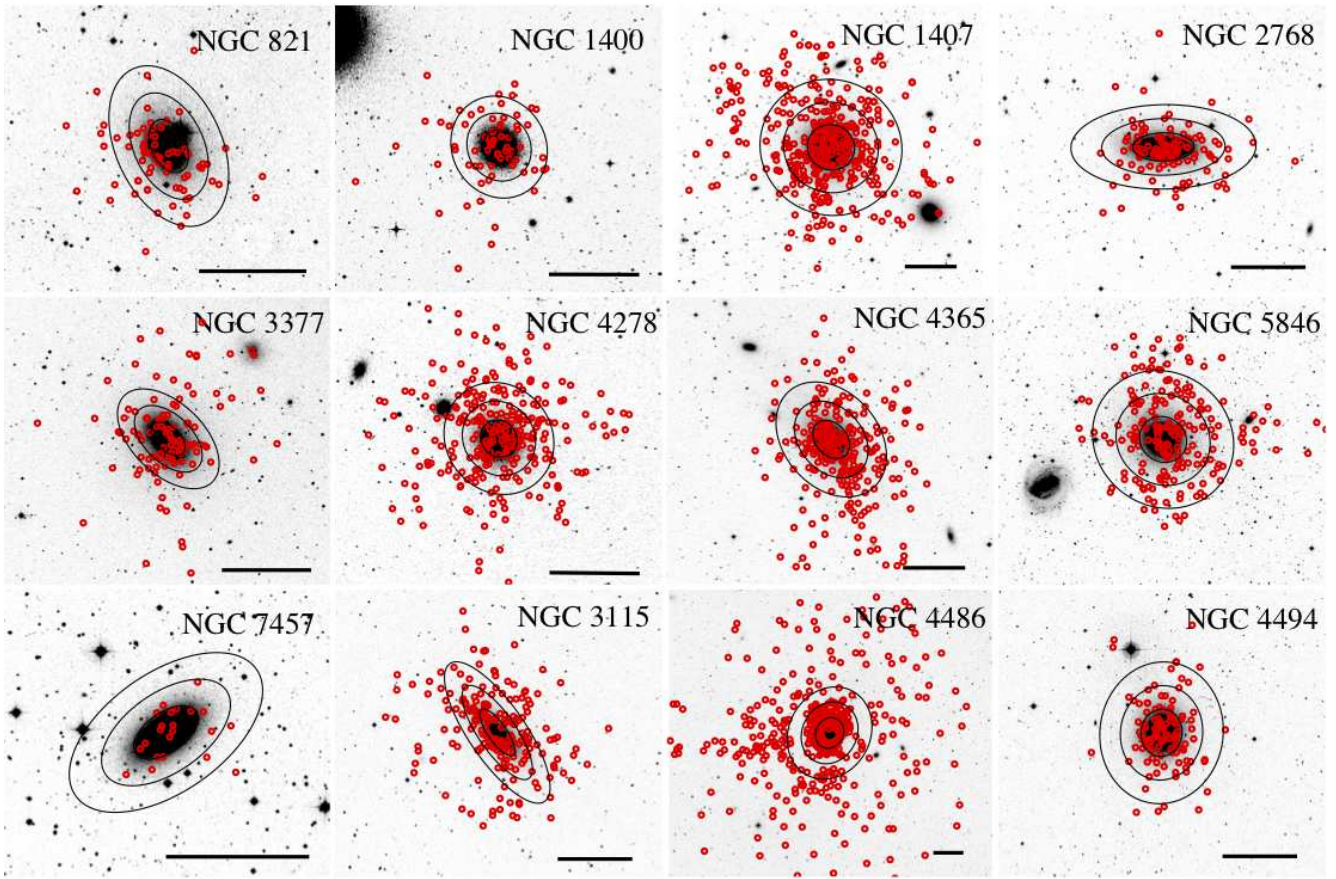} 
\caption{Digitalized Sky Survey (DSS) images of our galaxy sample. Red circles and black ellipses represent the spectroscopically confirmed GCs found in this work 
and the $2, 4, 6$ \Reff\ schematic isophotes corrected for the galaxy ellipticity respectively. The black line on the bottom-right spans $5$ arcmin in length.
North up and the East on the left. The elliptical galaxy south-west of NGC~$1407$ is NGC~$1400$ whose GC system is shown separately in this figure. 
The last three galaxies have been analysed in separate papers.}
\label{fig:RADEC}
\end{figure*}

The physical characteristics of the twelve galaxies are listed in Table~\ref{tab: survey_summary}, with their optical images shown in Figure~\ref{fig:RADEC}. 
This galaxy sample extends the study of extragalactic GC systems into a new regime, because it is representative of a wide range of luminosity, 
morphological type (from lenticulars to giant ellipticals) and environment (from field to clusters), with a velocity resolution three times better than typical 
previous studies. This improvement is shown in Figure~\ref{fig:magNGC} in which we compare the intrinsic properties of our dataset with previous
GC studies (that have employed various instruments including VLT/FLAMES, VLT/FORS2, Keck/LRIS or Gemini/GMOS). 
\begin{table*}
\begin{tabular}{@{}c c c c c c c c c c c}
\hline
Galaxy ID & Hubble& $V_{\rm sys}$  & \Reff & $(m-M)$ & $D$    & $M_{\rm K}$  &  $A_{\rm K}$ &  $PA_{\rm K}$    & ($b/a)_{\rm K}$ \\
      & Type     &  [km s$^{-1}$]  & [arcsec]       &   [mag]       & [Mpc]  &  [mag]       &    [mag]        &      [degree]       &                 \\
 (1)  & (2)      &  (3)            & (4)            & (5)     & (6)    &  (7)         &  (8)      &       (9)           &       (10)          \\
\hline
\hline   
  NGC~$0821$      &E$6$         & $1718$     & $51$  & $31.85$  &  $23.4$         & $-24.0$      &   $0.040$   & $31$         & $0.62$ \\
  NGC~$1400$      &SA$0$        & $558$       & $31$   & $31.05$&  $26.8$           &$-23.8$       &   $0.024$  &   $29$       &$0.90$ \\
  NGC~$1407$      &E$0$         & $1779$     & $72$  &  $32.30$ &  $26.8$         &$-25.4$       &  $0.025$   &  $70$        & $0.95$ \\
  NGC~$2768$      &E$6$         & $1353$     & $93$  &  $31.69$& $21.8$           &$-24.7$       &   $0.016$  &  $91$        & $0.46$ \\
  NGC~$3377$      &E$6$        & $690$        & $46$     &  $30.19$&  $10.9$       &$-22.7$        &   $ 0.013$ &  $46$       & $0.58$\\
  NGC~$4278$       &E$2$           & $620$       & $34$    &  $30.97$ & $15.6$     & $-23.8$       &   $0.010$ &   $219$       &$0.91$\\
  NGC~$4365$       & E$3$        & $1243$    & $60$   &     $31.84$ &  $23.3$     & $-25.2$       &  $0.008$ &    $41$      & $0.74$  \\
  NGC~$5846$       &E$0$        & $1712$     & $61$    &  $31.92$&  $24.2$         & $-25.0$      &    $0.020$ &  $233$       &$0.92$\\
  NGC~$7457$       &S$0$        & $844$       & $52$    &  $30.55$&  $12.9$         & $-22.4$      &   $0.019$  &  $125$      &$0.54$ \\
  \hline
  NGC~$3115$    &S$0$         & $663$     & $85$  & $29.87$ &   $9.4$               & $-24.0$       &   $0.017$   & $43$         & $0.45$ \\  
  NGC~$4486$    &E$0$         & $1284$     & $81$  & $31.18$  &  $17.2$         & $-25.3$        &   $0.008$   & $151$         & $0.86$ \\
  NGC~$4494$    &E$1$         & $1344$     & $53$  & $31.10$  &  $16.6$         & $-24.2$        &   $0.008$   & $173$         & $0.87$ \\
\hline  
\hline
\end{tabular}
\caption{General properties of our galaxy sample. The galaxy name (1) and Hubble Type (2) are from the NED database. The galaxy systemic velocity (3) and
ellipticity corrected effective radius (4) are from \citet{Cappellari11}, otherwise from NED and the RC3 catalogue \citep{Vaucouleurs} if not in \citet{Cappellari11}. 
The distance modulus (5) and the respective distance in Megaparsec (6) are from \citet{Tonry01} with a $-0.06$ correction as advocated by \citet{Mei}, respectively.
If the galaxy is in the  ACS Virgo Survey, we use the distances from \citet{Mei}.
We assume that NGC~$1407$ and NGC~$1400$ lie at the same distance, computed as the average of the respective \citet{Tonry01} distances. 
The $K$~band absolute magnitude (7) is from 2MASS apparent magnitude at the distances given in column 6 and corrected for the foreground Galactic extinction given
in column 8 (NED database). The photometric position angle (9) and axis ratio (10) are from 2MASS \citep{Skrutskie}. The last three galaxies have been analysed 
in separate papers (see text).}
\label{tab: survey_summary} 
\end{table*}

\section{Photometric observations and data analysis}
\subsection{Subaru data}
\label{sec:subarudata}

\begin{figure}
\centering
\includegraphics[scale=.5]{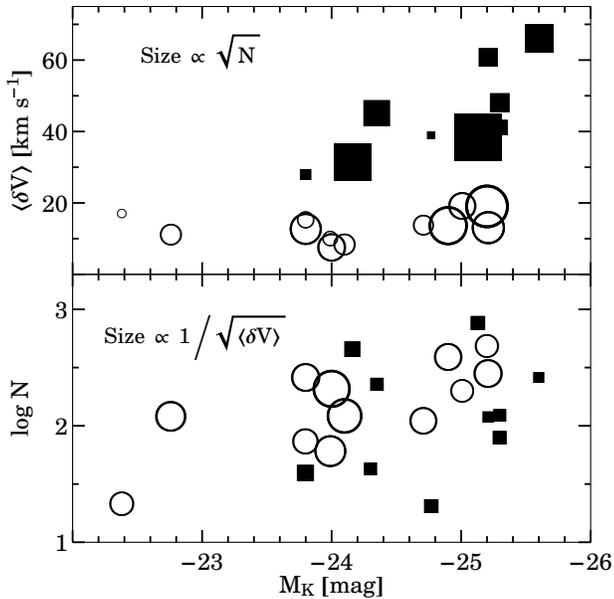} 
\caption{Accuracy of GC radial velocity measurements.  Open circles and open boxes represent our GC data and literature data, respectively.
Literature data will be discussed in Section \ref{sec:Merging}.
\textit{Top panel}. The median velocity uncertainty $\langle \delta V \rangle$ per GC system are shown as a function of absolute magnitude in the K band. 
Symbol sizes are $\propto \sqrt{N}$, where $N$ is the number of spectroscopically confirmed GCs per galaxy. \textit{Bottom panel}.
Log $N$ is shown as a function of the absolute magnitude in the K band, where symbol sizes are $\propto 1/ \sqrt{\langle \delta V \rangle}$.
Our data extend the range of galaxy luminosity probed with three times better velocity accuracy.}
\label{fig:magNGC}
\end{figure}

Multi-band photometric observations were carried out using the Subaru/Suprime-Cam instrument with a field of view of $34~\times~27$~arcmin$^2$ and a 
pixel scale of $0.202$ arcsec \citep{Miyazaki}. The galaxy sample was observed in the period between 2005 and 2010 with a standard Sloan Digital Sky Survey (SDSS)
$gri$ filter set. NGC~$2768$ and NGC~$4278$ were both downloaded from the SMOKA archive \citep{Baba} and were observed with a $R_{\rm C}iz$ and $BVI$ filter set, respectively. For NGC~$1407$ and
NGC~$4365$ we present existing Suprime-Cam photometry published in \citet{Spitler12} (see also \citealt{Romanowsky09}) and \citet{Blom}, respectively. For these
two galaxies both the imaging reduction and the catalogue extraction were performed with the methodology described in this paper. NGC~$7457$ is the 
only galaxy for which no Subaru imaging is available, and therefore we use WIYN/Minimosaic imaging in $BVR$ filters presented in \citet{Hargis} and 
we refer to this paper for a description of the data reduction. 
 
In Table \ref{tab: log_photo} the imaging observations are summarized.  The overall seeing conditions were mainly sub-arcsec. For NGC~$821$ ($g$ band),
NGC~$5846$ ($g$ band) and NGC~$4278$ ($B$ band) the data suffer from cloudy conditions and poor ($\ge 1$ arcsec) seeing. 

Subaru raw images were processed using the SDFRED data pipeline \citep{Ouchi} that yields standard flat field corrected images for each of the three
filters. Photometric point source catalogues were extracted using standard \texttt{IRAF/Daophot} aperture photometry routines. 
We summarise here the main steps of the data reduction and we refer to Section 3 of \citet{Blom} for a detailed description of the method. 
 
We obtain a raw list of object positions by running \texttt{IRAF/Daofind} on galaxy subtracted images in order to optimise the finding algorithm. 
The extraction threshold was typically set between $2$ and $4$ times the background depending on the filter and on the seeing conditions. 
The galaxy light was modelled with \texttt{IRAF/Ellipse} set to allow the position angle and ellipticity to vary. Next, we perform aperture photometry 
using \texttt{IRAF/Phot} on the preselected objects for a certain number of circular apertures from $1$ up to $15$ pixels (equivalent to $\sim 0.2$ arcsec to
$3$ arcsec for the Suprime-Cam pixel scale). The extraction radius was chosen in order to maximise the signal of the source and minimise the sky 
contribution. The extracted magnitude was corrected for the computed aperture correction using \texttt{IRAF/Mkapfile}. Photometric zeropoints were
estimated by boot-strapping the Suprime-Cam photometry to the Sloan Digital Sky Survey (SDSS) DR7 photometric system \citep{Abazajian} using
the brightest objects in common between the two datasets (typically with $17<i<21$). 
If not in SDSS. the zeropoints were calibrated using the flux from standard stars observed over the same night.
Finally, we use the reddening given in Table~\ref{tab: survey_summary} and the conversion table of  \cite{Schlegel} to derive the
Galactic extinction correction in our photometric bands. Hereafter, all magnitudes and colours are extinction corrected.

\subsection{HST data}
\label{sec:subarudata}
We use \textit{Hubble Space Telescope (HST)} archive images from the Advanced Camera for Surveys (ACS) and Wide Field
Planetary Camera 2 (WFPC2) to improve the quality of the photometric selection in the central regions of our galaxies.  
For most of them, we exploit existing photometric GC catalogues and we refer to the following authors for a detailed 
description of the data reduction and analysis: \citet{Spitler08} for NGC~$821$, \citet{Forbes06A} for NGC~$1407$ 
and NGC~$1400$,  \citet{Forbes96} for NGC~$5846$, \citet{Chomiuk} for NGC~$7457$ and \citet{Blom} for 
NGC~$4365$. 

We obtained, from the Hubble Legacy Archive, new \textit{HST}/ACS imaging for NGC~$3377$, NGC~$2768$ and 
NGC~$4278$, respectively. The ACS camera has a pixel scale of $0.05$ arcsec and a field of view of $3.36 \times 3.36$ arcmin$^2$.
 
The NGC~$3377$ imaging consists of one pointing in F475W ($\sim$ Sloan $g$) and F850LP ($\sim$ Sloan $z$) filters and
it was observed as part of the \textit{HST} project ID 10554. NGC~$2768$ (ID 9353) was imaged in F435W, F555W, F814W filters, 
equivalent to a $BVI$ configuration, respectively. Finally, the NGC~$4278$ (ID 10835) data consists of four pointings in F475W 
and F850LP filters that probe the galaxy up to $\sim 6$ arcmin from the centre (Usher et al. 2012, in preparation).

The \textit{HST} imaging was reduced and analysed using a custom built pipeline to find point-like sources and measure 
their magnitudes and half light radii. For details on the methods used by the pipeline including point spread function 
determination, we refer to \citet{Strader06} and to \citet{Spitler06}. The extracted magnitudes and sizes for GCs in 
NGC~$3377$ and NGC~$4278$ were compared with those published by \citet{Chies-Santos} for objects in the NGC~$3377$ 
pointing and in the two NGC~$4278$ pointings. Both magnitudes and sizes show good agreement without any evidence 
of statistically significant offset from the published data. 

\begin{figure}
\centering
\includegraphics[scale=.5]{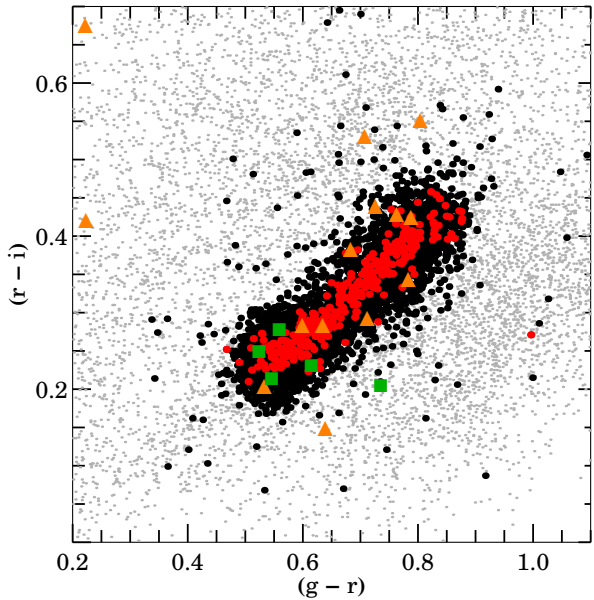} 
\caption{Colour-colour diagram of objects around NGC~$1407$. Grey and black points represent all the sources detected 
in the Suprime-Cam field-of view and all the photometrically selected GCs with $i<25$ and $i<24$, respectively. Spectroscopically confirmed 
sources are shown as red points (GCs), green boxes (Galactic stars) and orange triangles (galaxies) respectively.}
\label{fig:NGC1407}
\end{figure}

\subsection{Photometric GC selection}
\label{sec:photometric_selection}
Once the point-source catalogues have been constructed, they are photometrically selected to avoid contamination, 
such as unresolved galaxies and Galactic stars. As far as our new Subaru data are concerned, this process involves the following steps:
\begin{inparaenum}
\item selection in a colour-colour space; 
\item selection in a colour-magnitude space;
\item a visual check.
\end{inparaenum}
The GC selection in the ground-based imaging for NGC~$7457$ was carried out in \citet{Hargis} and we refer to this paper for a 
detailed description of their selection method. The steps listed above are subjected to variations in the procedures for the DEIMOS mask design.

\begin{table}
\centering
\begin{tabular}{@{}l c c c c}
\hline
Galaxy ID & Obs date & Filters & Exp. time & Seeing   \\
                  &      &  &      [sec] &     [arcsec]        \\
\hline
\hline   
NGC~$0821$ & 2008 Nov. & $gri$ & $960, 350, 220$ & $1.2, 1.0, 0.8$  \\
NGC~$1400$ & 2008 Nov. & $gri$  & $3240, 3600, 10800$ & $0.6, 0.6, 0.6$  \\
NGC~$1407$   & 2008 Nov. & $gri$ & $3240, 3600, 10800$  & $0.6, 0.6, 0.6$  \\
NGC~$2768$   & 2005 Mar. & $R_{\rm C}iz$  & $600, 100, 210$ & $0.6, 0.6, 0.6$ \\
NGC~$3377$    & 2008 Nov. & $gri$ & $500, 450, 375$ & $0.9, 0.7, 0.8$  \\
NGC~$4278$     &2002 Feb.  & $BVI$ &  $600, 450, 360$ & $1.4, 0.9, 0.9$\\
NGC~$4365$    & 2008 Apr.  & $gri$  & $650, 350, 300$ & $0.8, 0.8, 0.8$ \\
NGC~$5846$     &2009 Apr. &  $gri$ &  $2760, 1606, 1350$ & $1.0, 0.6, 0.8$\\
NGC~$7457$     & 2009 Oct. & $BVR$ & $6300, 6000, 7200$ & $0.7, 0.8, 0.7$ \\
\hline
\hline
\end{tabular}
\caption{Summary of the ground-based imaging observations. The galaxy name, observation date, filters employed, together with the 
respective exposure time and seeing are listed. All the observations were performed using Subaru/Suprime-Cam, except for NGC~$7457$ 
observed with WIYN/MiniMo \citep{Hargis}.}
\label{tab: log_photo}
\end{table}

GCs are known to populate a specific area of colour-colour diagrams \citep[e.g.,][]{Rhode,Faifer11,Chies-SantosAGE}. With a $gri$ filter set, this 
is enclosed within $0.4\la(g-i)\la 1.5$, $0 \la (r-i) \la 0.6$, $0.3 \la (g-r) \la 0.9$ where these boundaries run diagonally to the colour axes, as
shown in Figure~\ref{fig:NGC1407} for the galaxy NGC~$1407$. To take into account the dependence of the colour boundaries on the quality of the data, 
we flag as GC candidates all the objects deviating by less than 2$\sigma$ from these boundaries \citep{Spitler08}. 

Next, we apply a cut on the $i$ band magnitude ($I$ band for NGC~$4278$). 
Given the ongoing debate regarding the uncertain separation between GCs and ultra compact dwarfs \citep[UCDs,][]{Mieske,Brodie11}, we decided to set the upper 
brightness magnitude at $M_i \approx-11.6$ ($M_I \approx-12$), one magnitude brighter than the integrated magnitude of $\omega$ Cen, the brightest GC in the Milky Way. 
Nevertheless, in some cases we relax this criterion in order to include spectroscopically confirmed GCs that have magnitudes brighter than the set threshold. 
The separation between NGC~$1407$ and NGC~$1400$ objects will be discussed in Section \ref{sec:individualgalaxies}.

Next, we calculate the radius at which the number of GC candidates per unit area flattens out (see \S \ref{sec:GC spatial distribution} for the method), that is an estimate of the radius at which the contribution of the 
contaminants (Glactic stars and high redshift galaxies) becomes dominant. Therefore, we count out of the GC selection all the objects outside this background radius.
We have tested that the effect of the contaminants on the GC colour distribution is minimal $(<0.1)$ mag and that it does not affect considerably the GC colour bimodality \citep[see also,][]{Arnold}.

Finally, we perform a visual check to make sure that no outliers, such as extended sources or image artefacts, contaminate the final GC catalogue. 
Most of the outliers turned out to be close to the galaxy (within $1$ arcmin) where the galaxy light contamination and the crowded field makes the Subaru photometry
unreliable. Within this radius, the contribution of the \textit{HST} imaging becomes crucial. 

\begin{figure*}
\centering
\includegraphics[scale=.48]{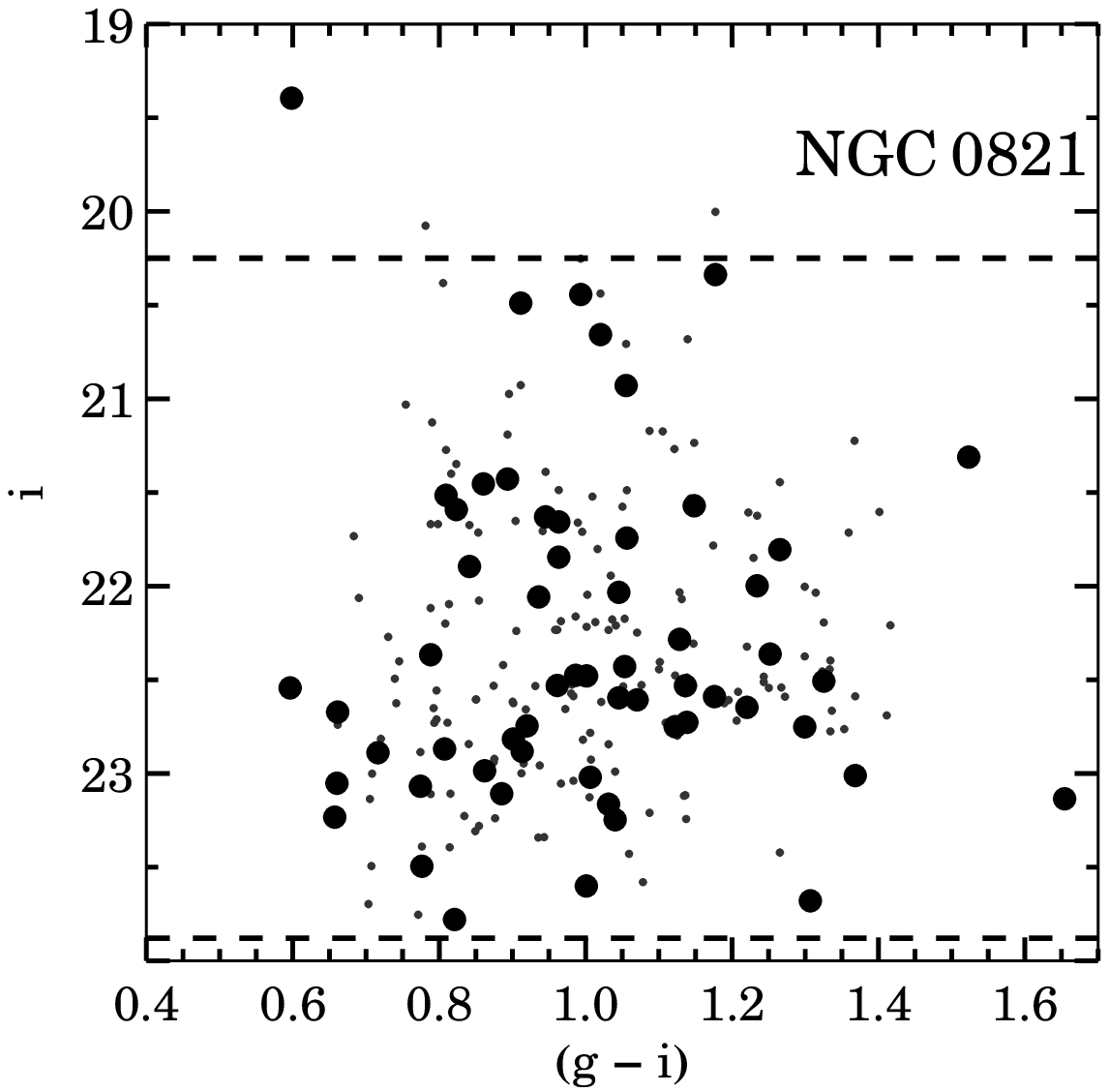} 
\includegraphics[scale=.48]{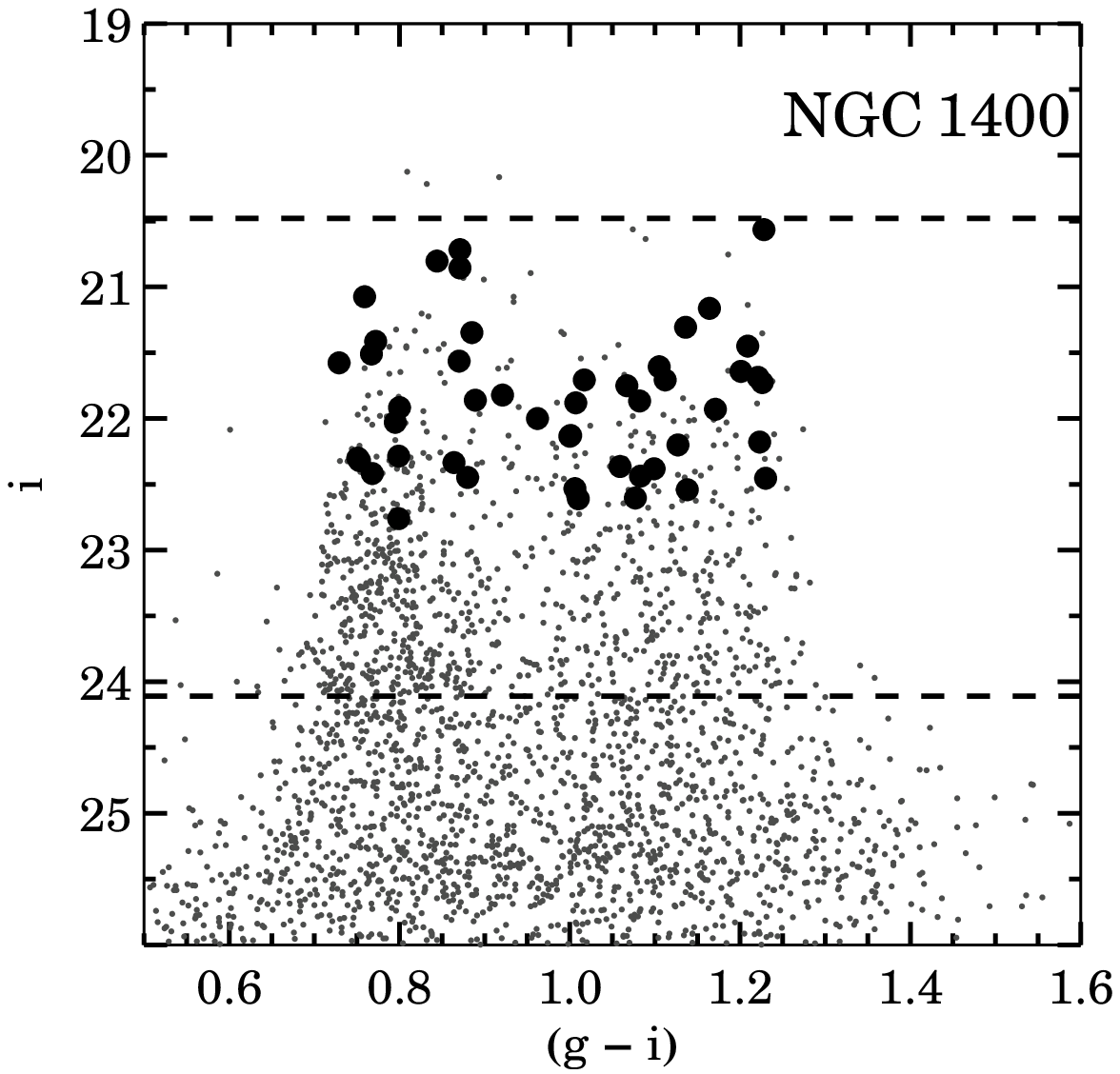} 
\includegraphics[scale=.48]{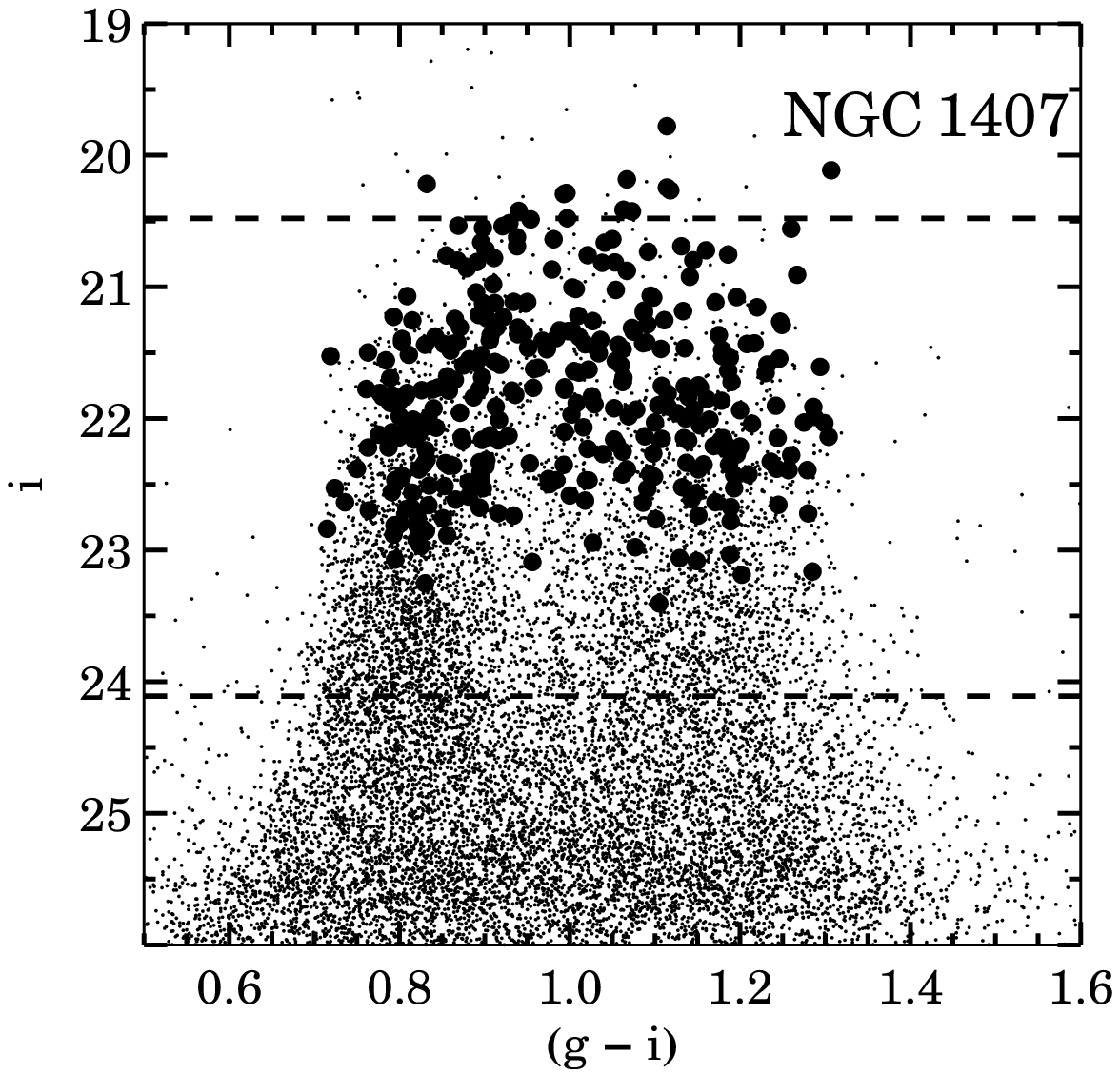} \\
\includegraphics[scale=.48]{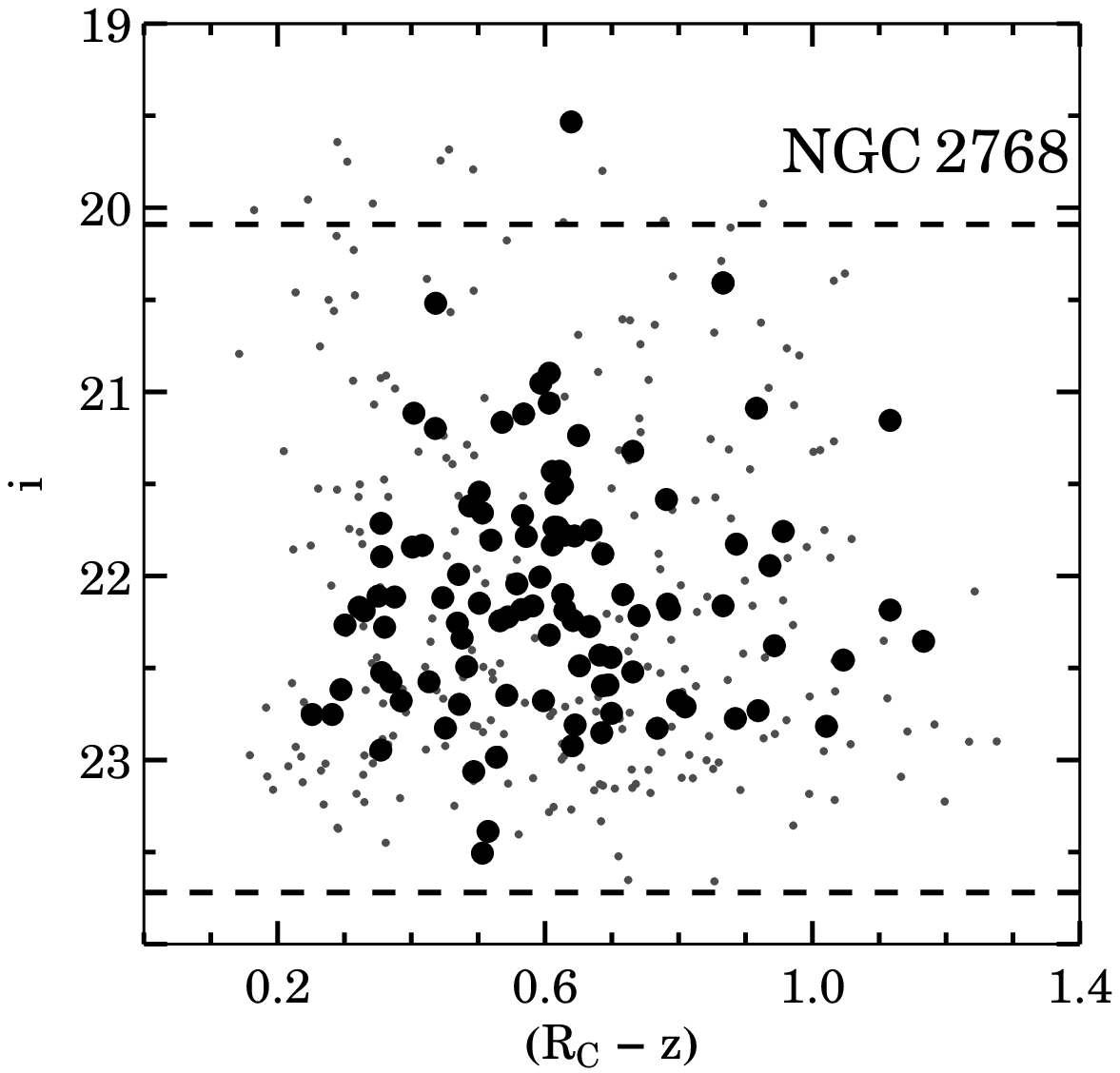} 
\includegraphics[scale=.48]{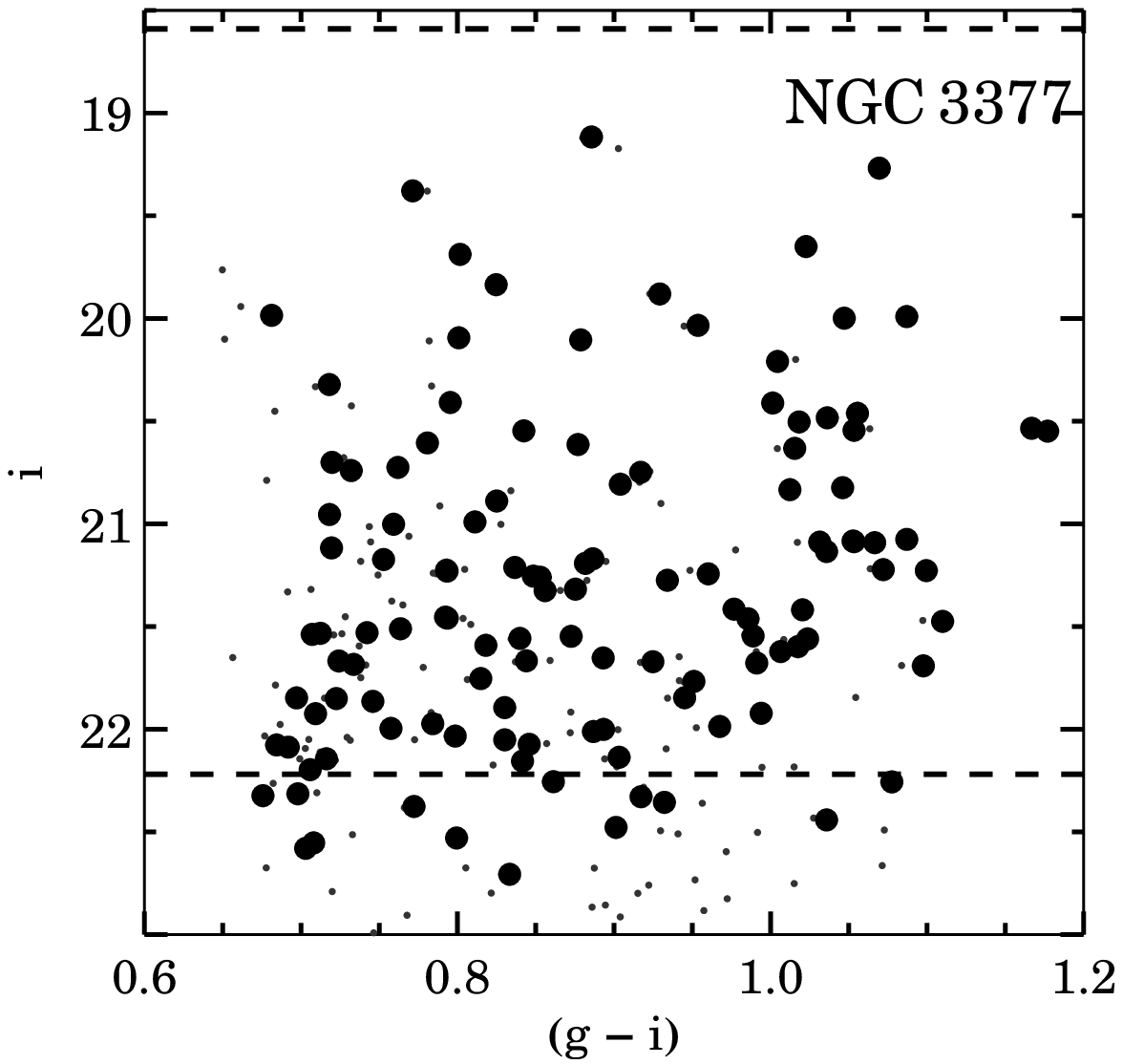} 
\includegraphics[scale=.48]{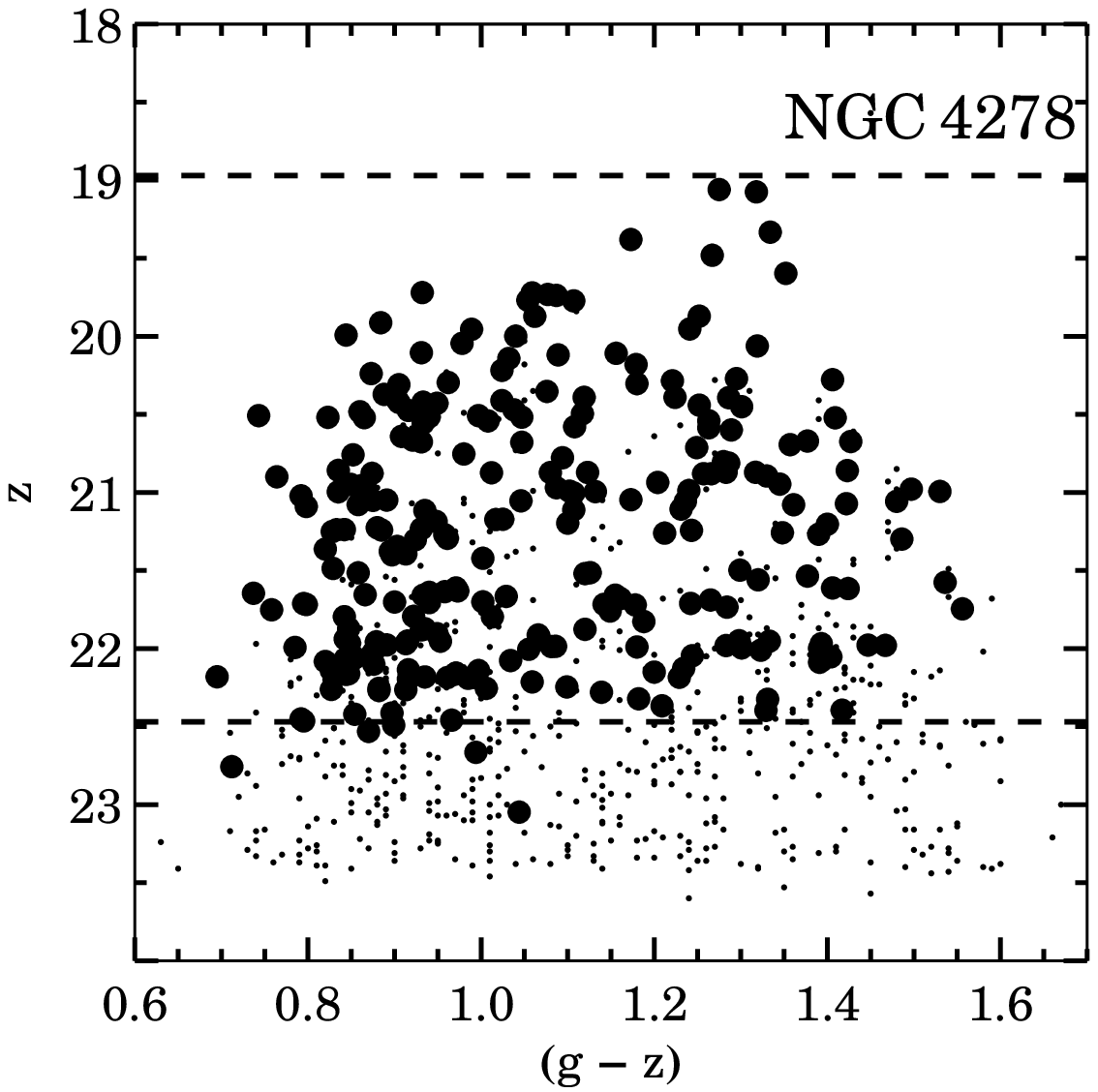} \\
\includegraphics[scale=.48]{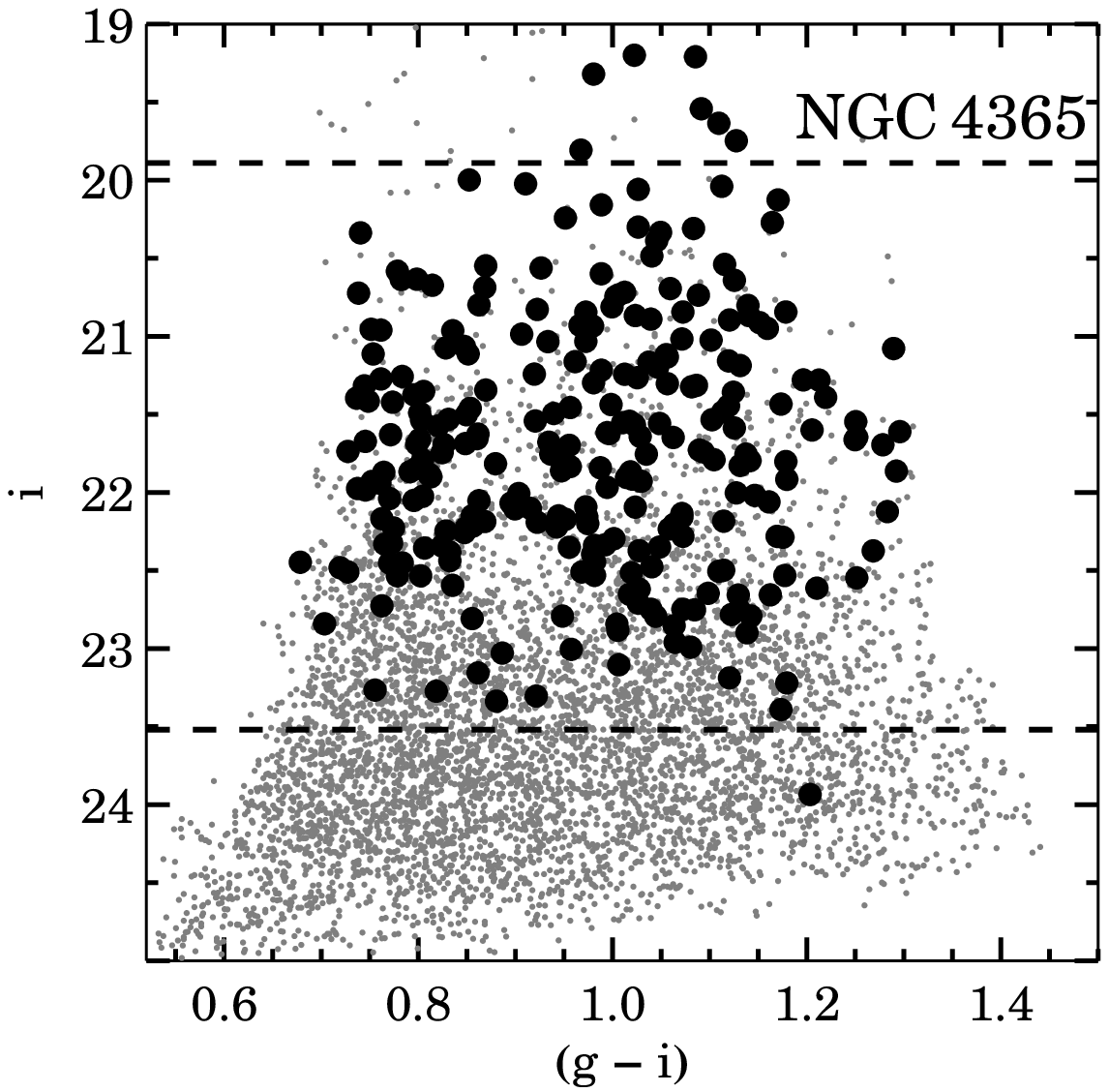} 
\includegraphics[scale=.48]{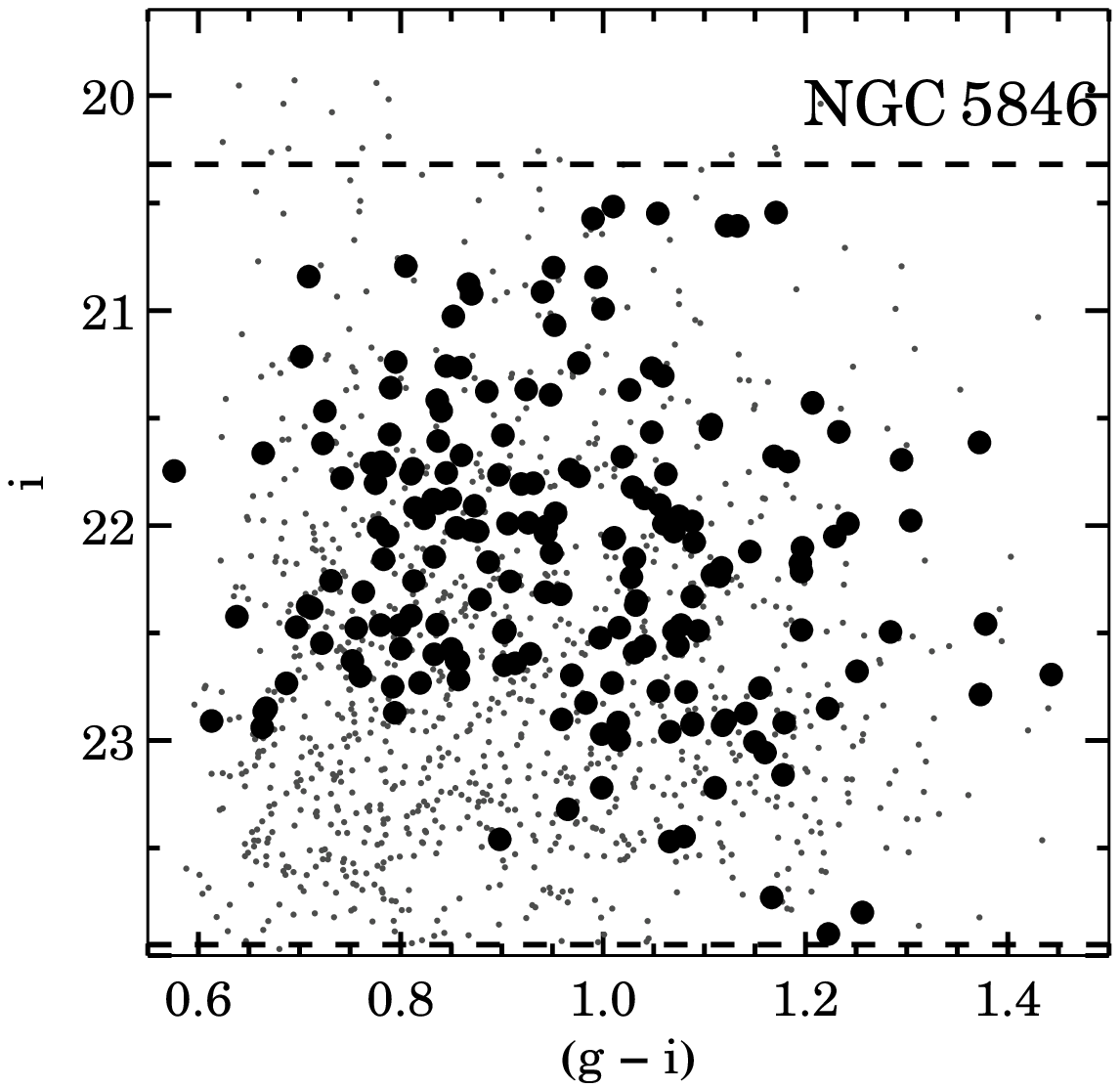} 
\includegraphics[scale=.48]{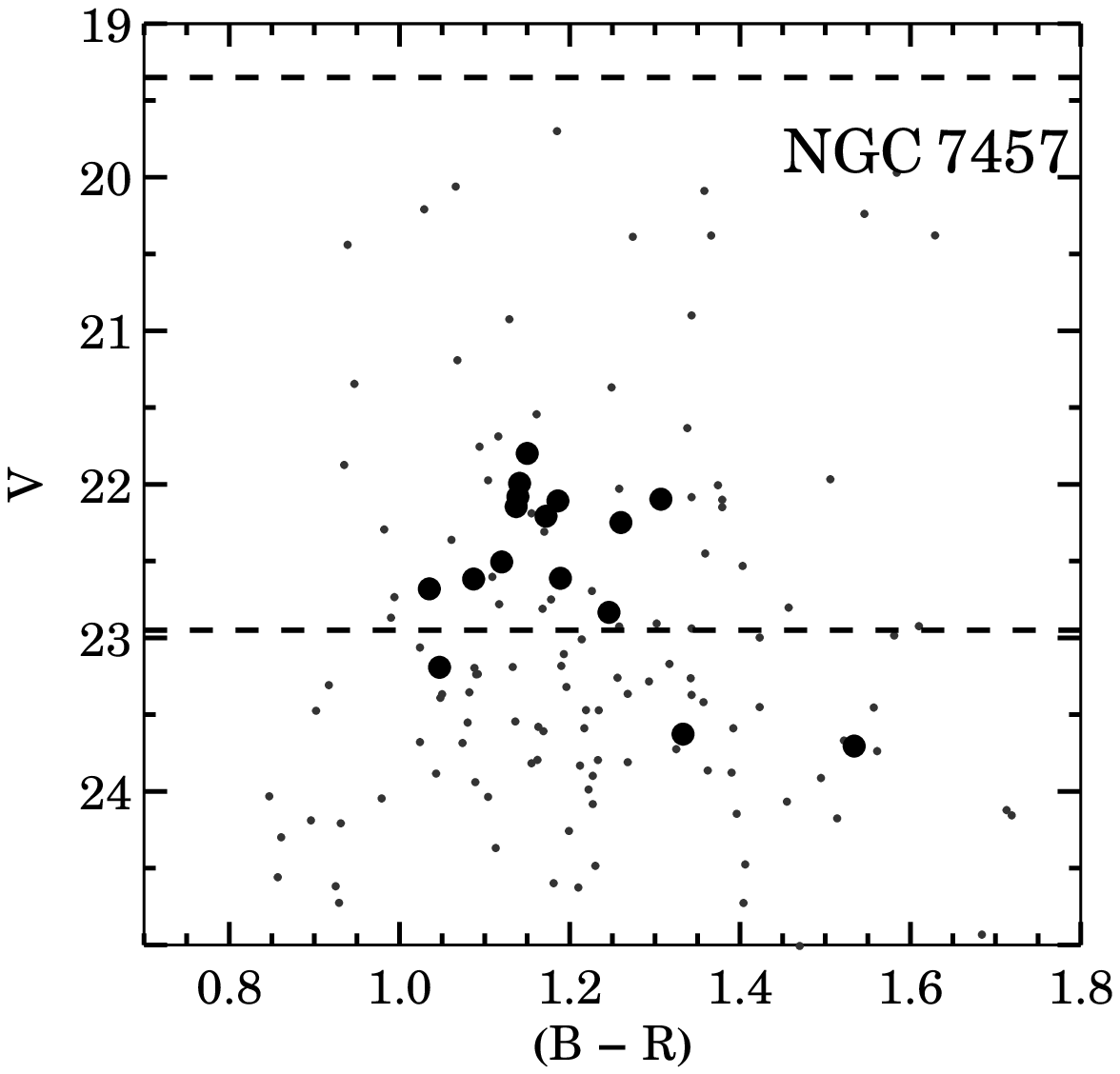} 
\caption{GC system colour-magnitude diagrams. Photometrically selected GCs and spectroscopically confirmed GCs are shown as small and large black points respectively. 
Upper and bottom dashed lines represent the $\omega$ 
Cen magnitude ($M_i \approx-11.6$) and the GC turnover magnitude ($M_{\rm TOM, i}\equiv-8$ mag) at the distances given in Table \ref{tab: survey_summary}.}
\label{fig:CM}
\end{figure*}

As far as the \textit{HST}/ACS imaging is concerned, it is worth noting that the diffraction limited quality of \textit{HST} imaging has the advantage of making extragalactic
GCs partially resolved for all of our galaxies. Therefore, the GC selection in our space-based imaging is also based on a size selection in addition to magnitude and 
colour-colour criteria (if available). In NGC~$3377$ and NGC~$4278$, for which only $g$ and $z$ imaging are available, we flag as GC candidates all the objects with
colour $0.7< (g - z) < 1.6$ and sizes $0.1<r_h <20$ pc. Such a choice is motivated by the clear drop off in the density of the objects outside the adopted colour cut, 
as observed in \citet{Blom}. For NGC~$2768$, we adopt the same size-cut as above and select objects within $0.4<(B - V)<1.2$ and $0.8<(V - I)<1.4$ and an upper 
magnitude of $M_I \approx -12$. In Figure~\ref{fig:CM}, the colour-magnitude diagrams of the GC candidates and of the spectroscopically confirmed GCs are shown. 

\subsection{GC colour bimodality}
\label{sec:GC bimodality}
To probe bimodality, we use a Kaye's Mixture Model algorithm (KMM, \citealt{Ashman94}) that returns colour peaks, variance and number of
objects in the detected subpopulations. KMM was run on all the $(g-i)$ distributions of the Subaru GC candidates brighter than the turnover 
magnitude $M_{\rm TOM}$. In this paper we adopt $M_{\rm TOM, i}\equiv8.00\pm0.03$, derived using the $i-I$ transformation from \citet{Faifer11} to calibrate 
the $I$ band TOM given by \citet{Kundu} into our photometric system. 

\begin{figure*}
\includegraphics[scale=.48]{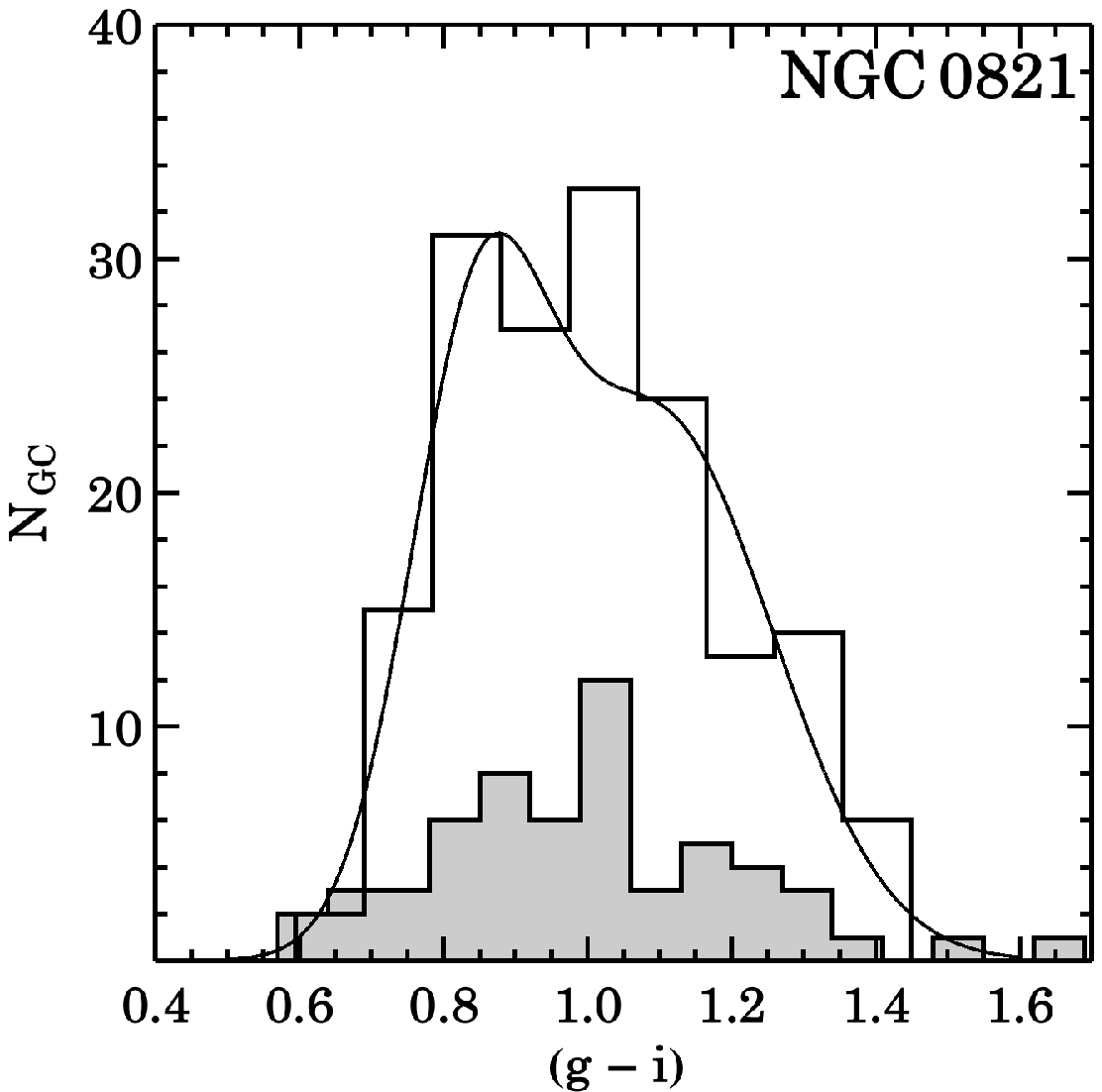} 
\includegraphics[scale=.48]{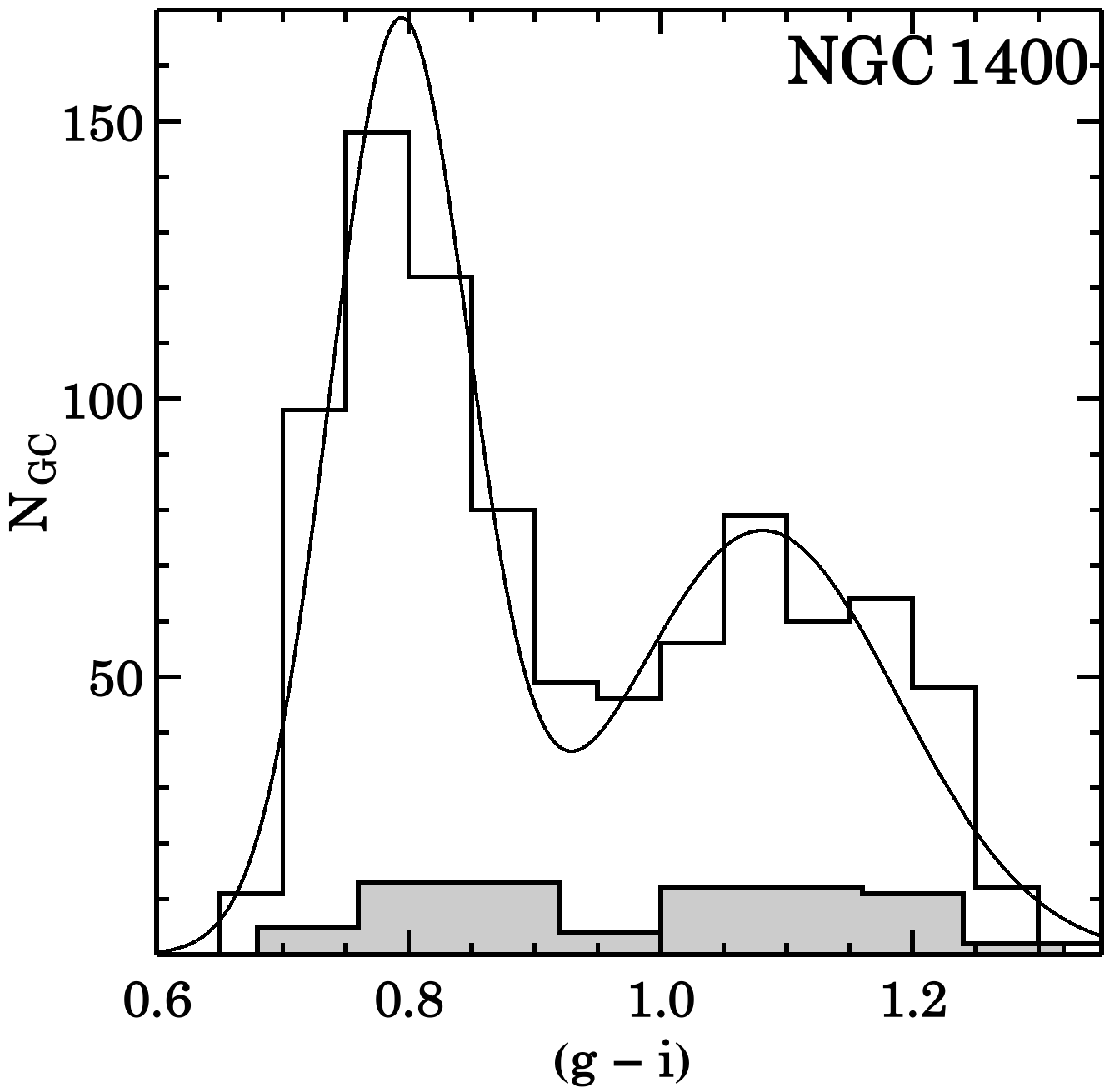} 
\includegraphics[scale=.48]{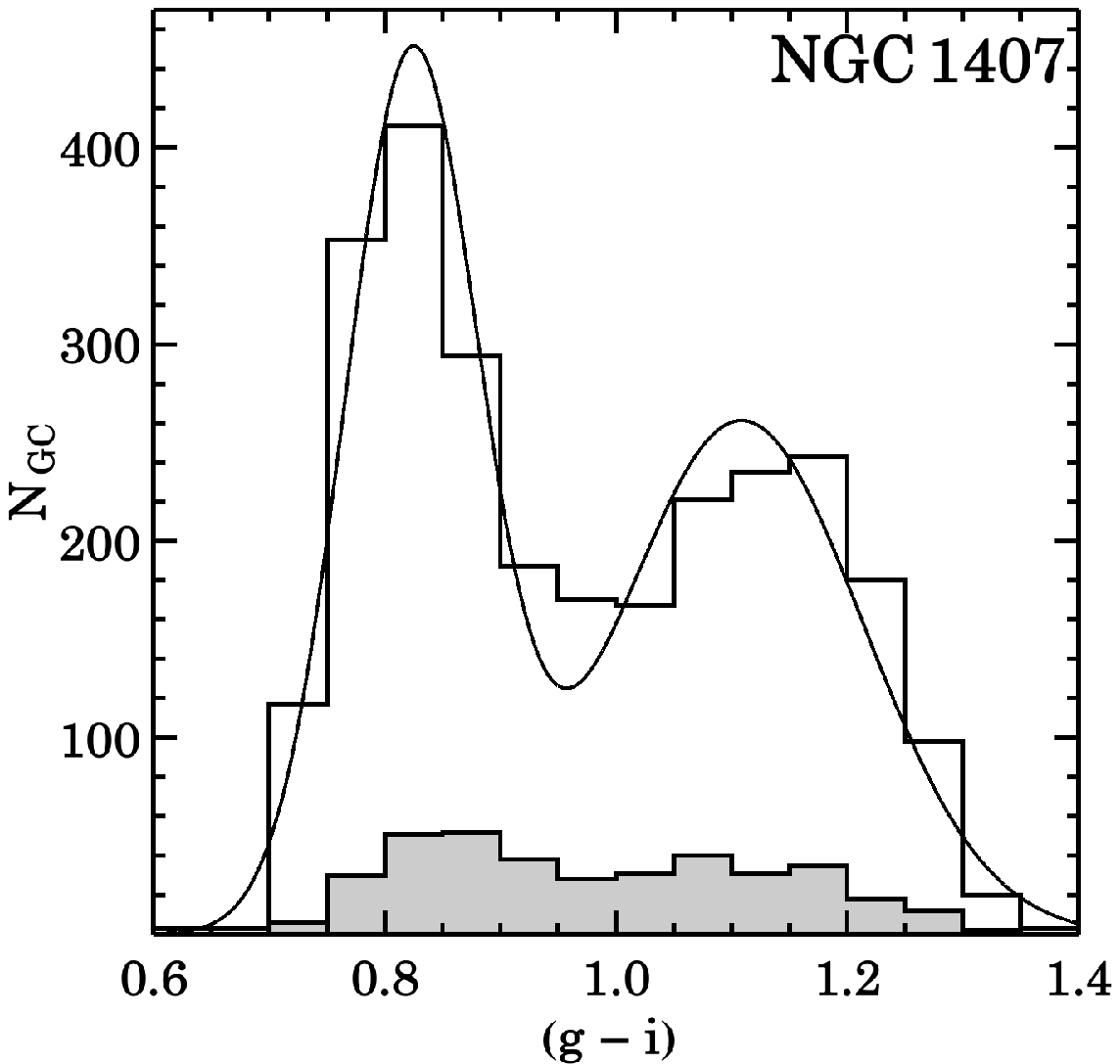} \\
\includegraphics[scale=.48]{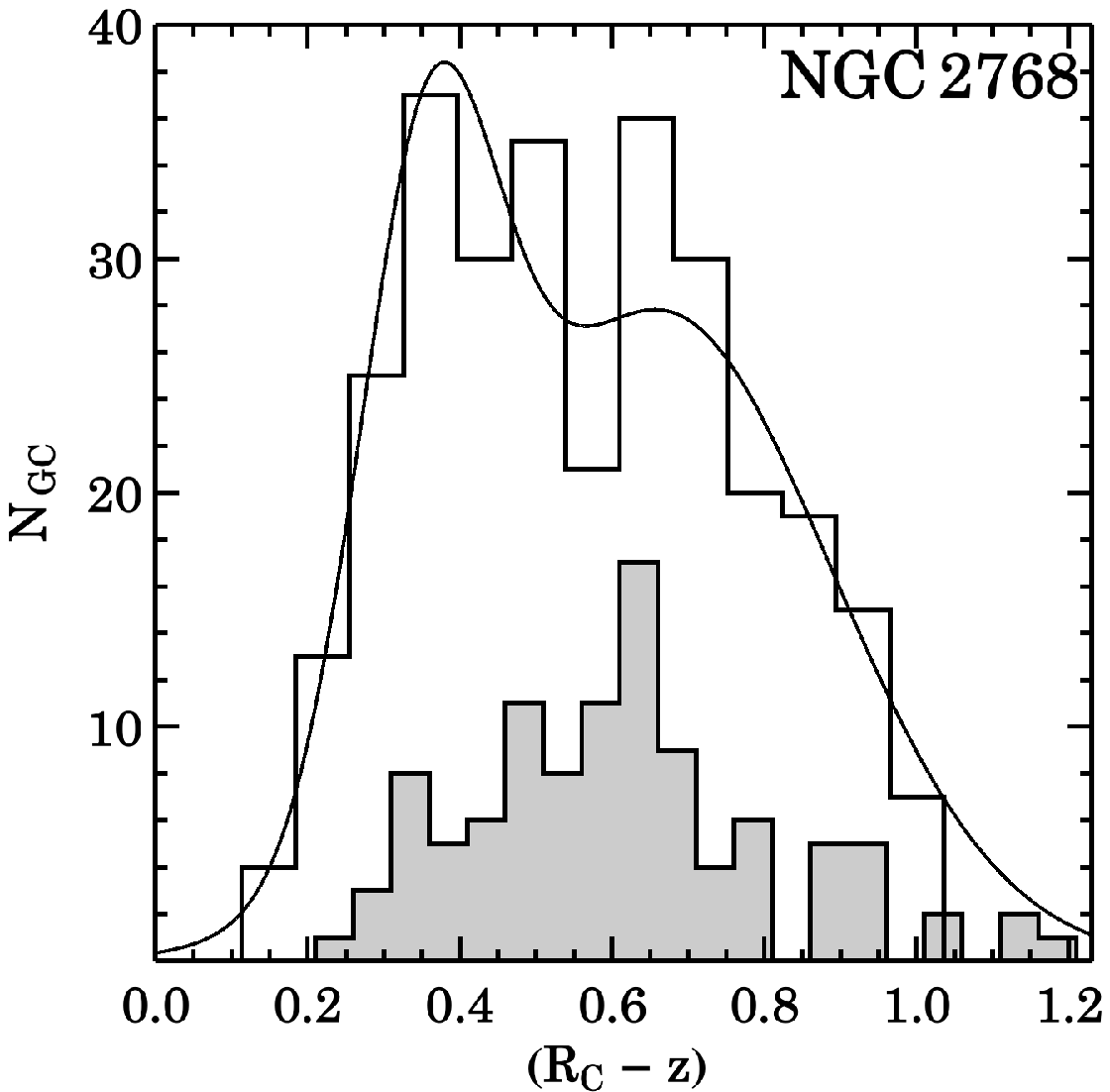} 
\includegraphics[scale=.48]{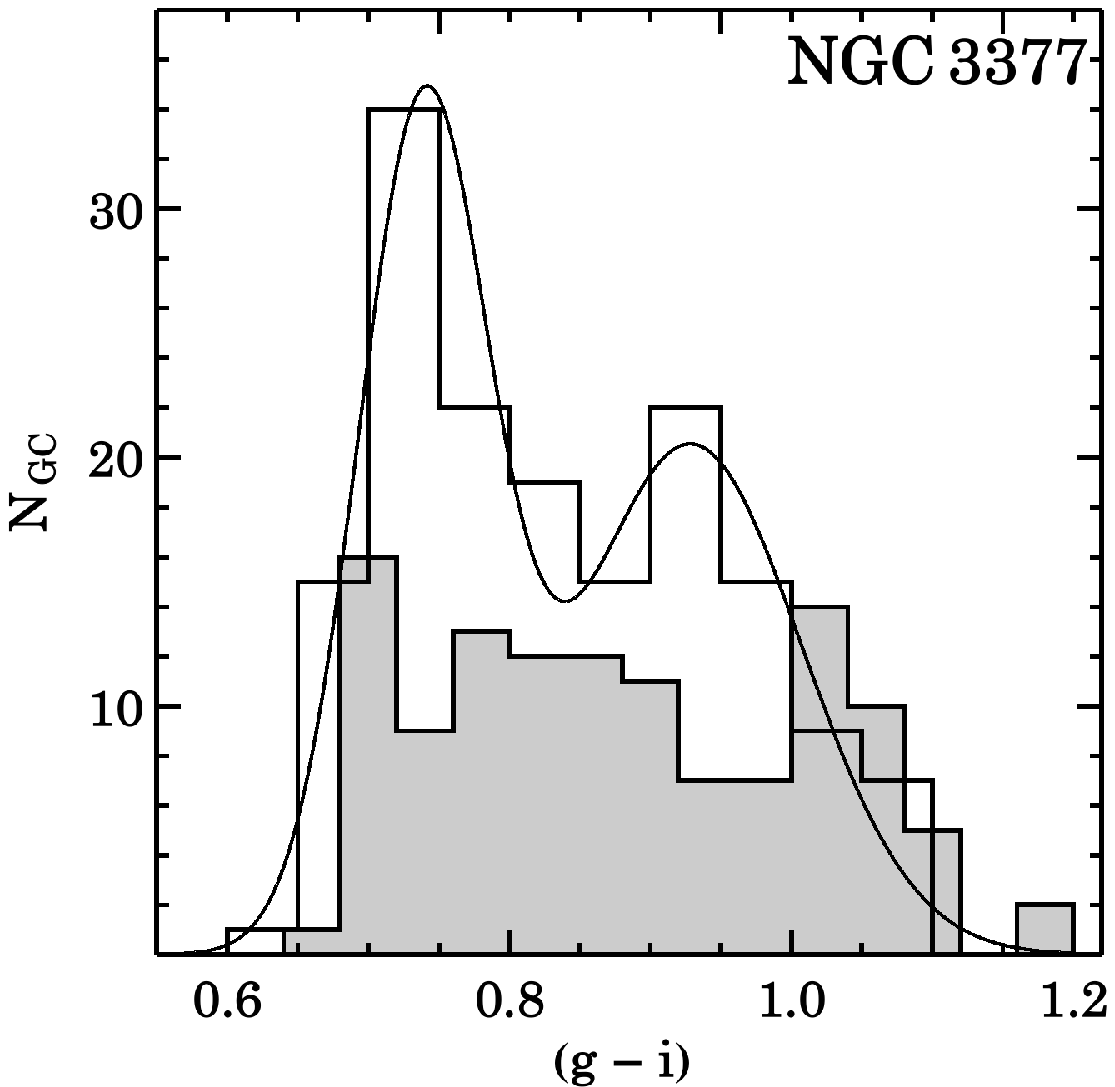} 
\includegraphics[scale=.48]{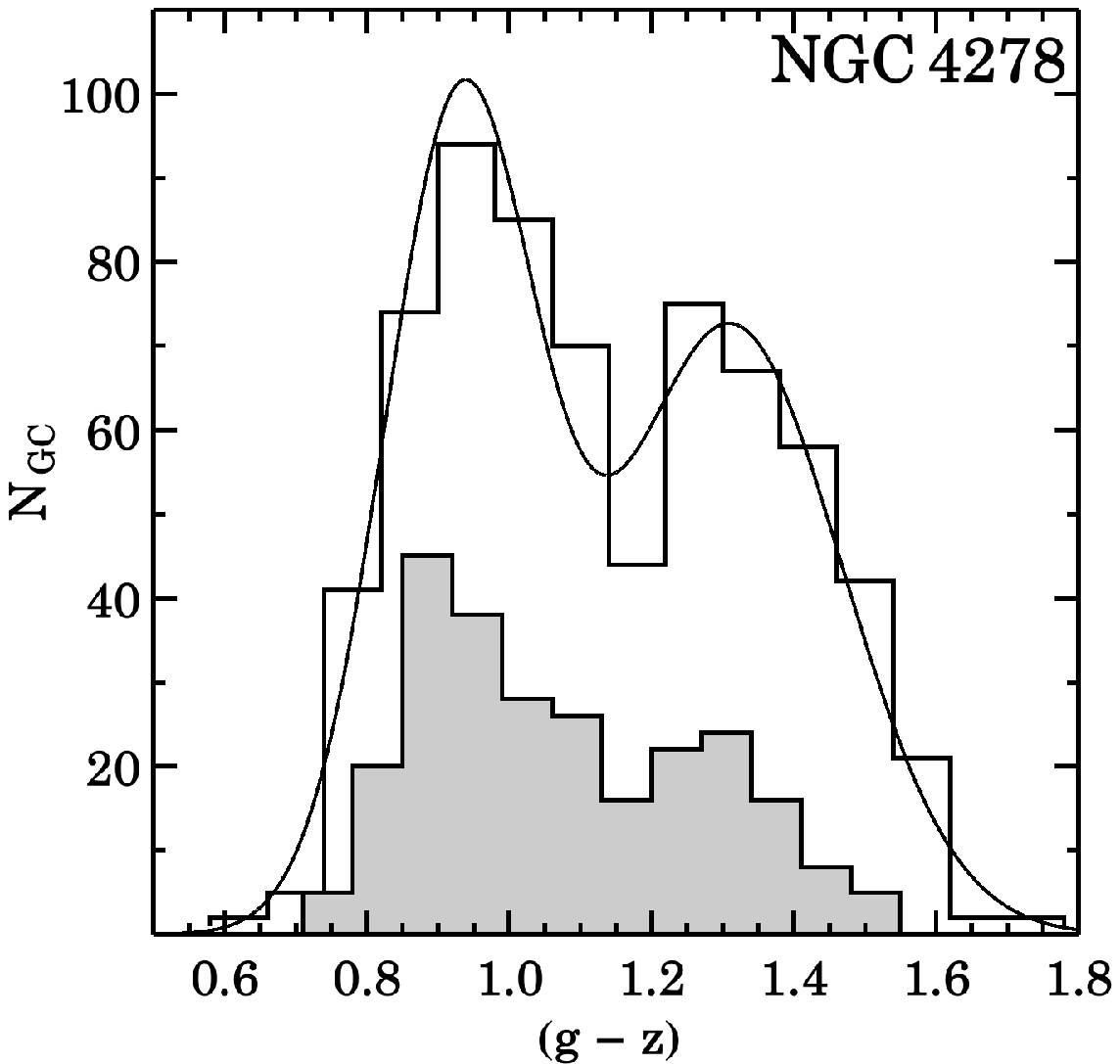} \\
\includegraphics[scale=.48]{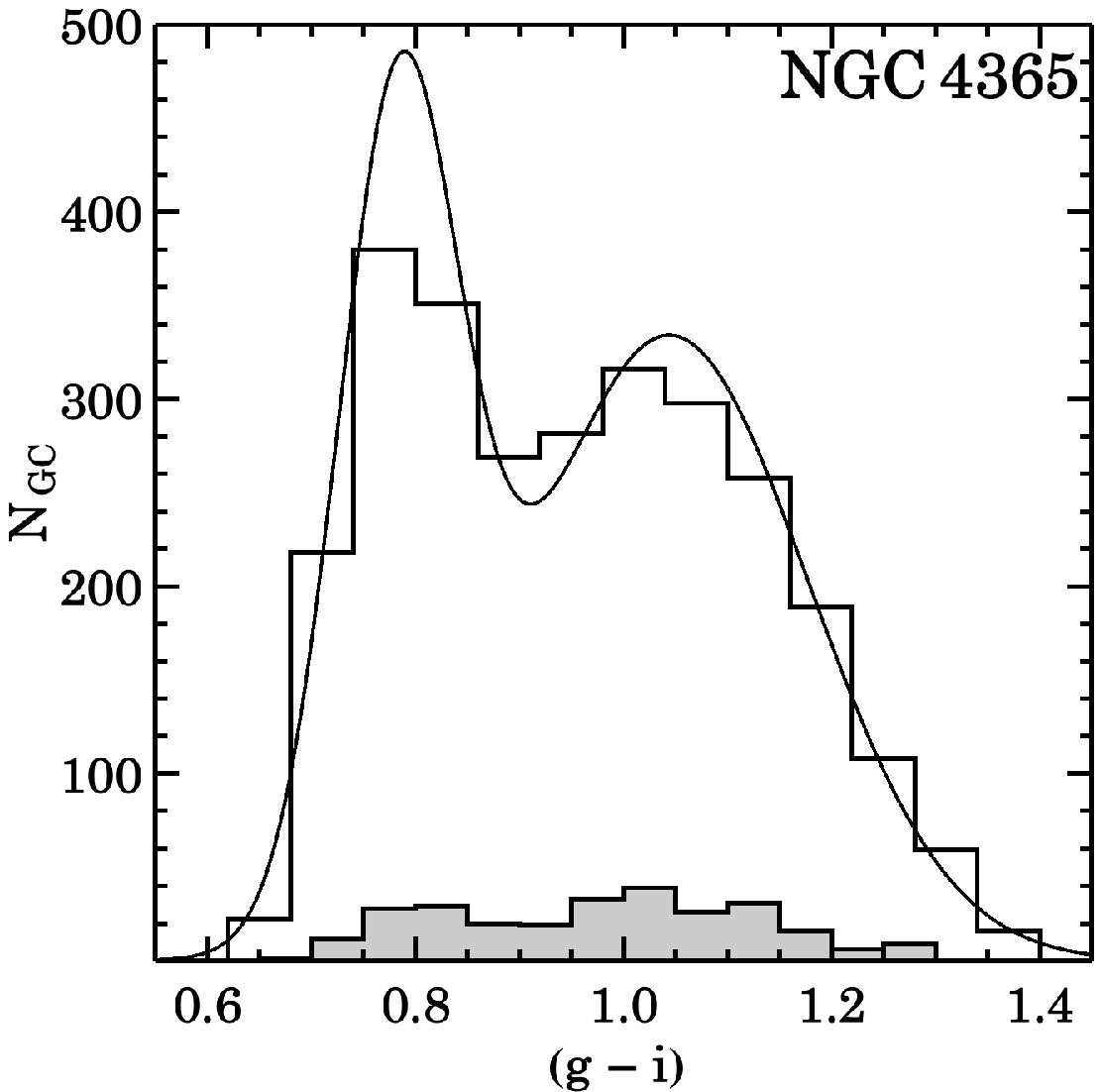} 
\includegraphics[scale=.48]{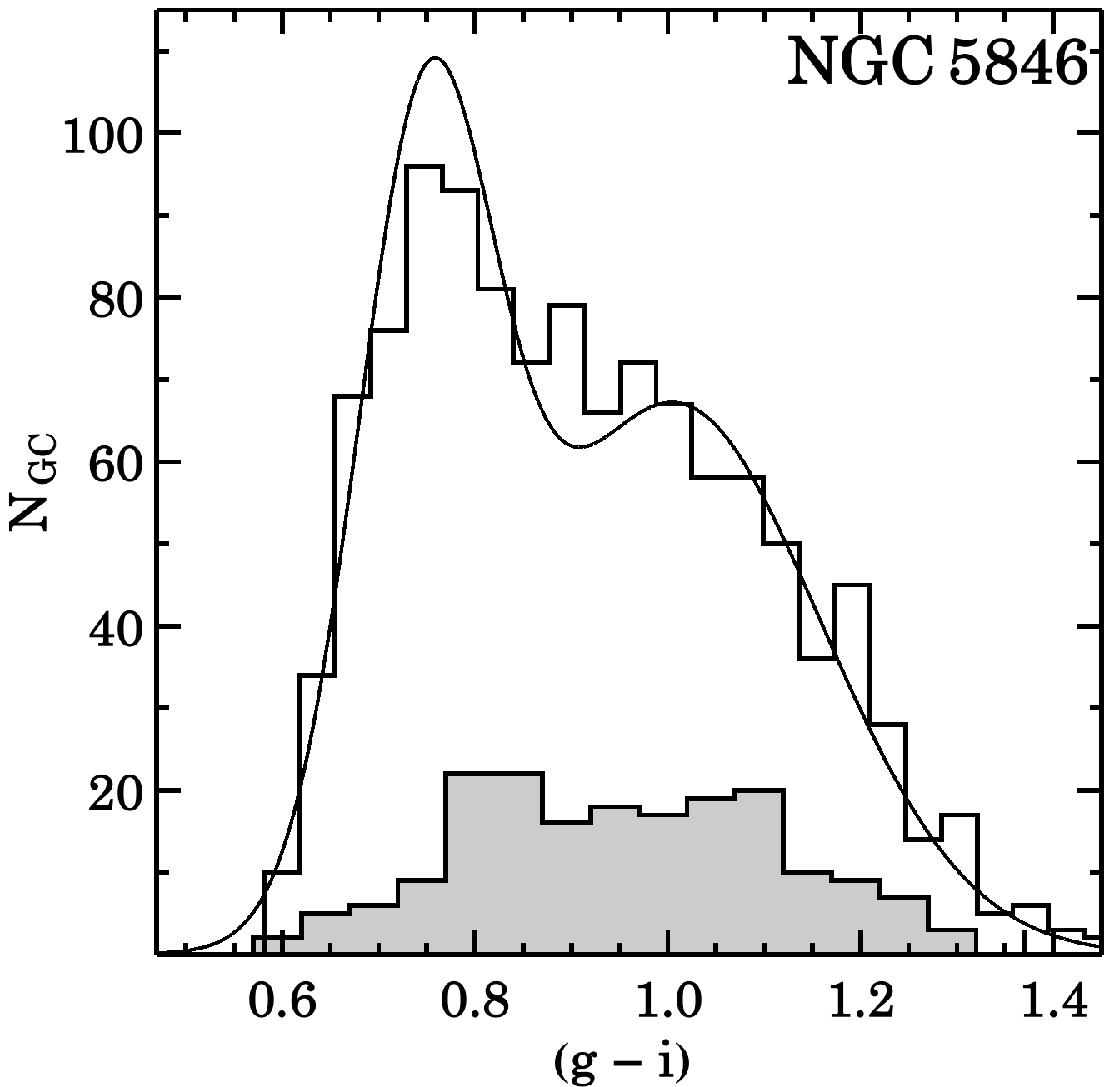} 
\includegraphics[scale=.48]{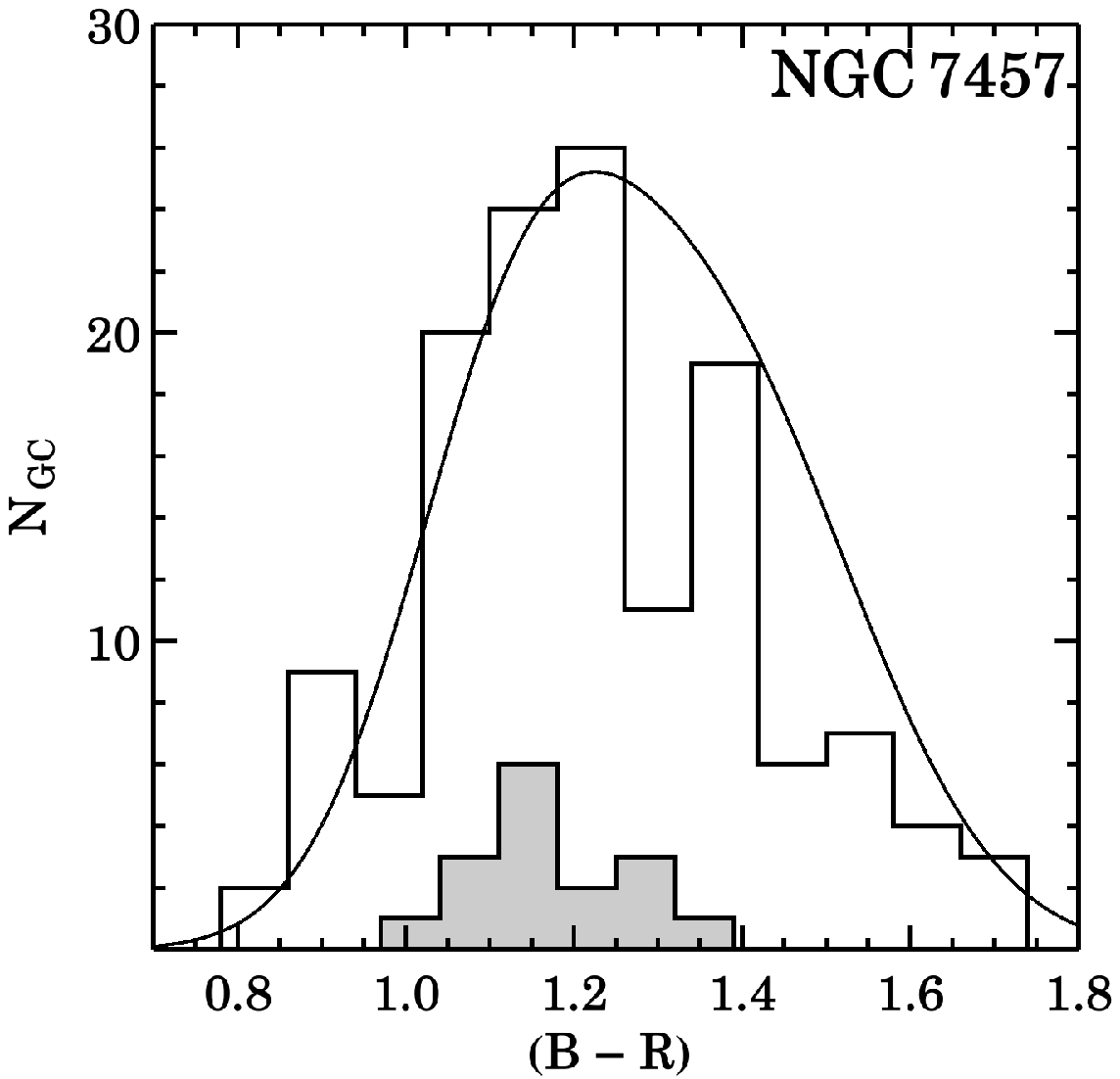} 
\caption{Colour histograms of our galaxy sample. GC candidates and spectroscopically confirmed GCs are shown as empty and grey 
filled histograms respectively. For NGC~$1407$, we only show GCs with $i<23$. Solid lines show the sum of the KMM fits of the two 
subpopulations. No significant bimodality was detected in NGC~$7457$.  The dearth of red GCs in NGC~$3377$ is due to the fact that 
we have excluded all the GC candidates within $1$ arcmin from the centre. The magnitude of these objects turned out to be corrupted by the underlying diffuse stellar light.}
\label{SpectroPhoto_Matrix}
\end{figure*}

For NGC~$2768$, NGC~$7457$ and NGC~$4278$, we study the $(R_{\rm C} - z)$, $(B - V)$ and $(g-z)$ colour distributions respectively. 
We perform a bisector fit \citep{Feigelson} to the bright GCs in common between the \textit{HST} and Subaru images, in order to convert \textit{HST} magnitudes of
the spectroscopically confirmed GCs into the respective Subaru photometric system. For NGC~$4278$, given the wide spatial coverage and better quality 
of its \textit{HST} data, we transform Subaru magnitudes of the confirmed GCs into the \textit{HST} photometric system. 
Results of the KMM analysis are presented in Table \ref{tab: KMM}. We find that eight out of nine galaxies show significant GC colour bimodality. 
GCs are divided into blue and red according to the local minimum of the best fit Gaussians in Figure~\ref{SpectroPhoto_Matrix}. 
\begin{table}
\centering
\begin{tabular}{@{}l c c c c l c}
\hline
Galaxy ID & $\mu_{\rm blue}$ & $\mu_{\rm red}$ & $\sigma_{\rm red}$ & $\sigma_{\rm red}$ & p-value & colour  \\
 \hline
\hline   
NGC~$0821$ & $0.84$ & $1.10$ & $0.09$  &  $0.15$ & $0.034$ & $(g-i)$ \\
NGC~$1400$ & $0.79$ & $1.05$ & $0.05$  &  $0.11$ & $10^{-10}$ & $(g-i)$ \\
NGC~$1407$   & $0.85$ & $0.77$ & $0.05$  & $0.1$&   $10^{-10}$& $(g-i)$ \\
NGC~$2768$  & $0.41$ & $1.10$ & $0.15$  & $0.22$&   $10^{-5}$& $(R_C-z)$ \\
NGC~$3377$  & $0.73$ & $ 0.93$ & $0.05$  & $0.08$&   $10^{-6}$& $(g-i)$\\
NGC~$4278$   & $0.93$ & $1.31$ & $0.10$  & $ 0.16$&   $10^{-10}$& $(g-z)$\\
NGC~$4365$   & $0.78$ & $1.04$ & $0.05$  & $0.13$&   $10^{-10}$& $(g-i)$\\
NGC~$5846$   & $0.74$ & $1.00$ & $0.15$  & $0.06$&   $10^{-5}$& $(g-i)$\\
NGC~$7457$   & $1.12$ & $1.36$ & $0.13$  & $0.17$&   $0.311$& $(B-R)$\\
\hline
\end{tabular}
\caption{KMM results. For each galaxy, $\mu_{\rm blue}$ and $\mu_{\rm red}$ represent the mean of the blue and red peak respectively, 
whereas $\sigma_{\rm blue}$ and $\sigma_{\rm red}$ are the Gaussian $\sigma$ for each peak. The last column gives the p-value, that is the
confidence level with which the hypothesis of an unimodal colour distribution can be rejected (larger confidence for smaller p-values). The last column gives 
the colour distribution used in the analysis. NGC~$7457$ is the only galaxy without significant bimodality detected.}
\label{tab: KMM}
\end{table}

\subsection{GC spatial distribution}
\label{sec:GC spatial distribution}
We construct projected surface density profiles for the GC subpopulations of our galaxy sample. GC candidates brighter than the turnover magnitude were
binned in circular annuli and corrected for areal incompleteness. The Subaru dataset was supplemented with \textit{HST} photometry, in order to 
probe the GC surface density in the inner regions. An important caveat to bear in mind is that we do not correct the Subaru dataset for magnitude incompleteness.
This effect becomes important especially for objects fainter than $i\sim 23.5$ in the innermost regions ($\sim 1$ arcmin) where the detection of sources is inhibited 
by the surface brightness contribution of the host galaxy \citep{Faifer11}. Conversely, the completeness of the \textit{HST} photometry in $z$ $(V)$ band is typically 
above $90$ $(70)$ per cent at the turnover magnitude even in the innermost regions, respectively \citep{Larsen01,Jordan07}.
Therefore, we select all the GCs brighter than the turnover magnitude (see Figure~\ref{fig:CM}) and we use circular radial bins in common between \textit{HST} and Subaru to 
correct and adjust the Subaru GC surface density to that of the \textit{HST} data. 
Although the matching between \textit{HST} and Subaru data points is arbitrary, this approach preserves the relative slopes of the spatial distribution of the two subpopulations
and allows us to verify whether or not blue and red GC subpopulations show different spatial distributions around the host galaxy as found from previous authors \citep[e.g.,][]{Bassino}.
\begin{figure*}
\includegraphics[scale=.48]{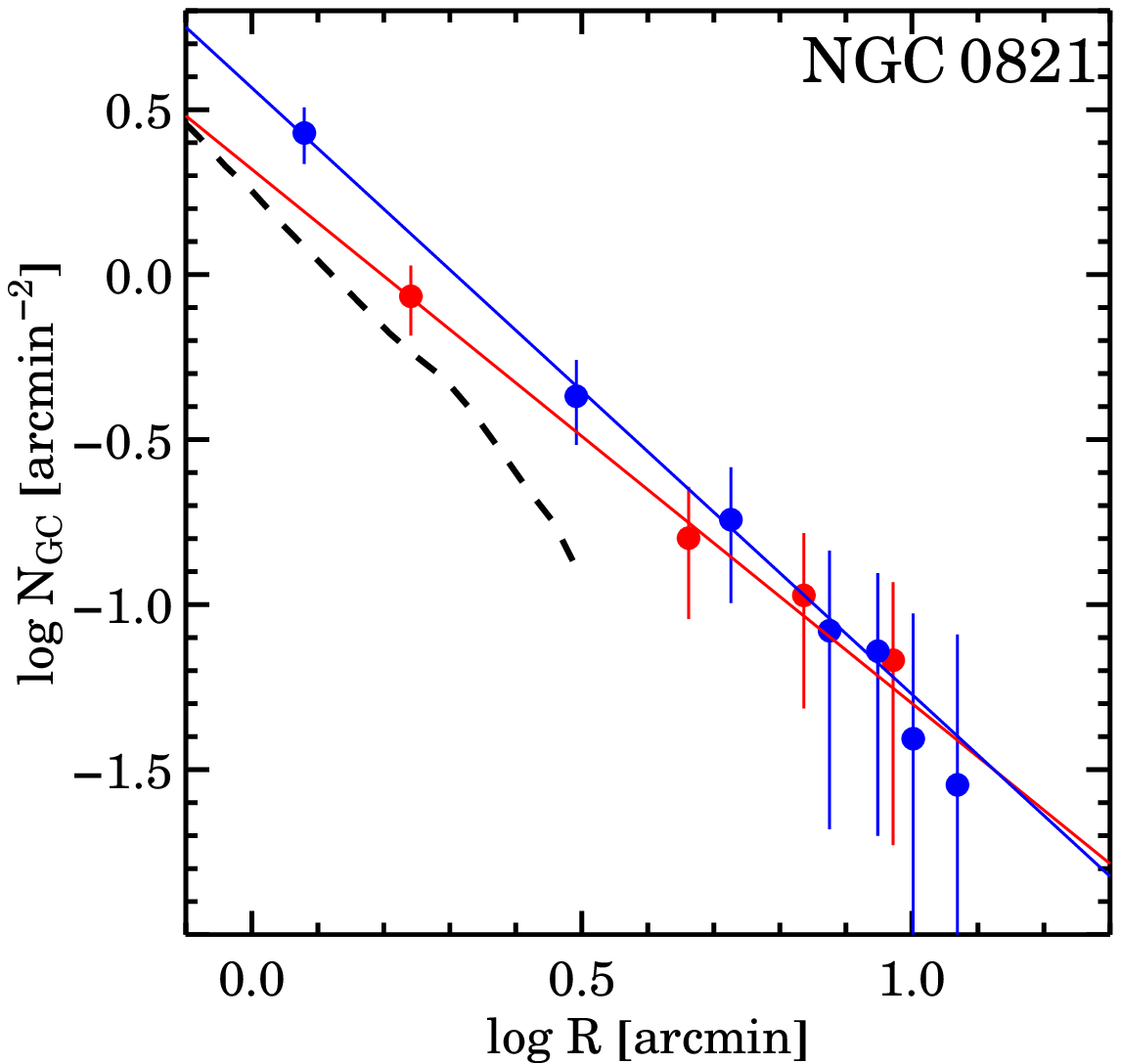} 
\includegraphics[scale=.48]{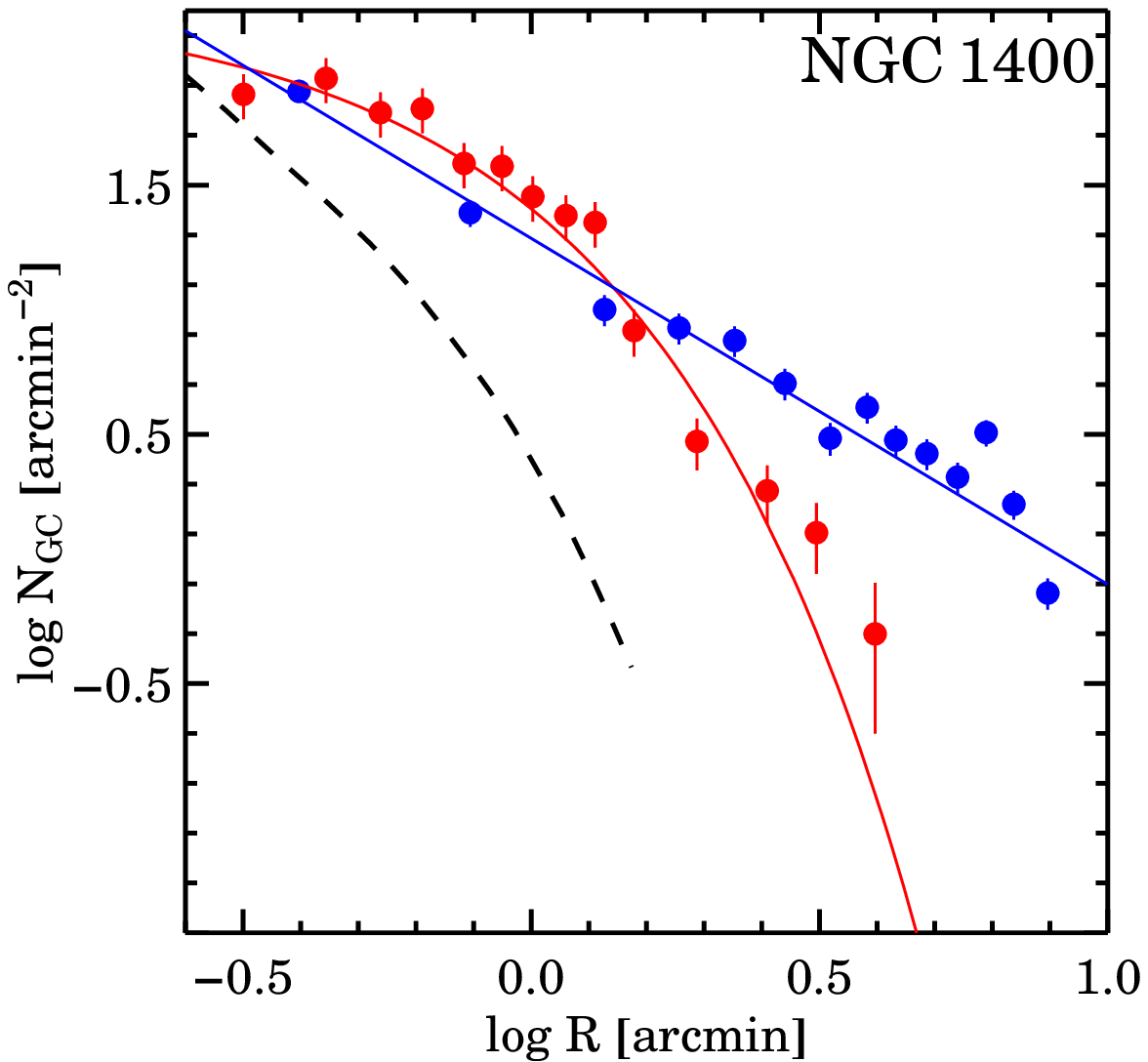} 
\includegraphics[scale=.48]{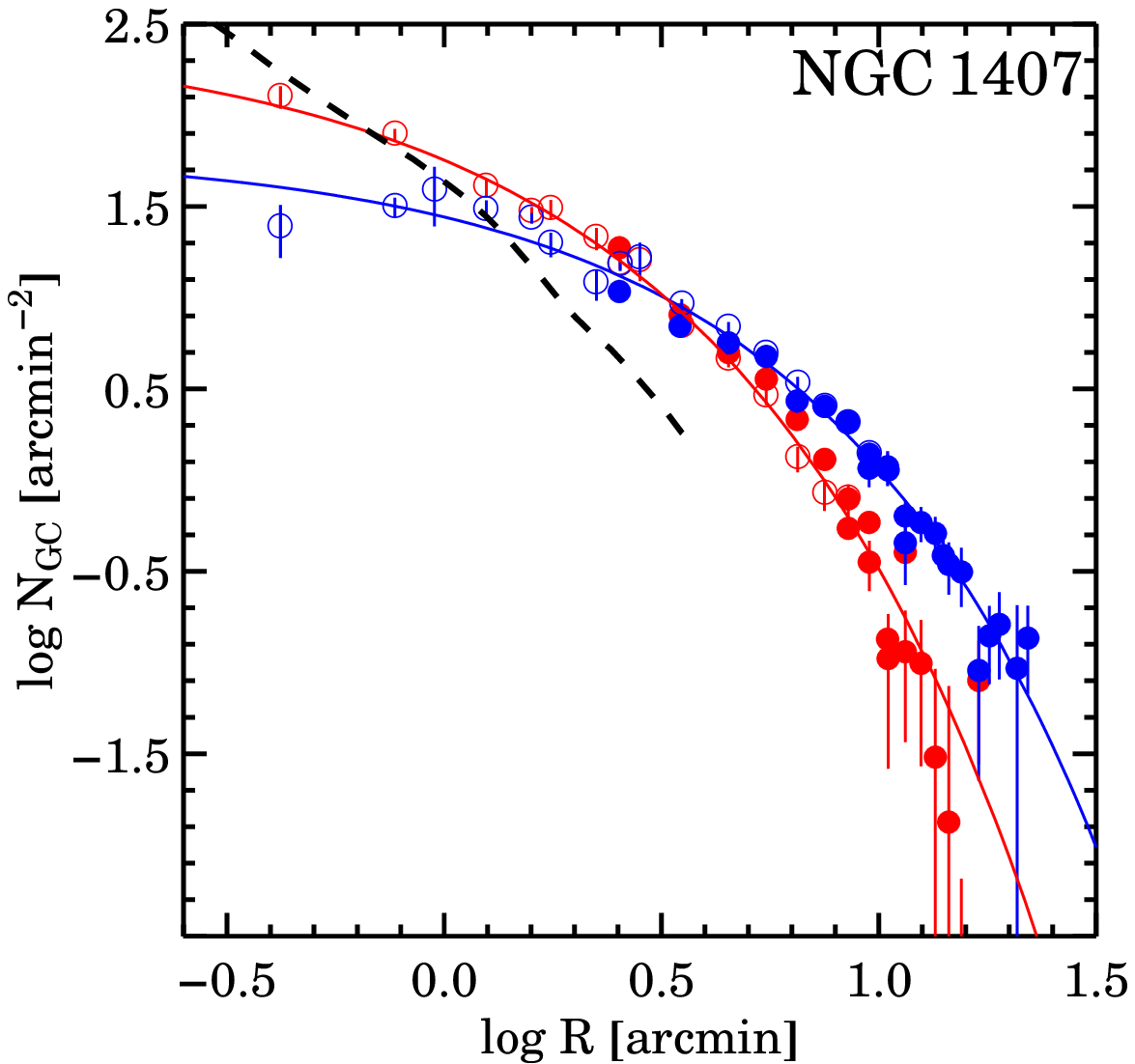} \\
\includegraphics[scale=.48]{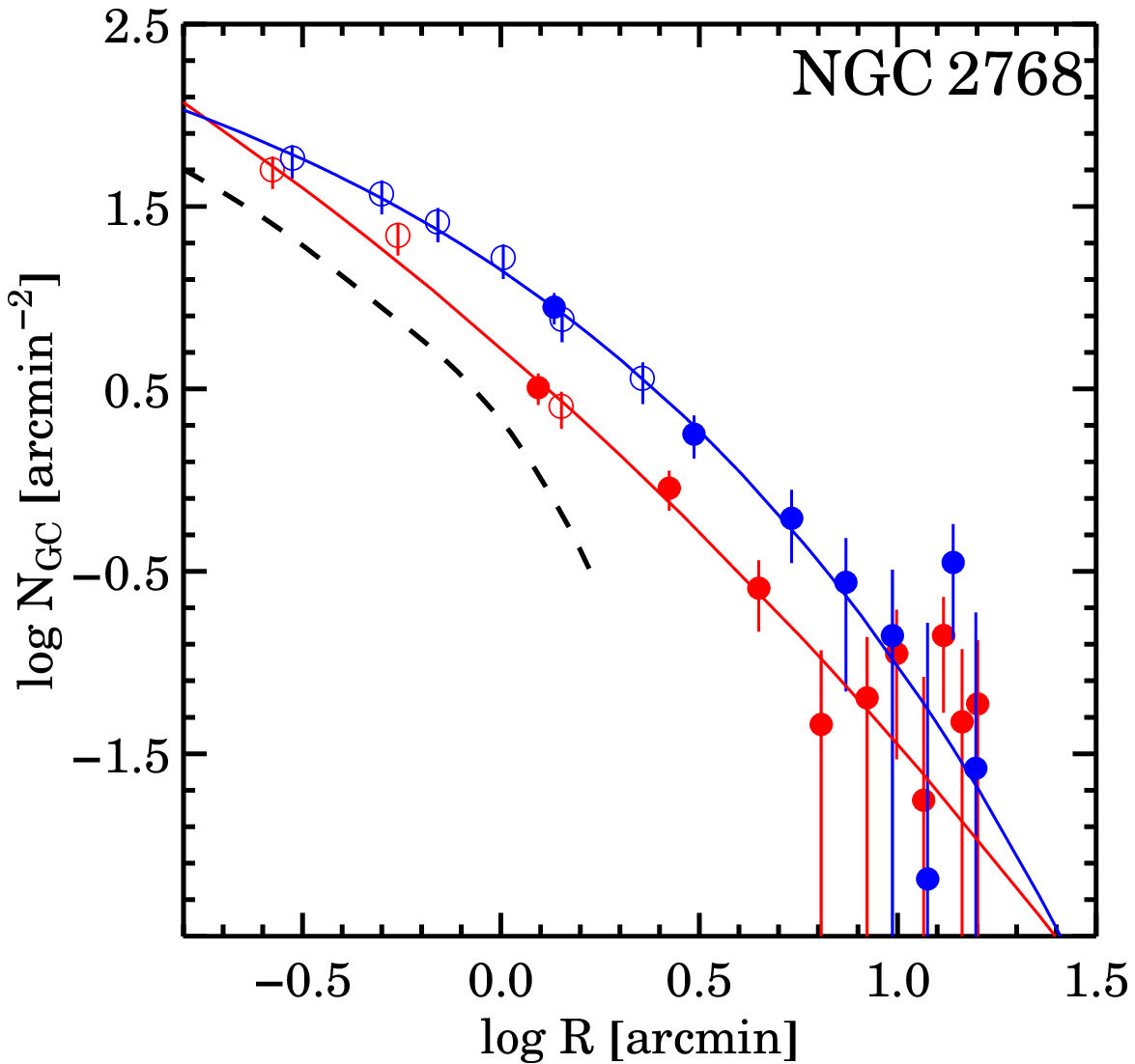} 
\includegraphics[scale=.48]{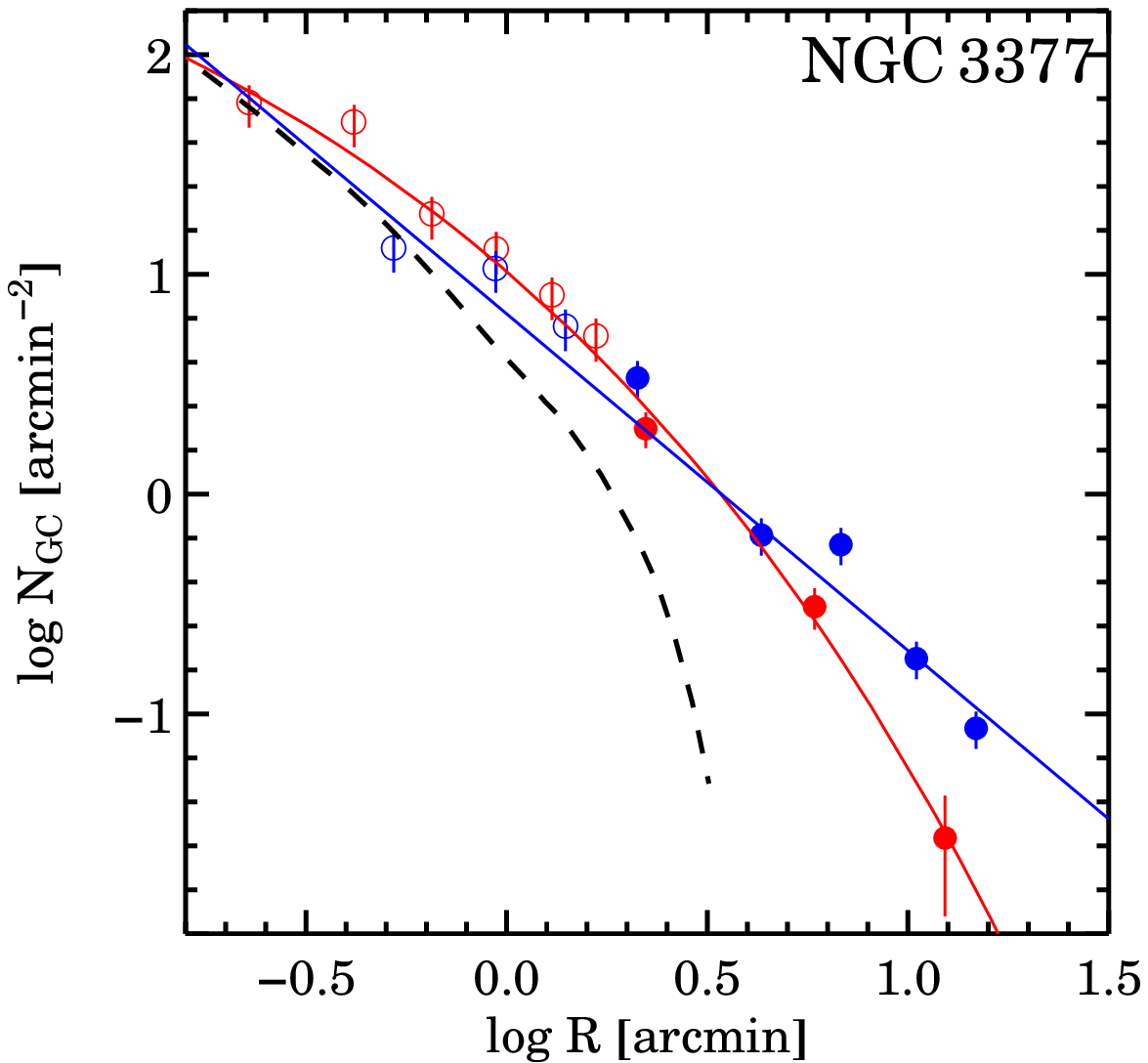} 
\includegraphics[scale=.48]{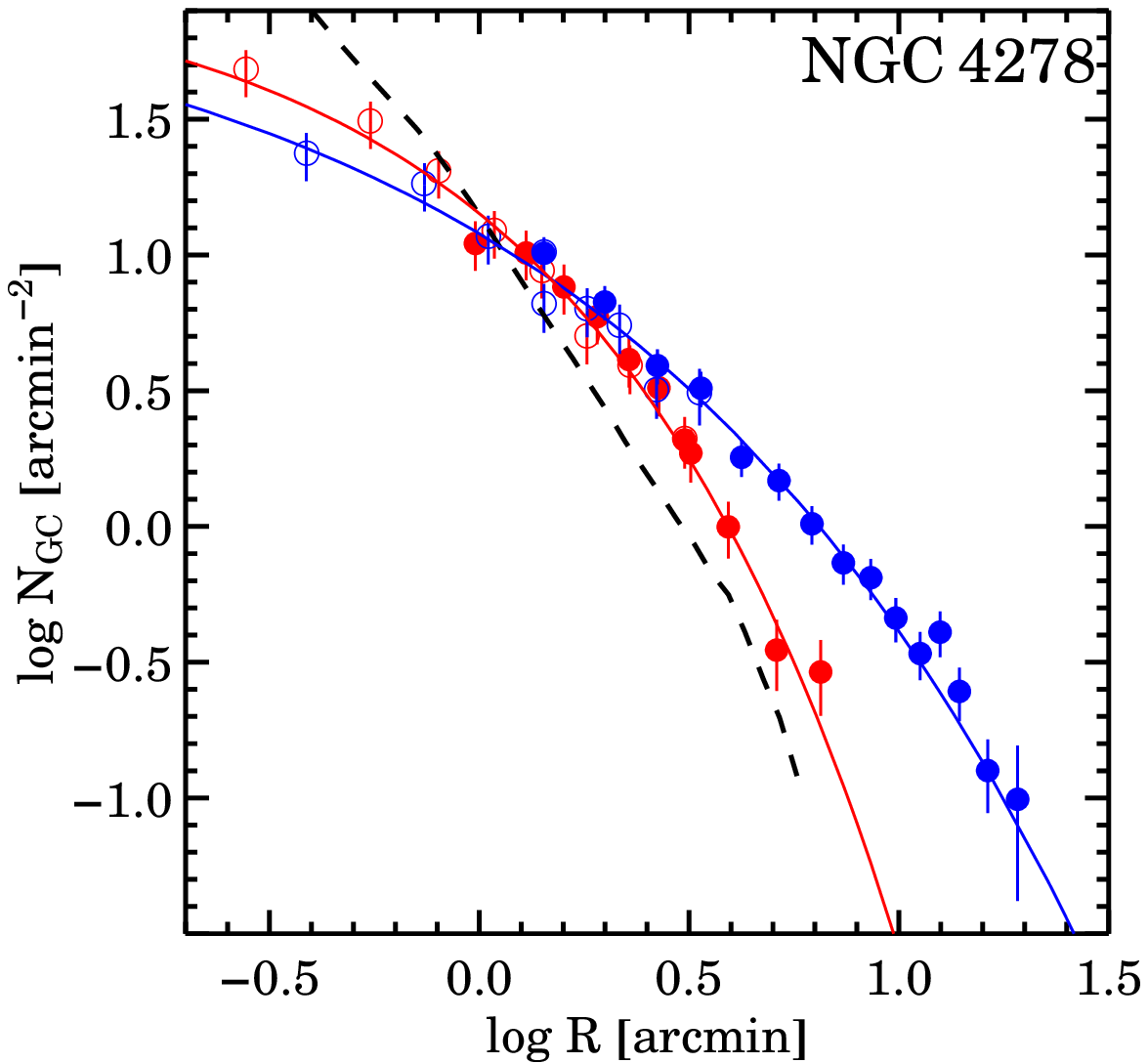} \\
\includegraphics[scale=.48]{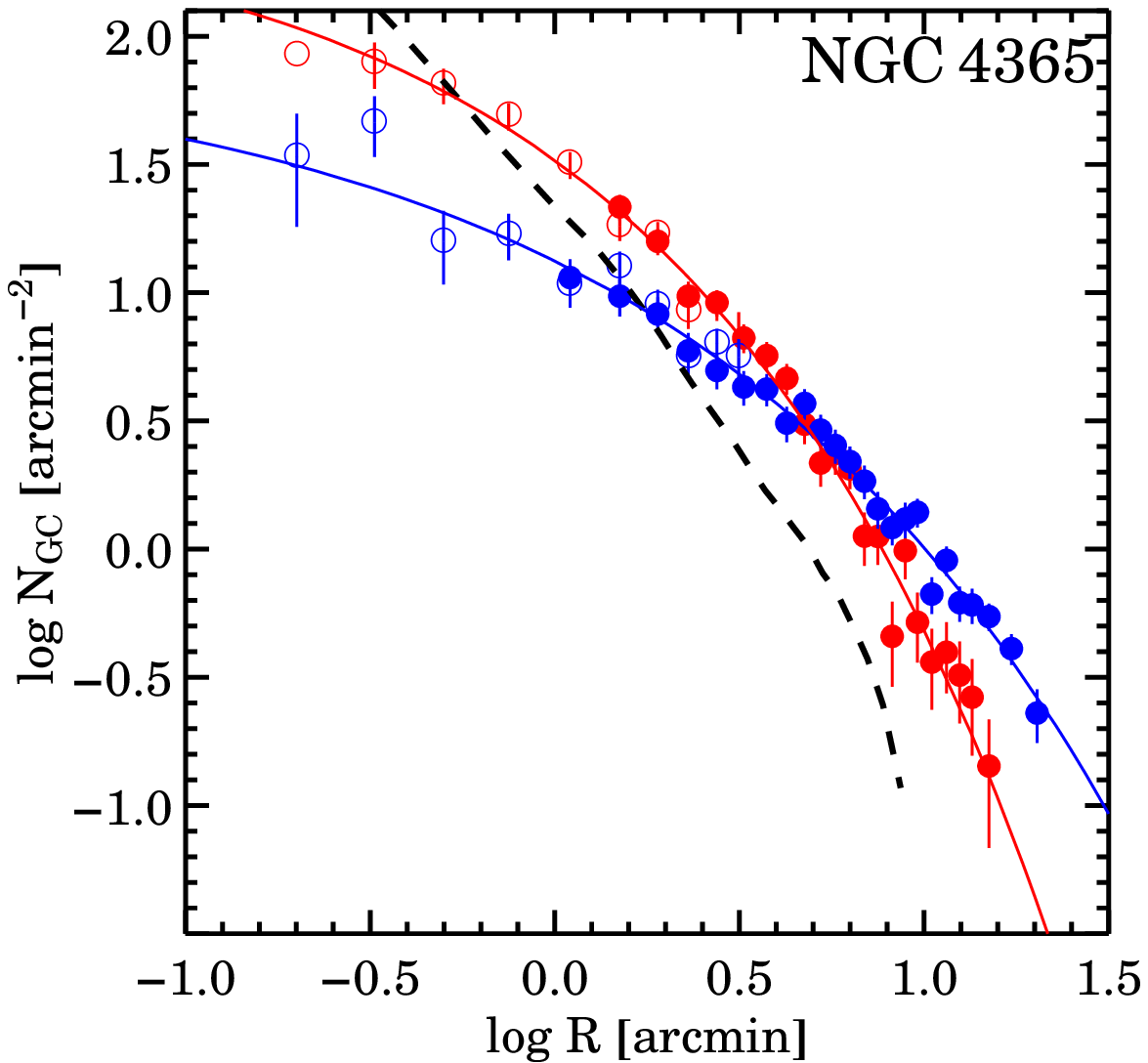} 
\includegraphics[scale=.48]{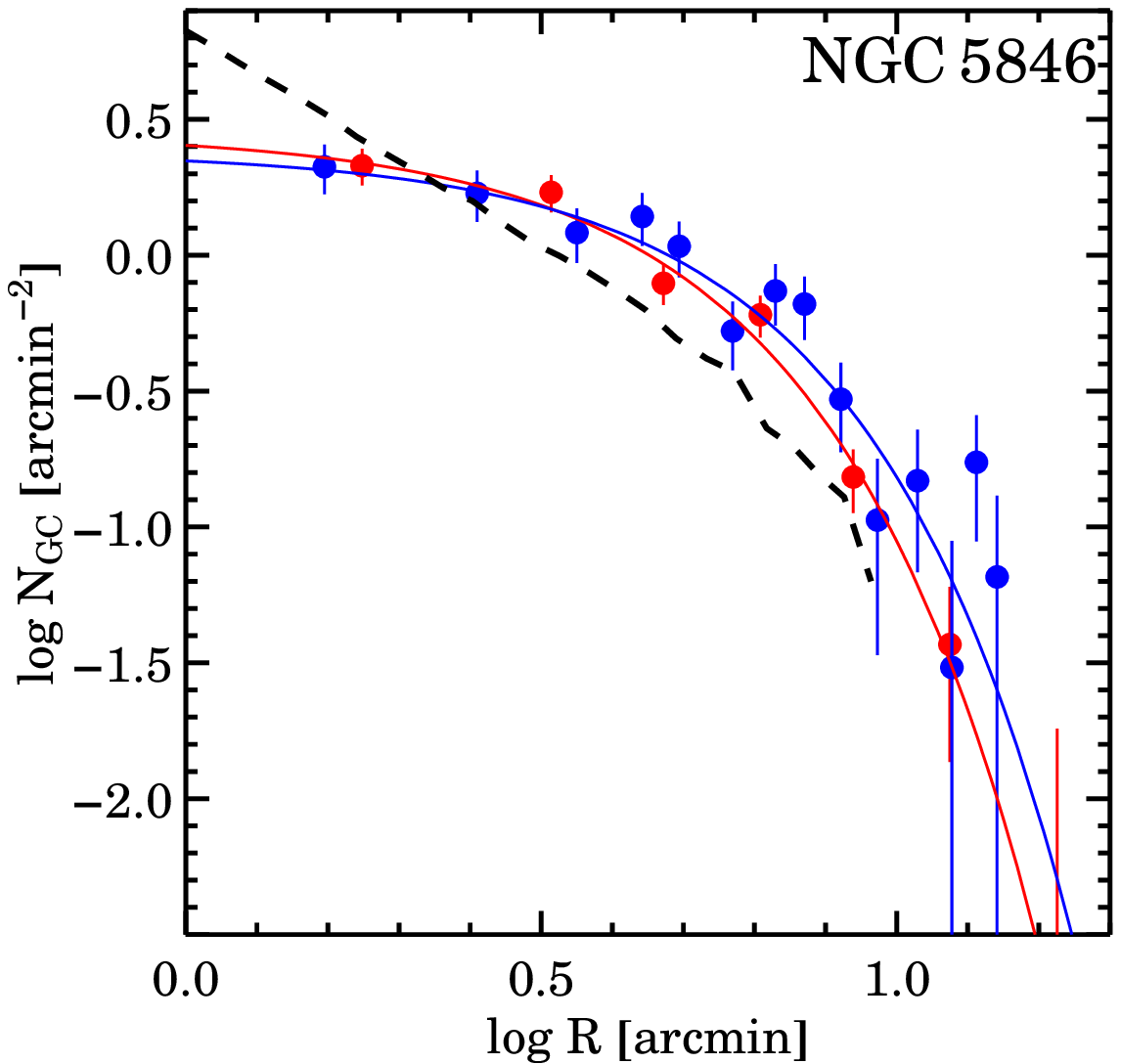} 
\includegraphics[scale=.48]{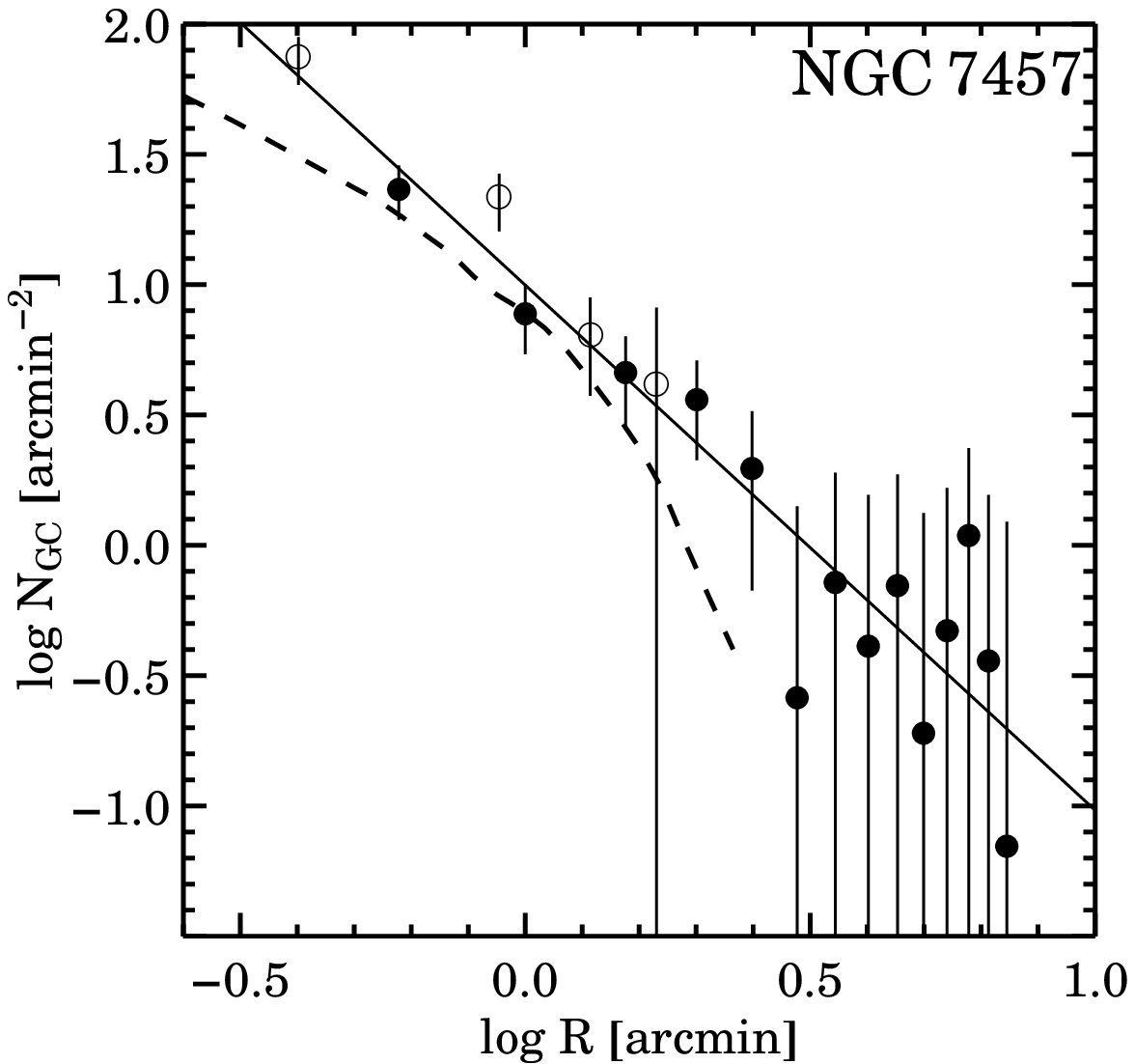} 
\caption{Surface density profiles of the GCs brighter than the turnover magnitude. Blue and red GC subpopulations are shown as blue and red filled points 
(if from Subaru imaging) and blue and red open circles (if from \textit{HST} imaging) respectively. S\'ersic fits or alternatively power-law fits to blue and red GC 
subpopulations are shown as blue and red lines respectively. For NGC~$7457$, the surface density and the fit to all GCs are shown as black points and 
black line respectively. Both the data points and the fits are background subtracted according to eq. \ref{eq:Sersic}. Also shown as dashed lines are the 
scaled and arbitrarily offset stellar surface brightness profiles obtained from Subaru imaging as described in \S \ref{sec:subarudata} .
For completeness, we also show the surface density profiles from \citet{Blom} for NGC~$4365$ and from \citet{Hargis} for NGC~$7457$, respectively.}
\label{fig:SD}
\end{figure*}

Next, we fit the resulting (\textit{HST} $+$ Subaru) GC density profile $N(r)$ with a S\'ersic law \citep{Sersic} similar to that commonly carried out for 
galaxy surface brightness profiles, but in practice we fit a variation of the S\'ersic profile from \citet{Graham} in order to quantify the background level:
\begin{equation}
N(r)=N_e \mbox{\rm exp} \left( -b_n \left[ \left( \frac{R}{\Reff} \right)^{1/n} -1 \right]  \right) + bg
\label{eq:Sersic}
\end{equation}
where $b_n = 1.9992n - 0.3271$, $n$ is the S\'ersic index, \Reff\ is the effective radius of the GC system, $N_e$ is the surface density 
at that radius and $bg$ is the background contamination level. 
In case a S\'ersic fit was not feasible due to small numbers we fit a power-law, i.e. $N(r) \propto r^{\alpha}$ that has been also used for similar
analyses \citep{Spitler08}. In Figure~\ref{fig:SD} the background subtracted GC surface density profiles for our galaxies are shown.
As found in other galaxies, the red GCs are more centrally concentrated than the blue GCs \citep[e.g.,][]{Geisler,Bassino,Faifer11,Strader11}.

We also compare the GC surface density to the galaxy surface brightness for each galaxy, obtained from the Subaru images using \texttt{IRAF/Ellipse} 
as described in \S \ref{sec:subarudata}. The surface brightness profiles were corrected for the local galaxy ellipticity and then shifted by an arbitrary 
constant for comparison purposes (different for each galaxy). Figure~\ref{fig:SD} shows that, qualitatively, the slope of the red GC subpopulation agrees with that of the surface 
brightness of the host galaxy.

\section{Spectroscopic observations and data analysis}
\label{sec:spectroscopic_observations}

Spectroscopic observations were performed with the DEep Imaging Multi-Object Spectrograph (DEIMOS, \citealt{Faber03}) mounted on the 10 m Keck-II telescope.
Galaxies were targeted in the period between 2006 and 2011, employing a different number of masks for each galaxy depending on the richness of the GC system. 
Objects selected for spectroscopic follower were chosen according to their likelihood of being GCs, i.e. giving priority to the objects selected with methods as  
Section \ref{sec:photometric_selection}. The spectroscopic observations are summarized in Table~\ref{tab: log_kine}.

The large collecting area of the 10-meter Keck primary mirror combined with the $\sim16\times5$ arcmin$^2$ of DEIMOS is the ideal combination to investigate the outskirts of 
galaxies where GCs are expected to be one of the best tracers of the total galaxy potential. For all the galaxies, DEIMOS was set up with the 1200 l/m
grating centred on $7800$ \AA\ together with $1$ arcsec wide slits, allowing coverage of the region between $\sim 6500$ -- $8700$ \AA\ with a resolution 
of $\Delta \lambda \sim 1.5$~\AA. Raw spectra were reduced using the DEIMOS/spec2d reduction pipeline provided online, that produces calibrated 
and sky subtracted spectra for each slit \citep{Newman12,Cooper12}. 

We estimate the radial velocity of the GCs by measuring the Doppler shift of Calcium Triplet (CaT) absorption lines that characterises
the infrared part of their spectra at 8498~\AA, 8542~\AA, 8662~\AA, respectively.
We measure radial velocities with \texttt{IRAF/Fxcor} that performs a cross-correlation between the Fourier transformed
science spectrum and $13$ template Galactic star Fourier transformed spectra. The template spectra were observed
with the same DEIMOS setup used for scientific spectra and they cover a wide range of spectral type, luminosity and metallicity (from F to M type).
Fxcor was configured to have all the science and template spectra in the same wavelength range from $8300$ to $8900$ \AA\ with the same 
DEIMOS spectral resolution. The radial velocity for each object was estimated as the mean of the radial velocity resulting from the correlation 
with each template star. The respective errors were evaluated by adding in quadrature the default error given by fxcor as described in 
\citet{Tonry79} to the standard deviation among the stellar templates, which is an estimate of the systematics.

\begin{table}
\centering
\begin{tabular}{@{}l c c c c c c}
\hline
Galaxy ID & Masks & Exp. time & N$_{\rm GCs}$ & N$_{\rm stars}$ & N$_{\rm gal}$   \\
                  &      &    [hr] &  &      \\
\hline
\hline   
NGC~$0821$ & $6$ & $9.2$ &   $61$ & $10$ & $4$\\
NGC~$1400$ & $4$ & $9.0$  &   $72$& $6$& $27$ \\
NGC~$1407$   & $9$ & $20$ &  $369$& $5$& $14$\\
NGC~$2768$   & $5$ & $12.7$ & $109$& $57$& $9$\\
NGC~$3377$    & $4$ & $8.3$ &  $126$& $16$& $74$ \\
NGC~$4278$     & $4$ & $ 8.8$ &   $256$& $44$& $33$\\
NGC~$4365$    & $6$ & $9.0$  &  $269$& $6$& $49$\\
NGC~$5846$     & $6$ & $9.1$ &   $195$& $32$& $4$\\
NGC~$7457$     & $2$ & $4.2$ &   $21$& $14$& $4$\\
\hline
NGC~$3115$    & $5$ & $14$  & $190$ & $29$ & 0\\
NGC~$4486$     & $5$ & $5.0$ &  $737$& $116$& $27$\\
NGC~$4494$     & $5$ & $4.6$ &  $117$ & $34$ & $108$ \\
\hline
\hline
\end{tabular}
\caption{Summary of the spectroscopic observations for our twelve galaxies. The table lists the galaxy name and the total number of the DEIMOS masks used. To take into account
the different seeing conditions over different nights, we show the effective exposure time weighted by the mean seeing conditions during the observation time as done in \citet{Coccato}. 
Also shown are the total number of spectroscopically confirmed GCs (including marginal GCs), Galactic stars and background galaxies respectively.  
The datasets for NGC~$3115$ and NGC~$4486$ also include GCs from external datasets as described in \citet{Arnold} and \citet{Strader11} respectively.}
\label{tab: log_kine}
\end{table}
\subsection{Kinematic selection criteria}
\label{sec:spectroscopic_selection}
\begin{figure*}
\centering
\includegraphics[scale=.48]{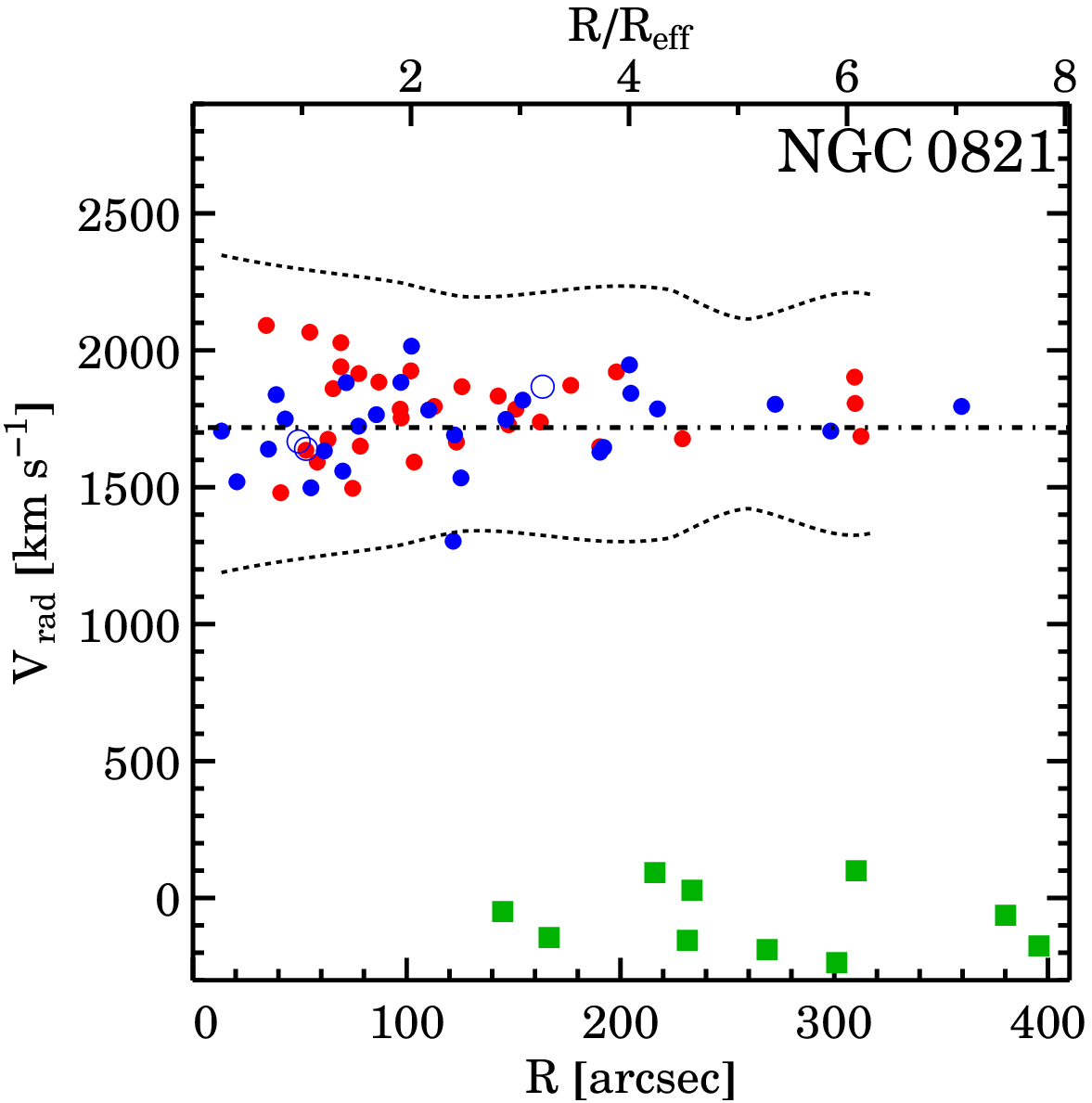} 
\includegraphics[scale=.48]{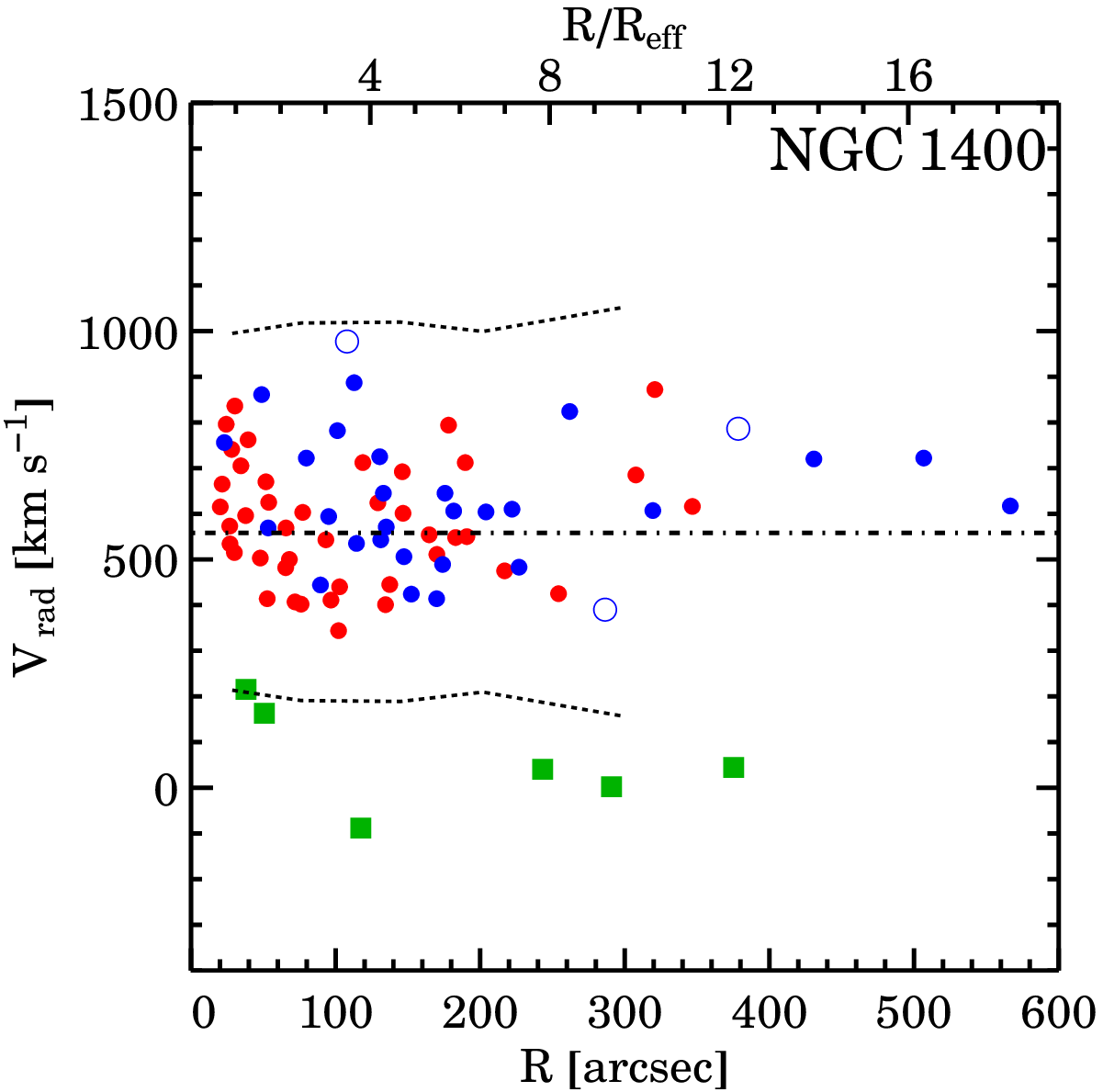} 
\includegraphics[scale=.48]{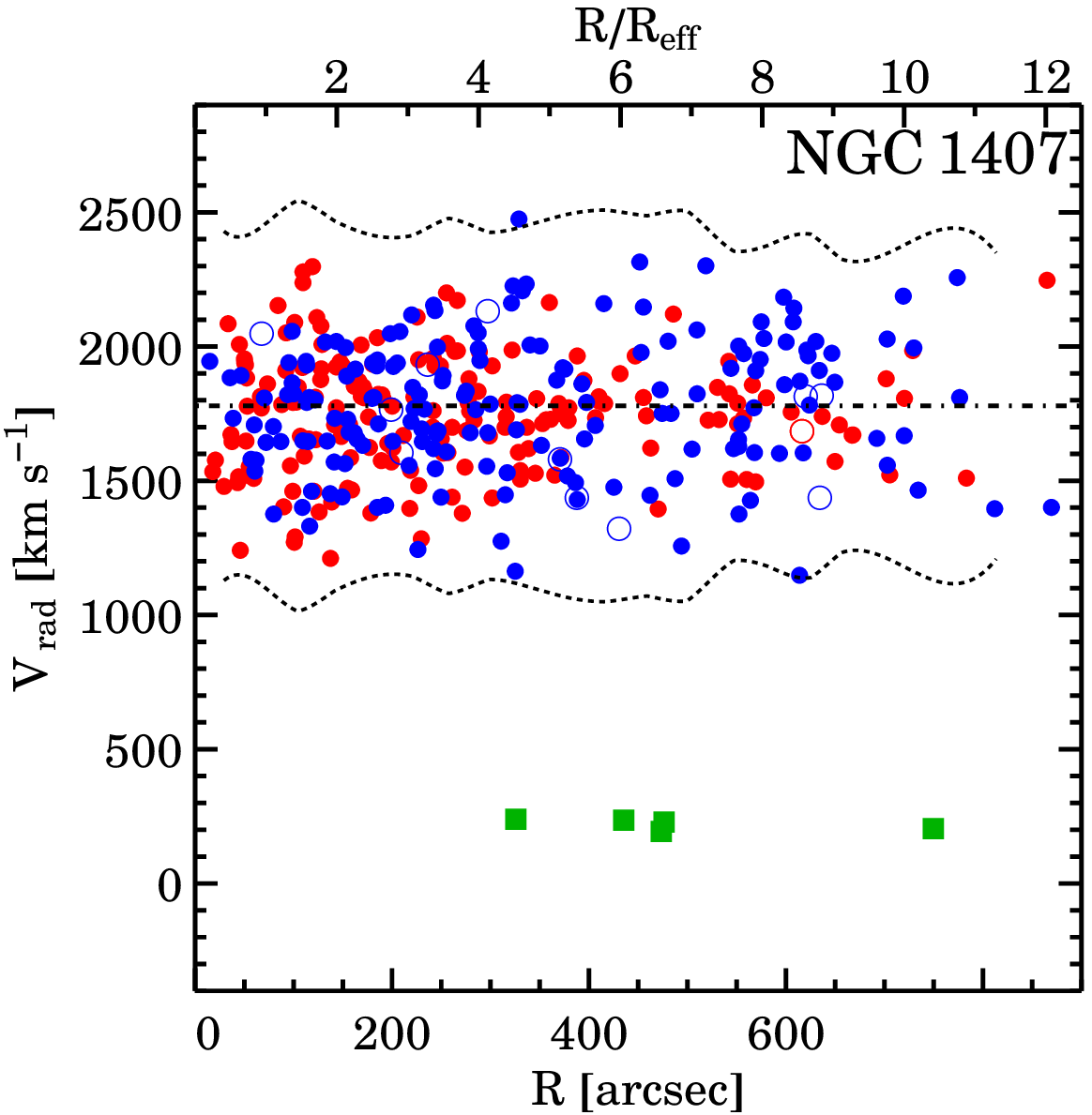} \\
\includegraphics[scale=.48]{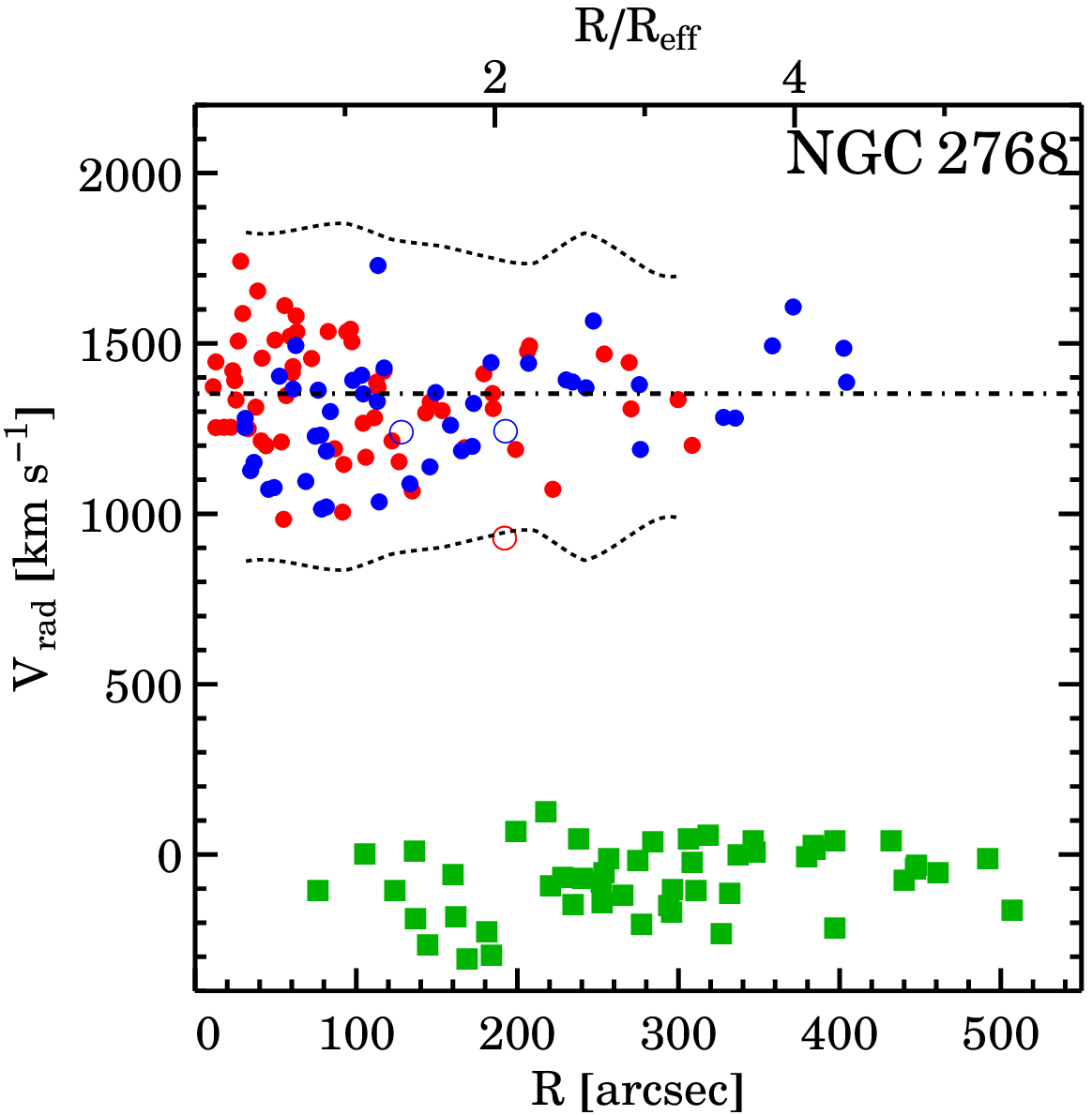} 
\includegraphics[scale=.48]{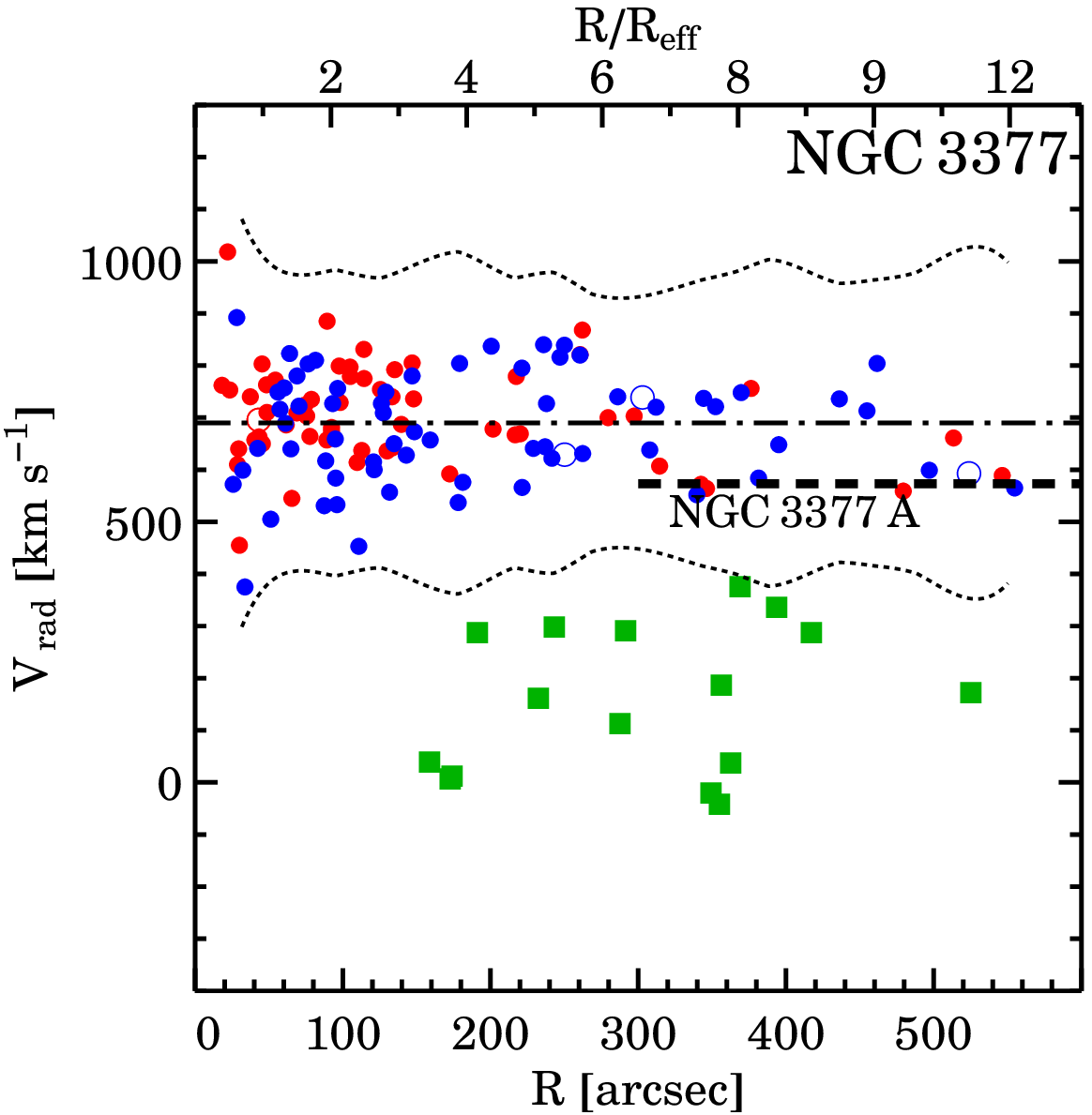} 
\includegraphics[scale=.48]{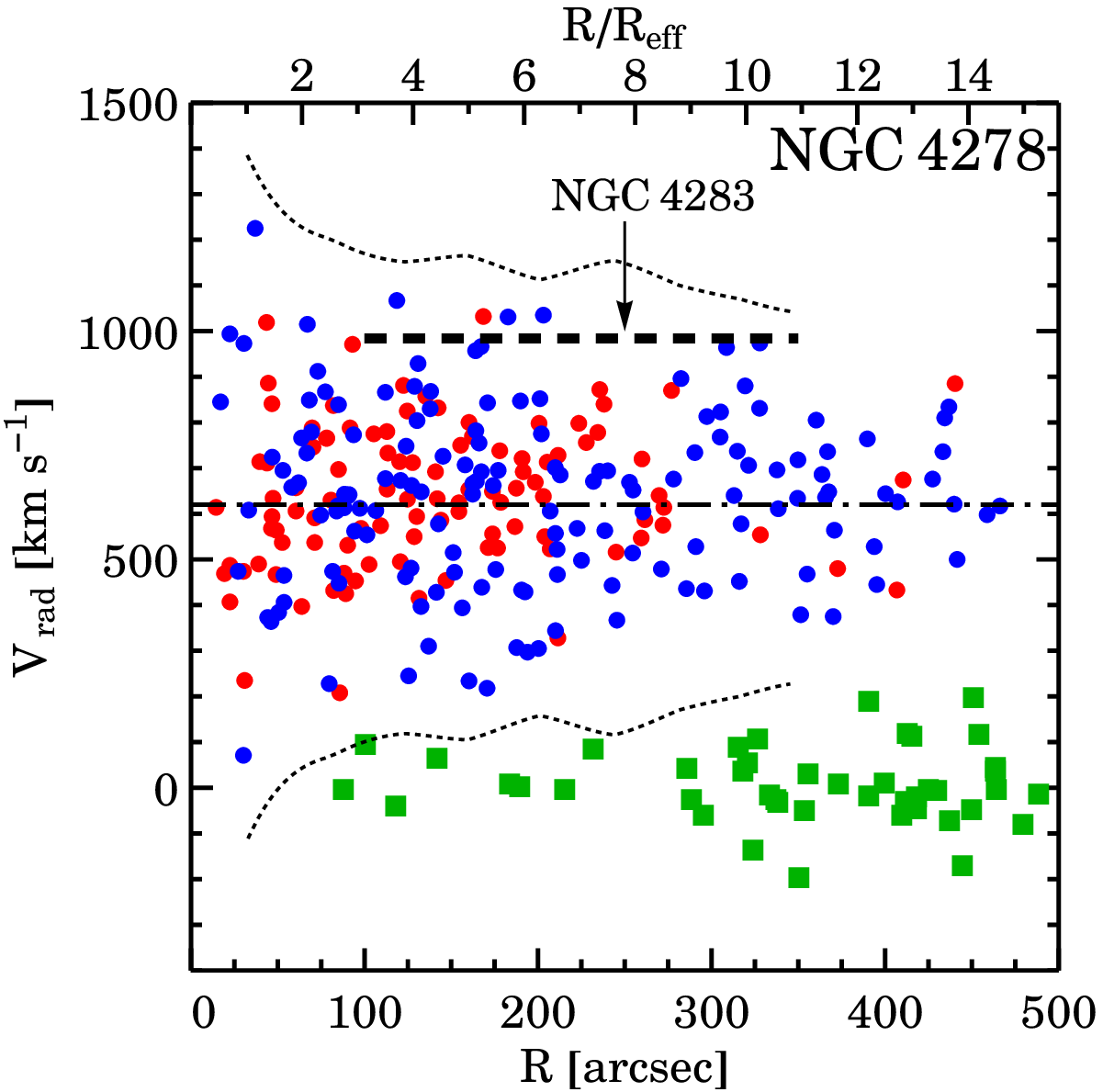} \\
\includegraphics[scale=.48]{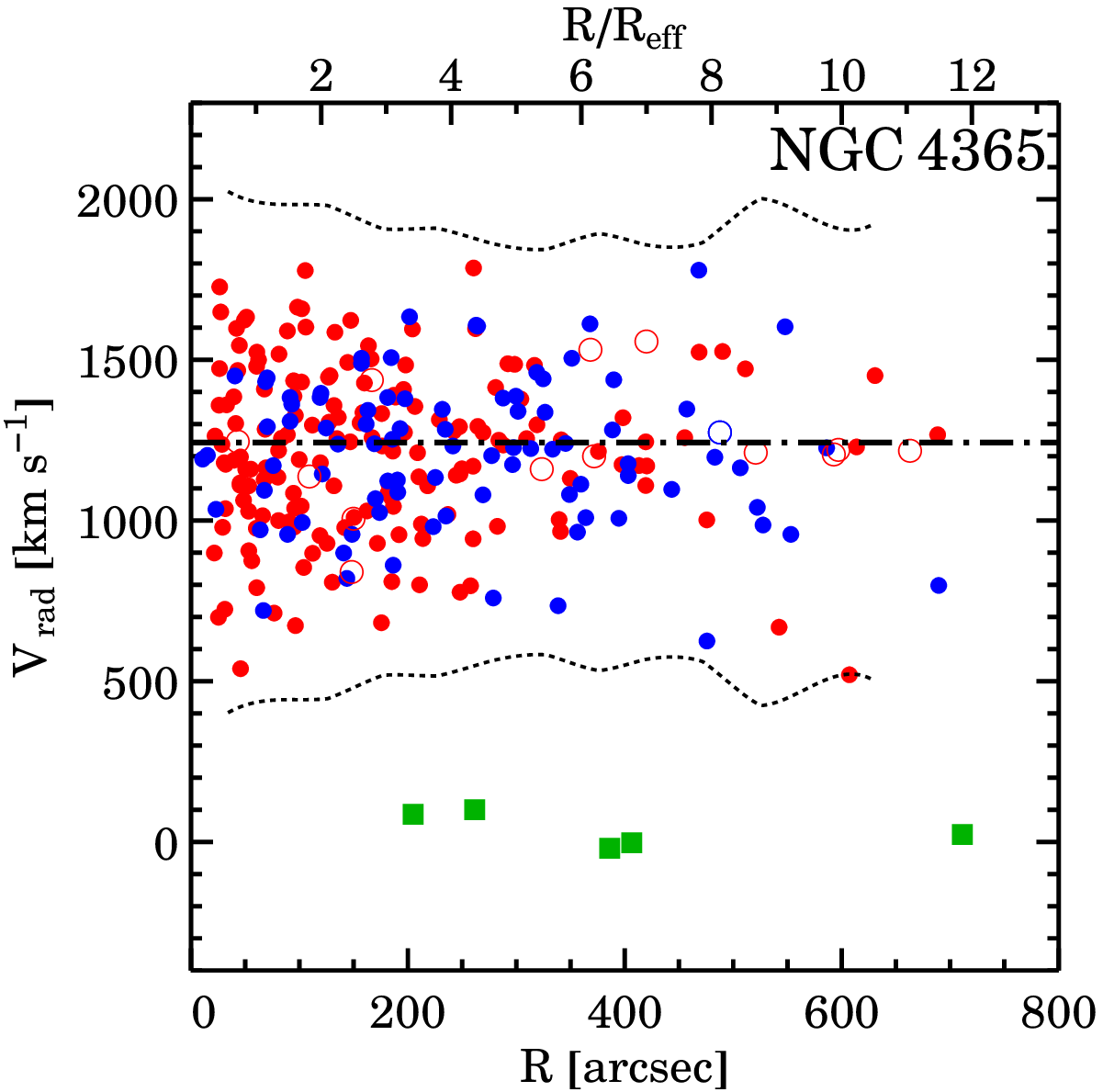} 
\includegraphics[scale=.48]{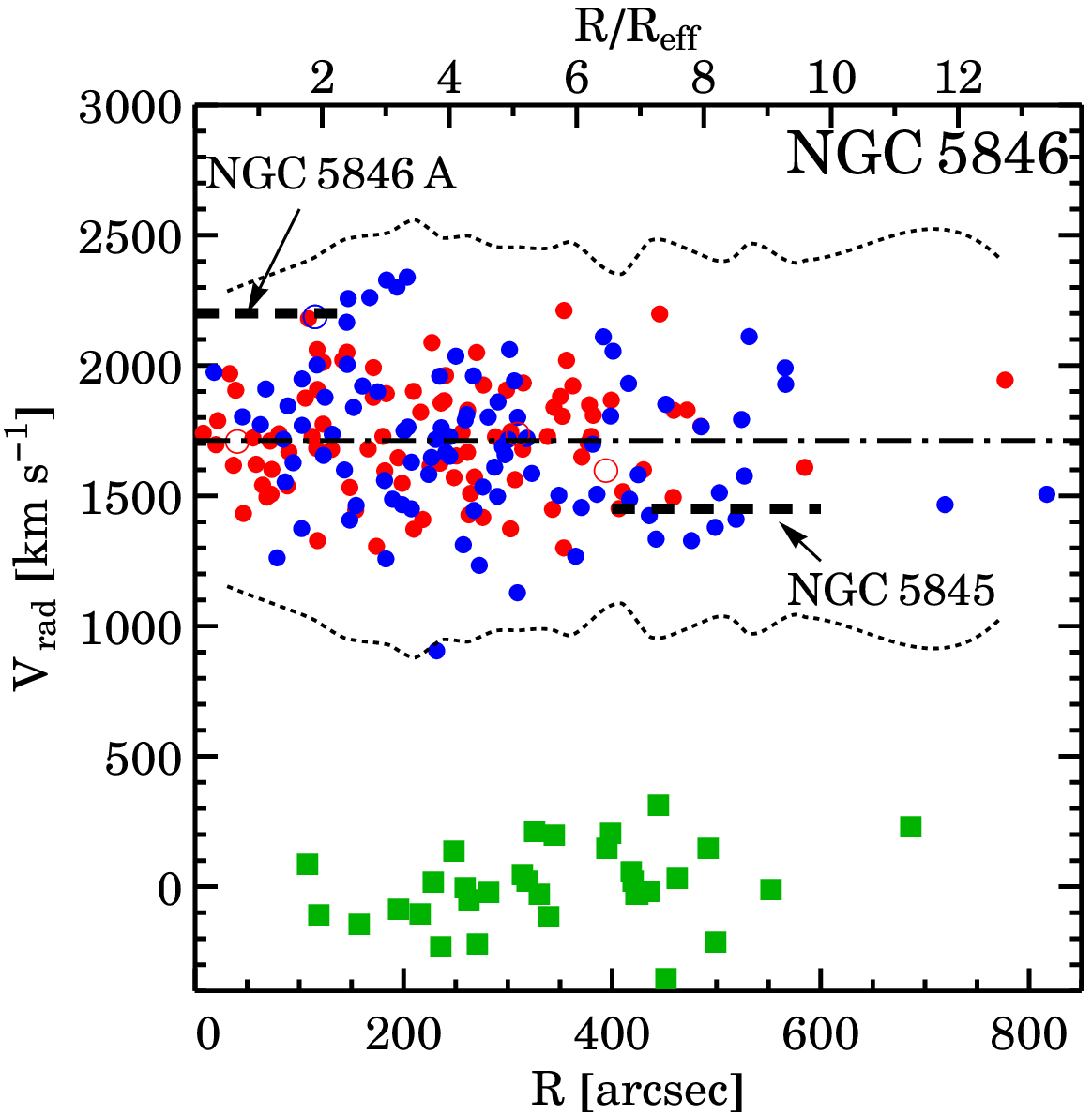} 
\includegraphics[scale=.48]{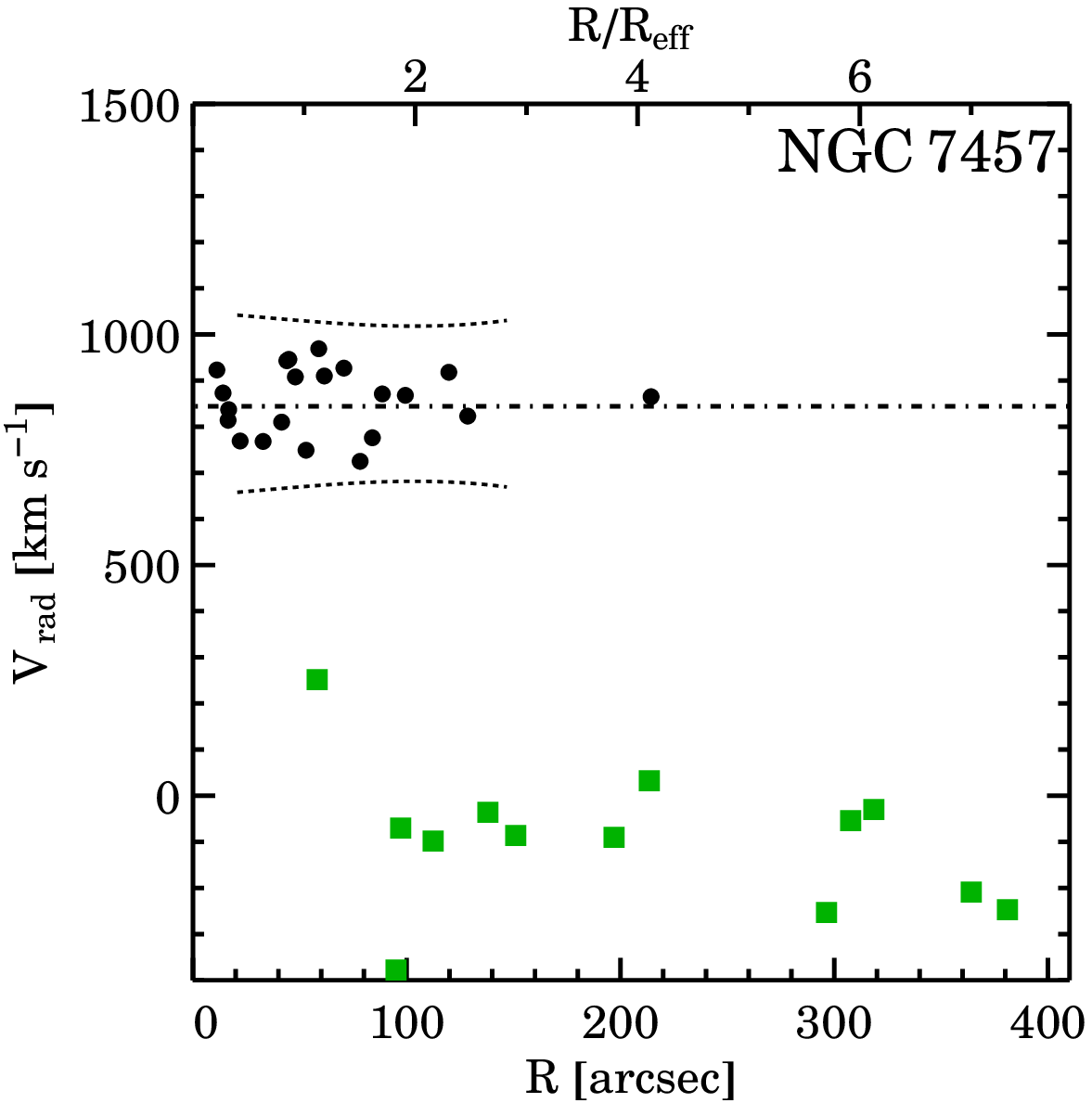} 
\caption{GC radial velocity distributions with galactocentric radius. Confirmed GCs, marginal GCs and Galactic stars are shown as filled points,
open points and green boxes respectively. GCs are colour coded according to their subpopulation (blue or red) membership. 
Galactocentric radii have been translated into effective radii on the top axis. Dotted curves and dot-dashed lines are 
the clipping envelope as defined by the friend-less algorithm (see text) and the galaxy systemic velocities from Table \ref{tab: survey_summary}, respectively. Also shown as thick 
dotted lines are the systemic velocities of the galaxies that might contaminate the GC system of the target galaxy. GCs have a mean velocity similar to the galaxy systemic velocity.}
\label{fig:RVs}
\end{figure*}

Once the final radial velocities of the GC candidates are obtained, we redshift-correct the spectra and perform a visual check 
to verify that the CaT lines are real and that they lie at the expected rest wavelength. Our selection criteria require the presence 
of at least two ``visible'' CaT lines (typically the brightest $8542$~\AA\ and $8662$~\AA\ lines) and of the H$\alpha$ 
absorption line at $6563$~\AA, if probed in the redshift corrected spectra. The visual 
analysis is performed by at least two members of the team and it produces a final spectroscopic consensus catalogue. Spectra 
that show a radial velocity consistent with a GC, but for which it was not possible to reach a consensus, were flagged as ``marginal'' 
and are not included in the kinematic analysis. These objects have usually low signal-to-noise spectra or suffer from bad sky-line 
subtraction making the line identification subjective.

Besides background galaxies that can be usually spotted by the emission lines in their spectra, the main outliers of our spectroscopic
selection are Galactic stars because they show a stellar spectrum with radial velocity $V_{\rm rad} \approx 0 \pm 500$\kms. 
In most cases GCs and Galactic stars are well-separated in velocity, but for galaxies with $V_{\rm sys}<1000\kms$ Galactic stars might 
introduce a low velocity tail in the observed GC candidate velocity distribution. We decided to use a 
friendless algorithm introduced in \citet{Merrett} that flags objects deviating by more than $n\times\sigma$ from the velocity 
distribution of their $N$ nearest neighbours. We use $n=3$ and $10<N<20$ (depending on the galaxy) to exclude both stars 
and possible outliers that lie outside the $3\sigma$ envelope from the kinematic analysis. 

In Figure~\ref{fig:RVs} we show the distribution of all the spectroscopically confirmed GCs and Galactic stars in a radius-velocity phase space. 
In this plot and hereafter, galactocentric distances are expressed as a equivalent radius, that is defined as:
\begin{equation}
R=\sqrt{qX^2 + \frac{Y^2 }{q}}, 
\label{eq:Rc}
\end{equation}
where $q$ is the axis ratio defined as the ratio of minor over the major axis of the galaxy (see Table \ref{fig:RADEC}), $X$ and $Y$ are Cartesian 
coordinates of an object in the galaxy rest frame, with the origin at the galaxy centre, and with $X$ and $Y$ aligned along the photometric major axis and minor axis respectively. 
The final spectro-photometric catalogues of our galaxy sample are available on-line and they include spectroscopically confirmed GCs, ``marginal'' GCs, Galactic stars and galaxies.

\begin{figure*}
\includegraphics[scale=.45]{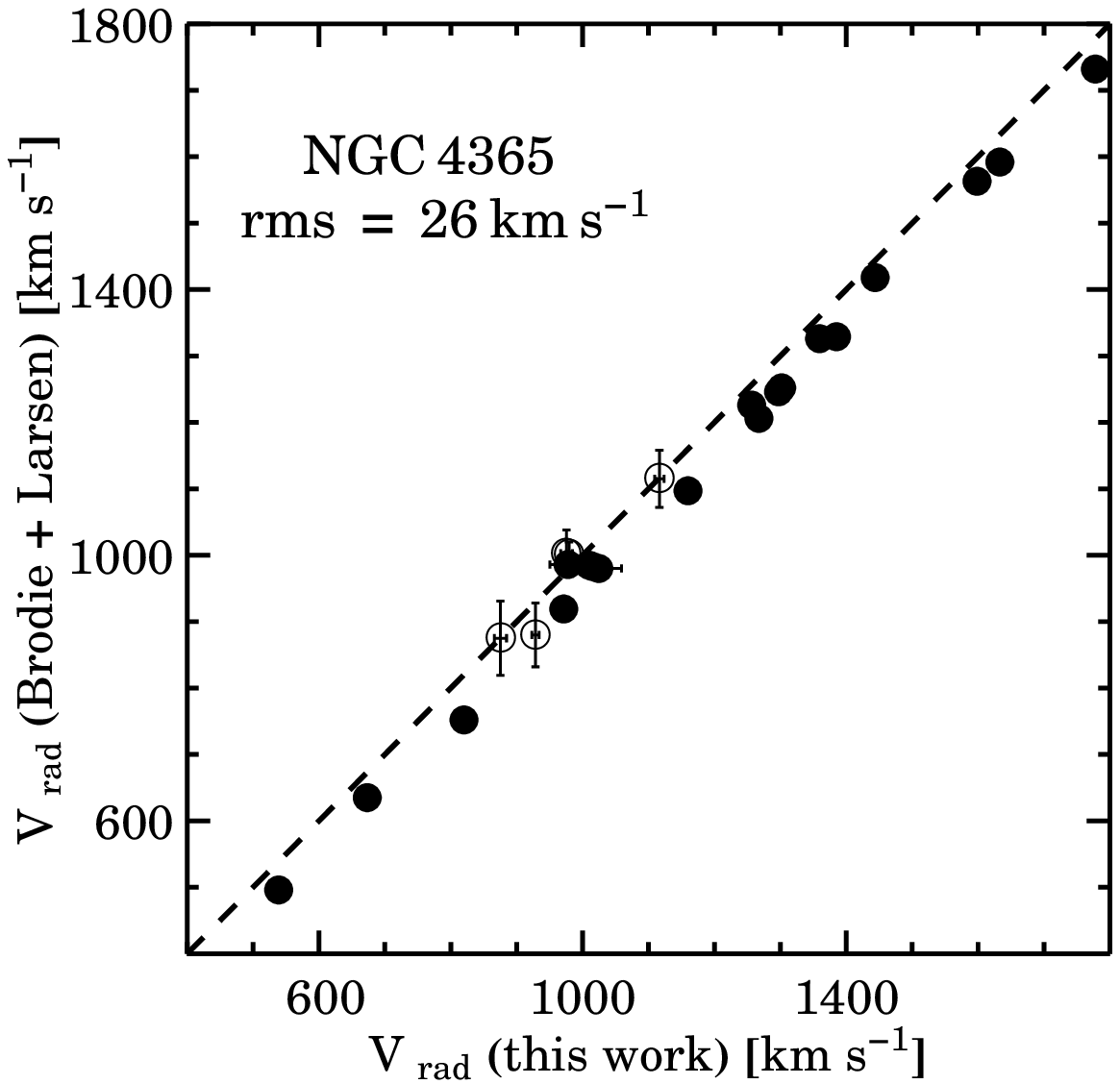} 
\includegraphics[scale=.45]{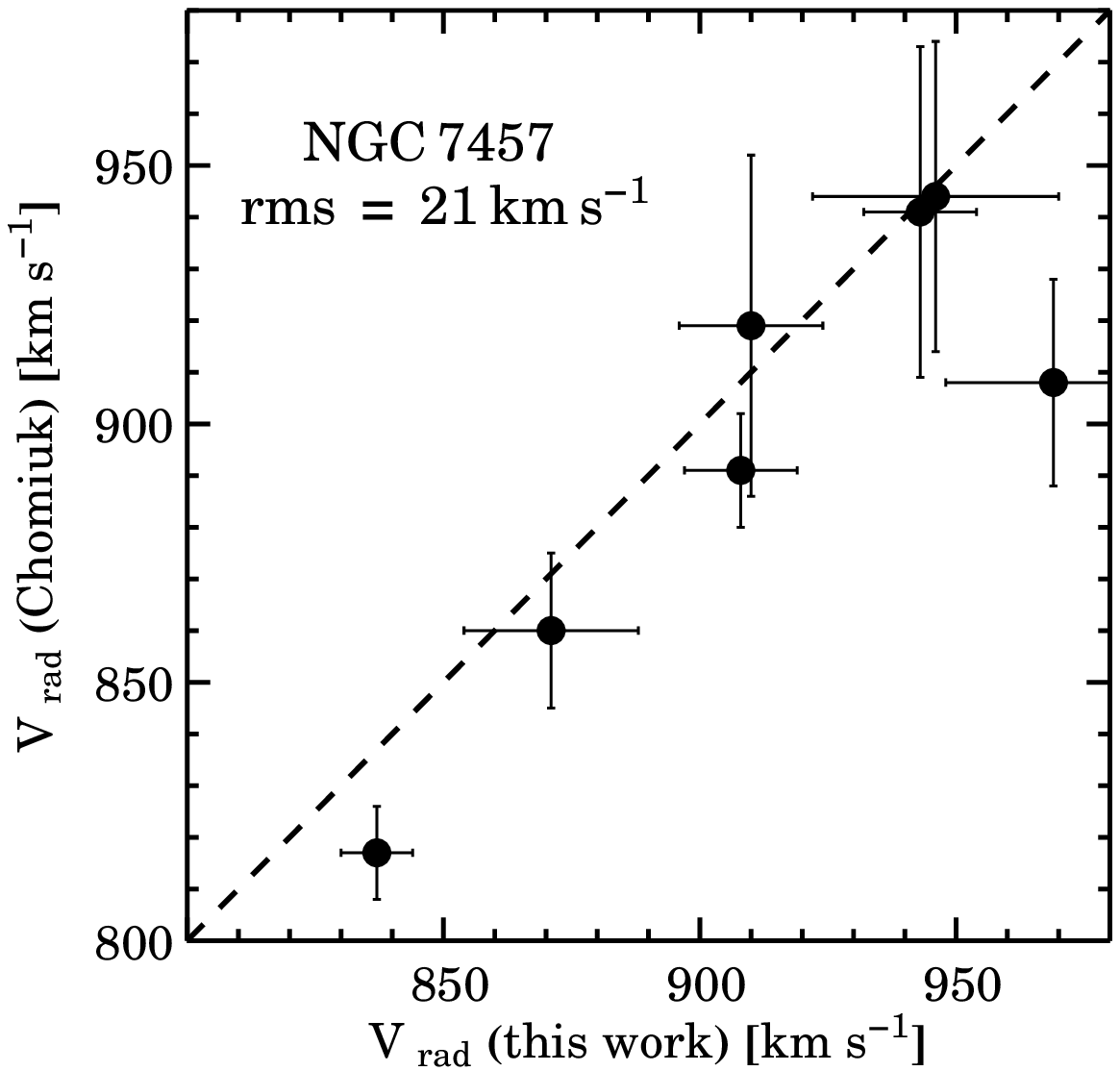} 
\includegraphics[scale=.45]{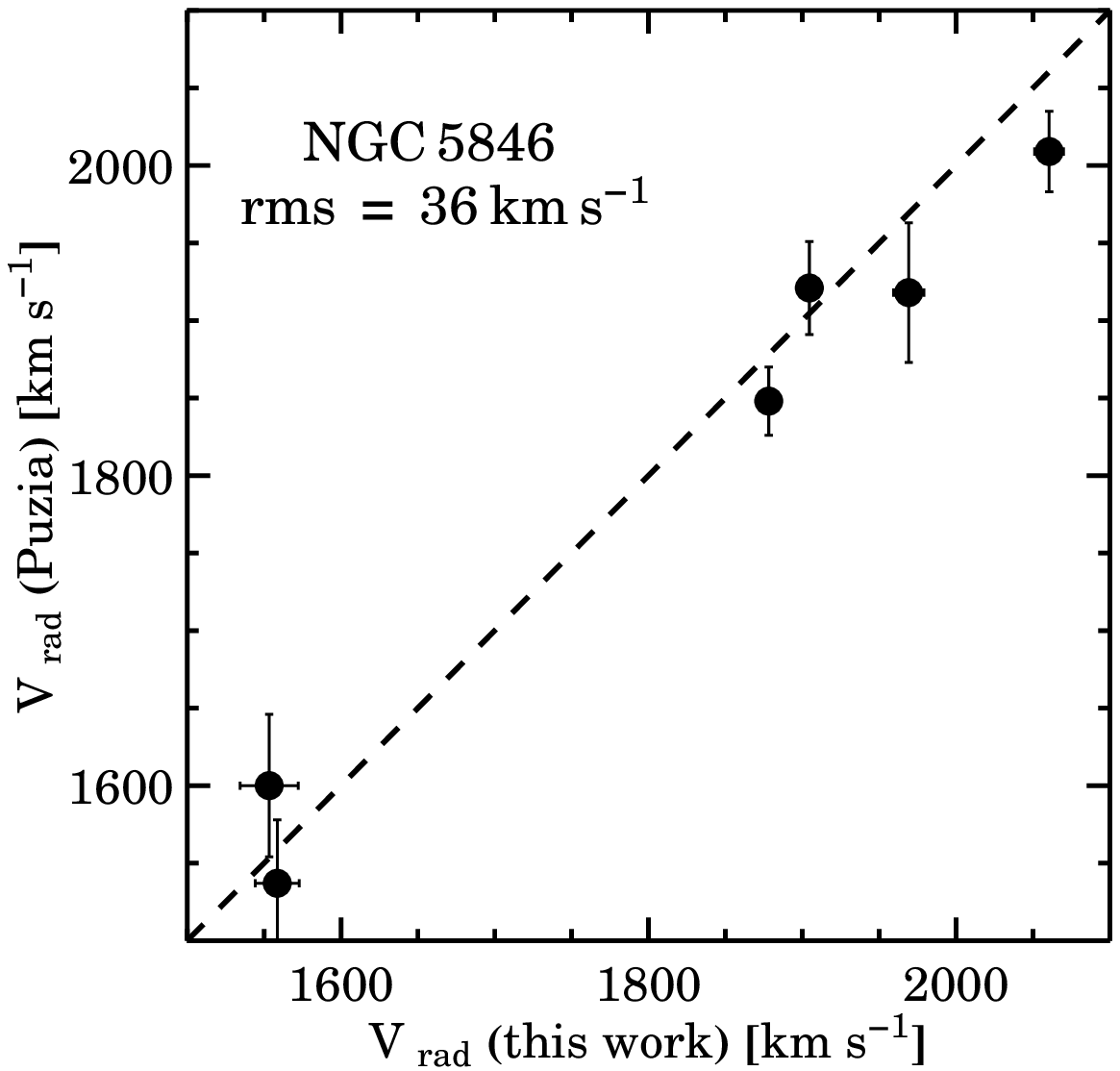} 
\caption{Comparison of our GC radial velocities with previous studies. For each galaxy the root-mean-square of the velocity difference between the two samples is also shown. 
Literature datasets come from \citet{Brodie05} (black points)  and \citet{Larsen03} (open circles) for NGC~$4365$; \citet{Chomiuk} for NGC~$7457$; \citet{Puzia04} for NGC~$5846$. 
The one-to-one line is shown as a dashed line.}
\label{fig:RepSingle}
\end{figure*}

\subsection{Repeated GC measurements}
\label{sec:Repeated}

We searched through the literature for GCs observed in our surveyed galaxies, finding existing datasets for three galaxies. In summary, 
we have re-observed:
\begin{inparaenum}
\item 6 GCs of the 26 GCs confirmed by \citet{Puzia04} in NGC~$5846$ using VLT/FORS2;
\item 7 GCs of the 13 GCs found by \citet{Chomiuk} in NGC~$7457$ using Keck/LRIS ;
\item 24 GCs of the 33 GCs resulting from the combination of the \citet{Brodie05} and \citet{Larsen03} catalogues of NGC~$4365$, both observed with Keck/LRIS.
\end{inparaenum}

\begin{figure}
\hspace{-0.1cm}
\centering
\includegraphics[scale=.5]{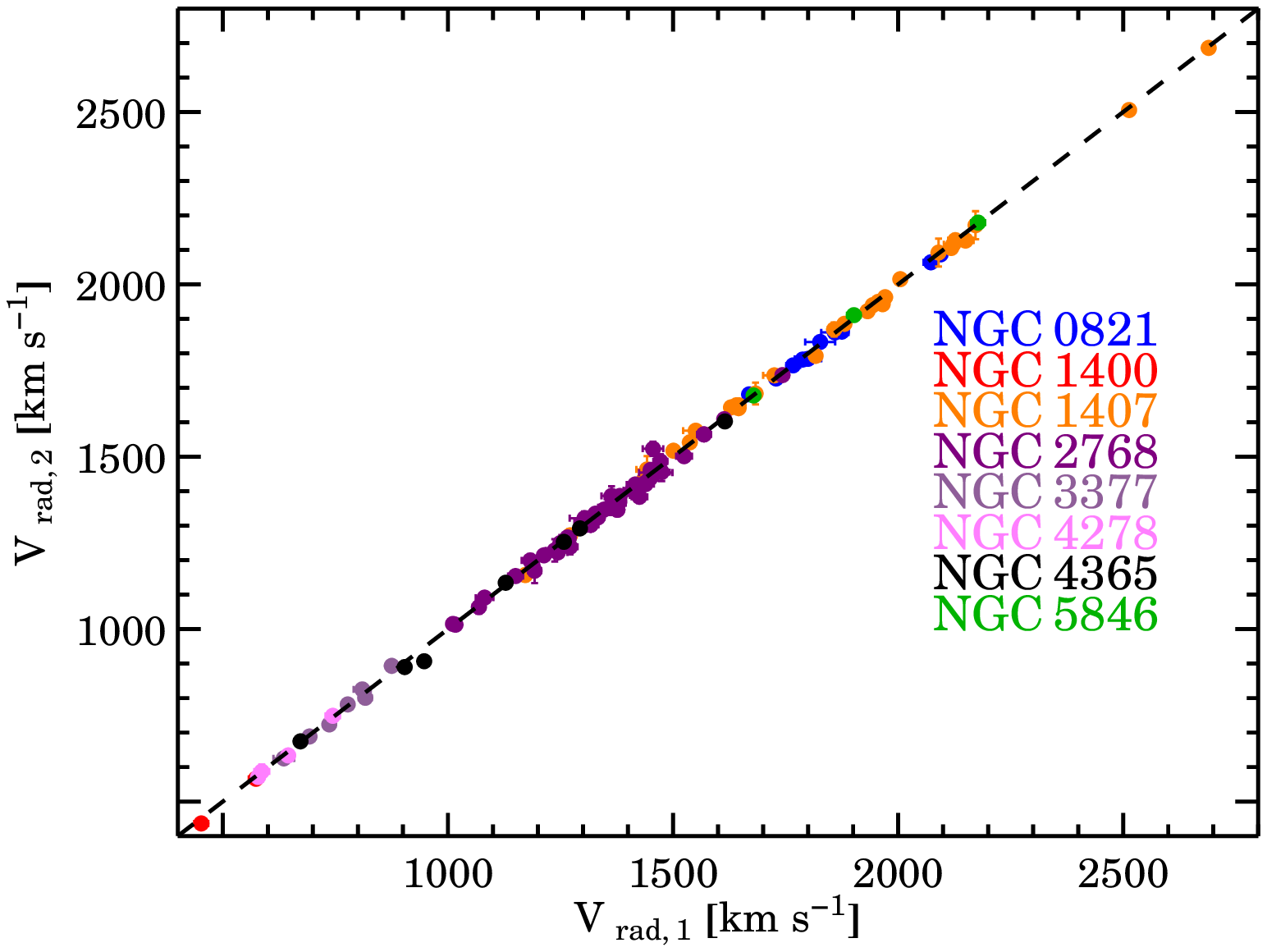}
\includegraphics[scale=.5]{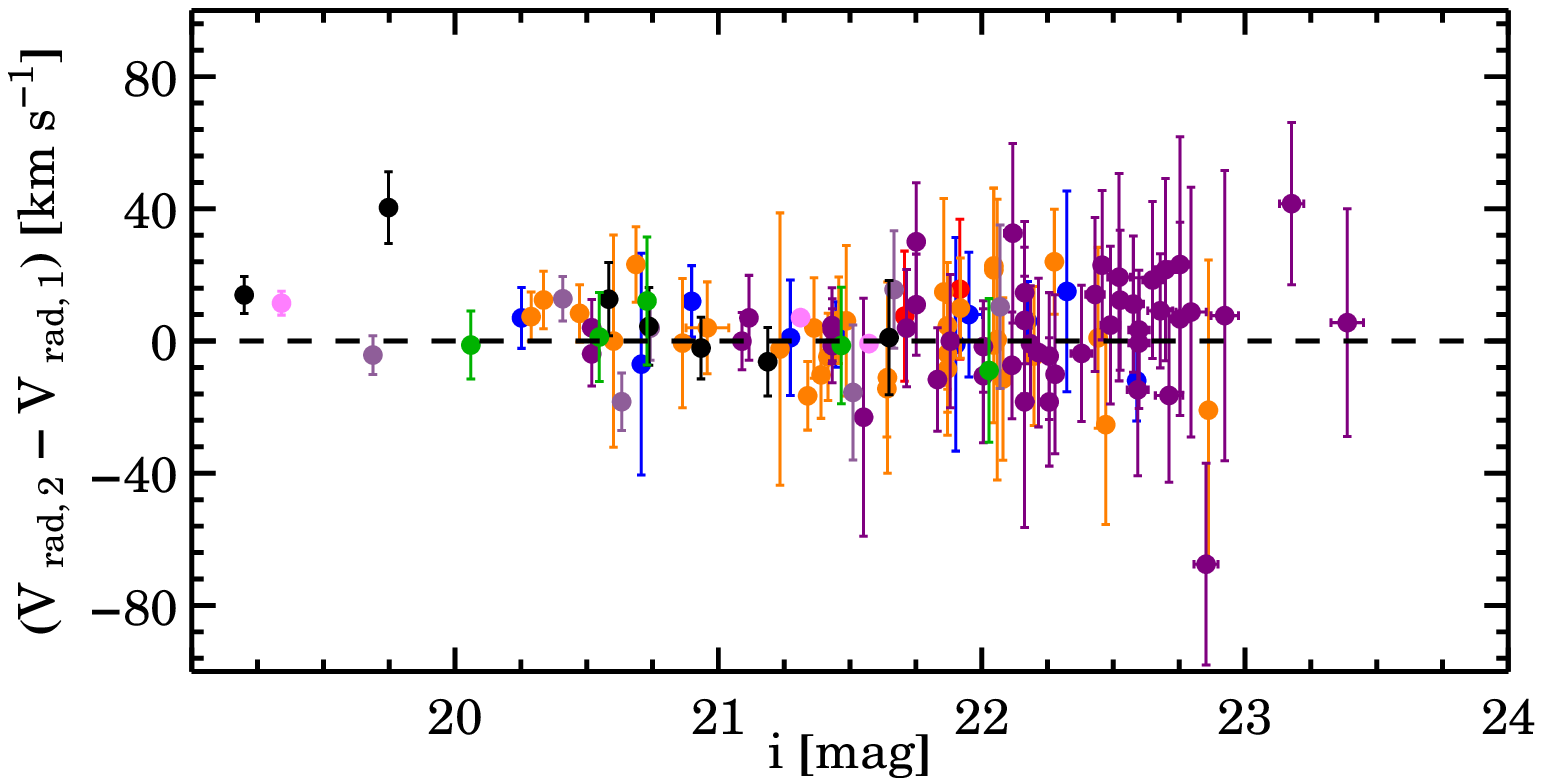}
\caption{Repeated GC radial velocity measurements. In the \textit{top panel} we show the radial velocity of a GC $V_{\rm rad, 1}$ against 
the radial velocity of the same GC observed in a different night $V_{\rm rad, 2}$. Data points are colour coded according their host galaxy membership 
listed on the right. The dotted line is a one-to-one line. In the \textit{bottom panel} we show the difference $V_{\rm rad, 1} - V_{\rm rad, 2}$ as a function
of the $i$ band magnitude of the object. The dotted line shows the constant $V_{\rm rad, 2} - V_{\rm rad, 1} = 0$ to guide the eye. 
Radial velocities from different nights are in good agreement with each other, with an overall root-mean-square (rms) of $15 \pm 1 \kms$.
No significant trend is seen as a function of GC magnitude.}
\label{fig:Repeated}
\end{figure}

In Figure~\ref{fig:RepSingle} we compare our DEIMOS dataset with literature studies.
There is a rough agreement between external datasets and our DEIMOS radial velocities, although the root-mean-square of the velocity difference between 
the two datasets is larger than what was found with DEIMOS repeated measurements. 

We supplement our GC catalogues with external GC radial velocities corrected for the mean offset between the literature and our DEIMOS measurements. 
This offset is  $-15 \kms$ for both NGC~$7457$ and NGC~$5846$. In NGC~$4365$,  the offset between DEIMOS radial velocities and \citet{Brodie05} is effectively zero,
and we only correct the dataset of \citet{Larsen03} by $-41\kms$.

We test the reliability of our spectroscopic measurements by comparing the radial velocity of two or three GCs observed 
over different nights. Overall, we have collected 118 repeated GC radial velocities and these are shown in Figure \ref{fig:Repeated}. 
One-third of the total repeated sample is from a NGC~$2768$ mask that was observed twice with a one night offset. We find that 
repeated GC measurements are in good agreement in all the galaxies. The root-mean-square (rms) of the velocity difference between 
two observations is $\delta V_{\rm rms} = 15 \pm 1 \kms$.

\subsection{Kinematic analysis}
\label{sec:kinematics_analysis}
We study the kinematic properties of our GC systems using a maximum likelihood approach. We summarize here the main points of this analysis and refer to \citet{Foster11} for the details. 
We divide the data in radial bins and then we fit the amplitude of the rotation ($V_{\rm rot}$), the velocity dispersion ($\sigma$) and the kinematic position angle ($PA_{\rm kin}$) 
simultaneously. The bin size varies from galaxy to galaxy (wider for larger datasets) and it was set to have roughly the same number of objects per bin. 
For the $j^{\rm th}$ radial bin we minimise the $\chi^2$ function:
\begin{equation}\label{eq:GC_LR}
\chi^2_{j}\propto\sum^{i=N_j}_{i=1} \left[\frac{(V_{{\rm rad},i}-V_{{\rm mod},i,j})^2}{(\sigma_j^2+(\Delta V_{{\rm rad},i})^2)}+\ln (\sigma_j^2+(\Delta V_{{\rm rad},i})^2)\right],
\end{equation}
where:
\begin{equation}
\label{eq:Vobs2}
V_{{\rm mod},i,j}=V_{{\rm sys}} \pm \frac{V_{{\rm rot},j}}{\sqrt{1+\left(\frac{\tan(PA_{i}-PA_{{\rm kin},j})}{q_{{\rm kin},j}}\right)^2}}.
\end{equation}
In eq. \ref{eq:Vobs2}, $PA_i$, $V_{{\rm obs},i}$ and $\Delta V_{{\rm obs},i}$ are the position angle, recession velocity and uncertainty on the recession velocity for 
the $i^{\rm th}$ GC, respectively. The axis ratio of GC systems is challenging to derive directly because of low number statistics, but it has been constrained 
in galaxies with large photometric datasets (\citealt{Kissler-Patig}, \citealt{Strader11}, \citealt{Blom}).
Therefore, the kinematic axis ratio $q_{\rm kin}$ of the GC system was assumed to be equal to the respective photometric axis ratio of the galaxy light 
(see Table \ref{tab: survey_summary}). The galaxy systemic velocity $V_{\rm sys}$ was fixed to the values given in Table~\ref{tab: survey_summary} because
the GC mean velocity is in good agreement with the galaxy systemic velocity itself.
Uncertainties on the fits to the GC kinematics are obtained using a bootstrapping method similar to that used by \citet{Cote01}. We obtain 1000
 ``mock'' GC kinematic samples for each galaxy, by sampling with replacement from our measured distribution. 
Think kind of kinematic modelling tends to overestimate the rotation amplitude when the kinematic position angle is a free parameter. 
The correction for this bias is described in Appendix~\ref{appendix}.
 
\begin{figure}
\centering
\includegraphics[scale=.42]{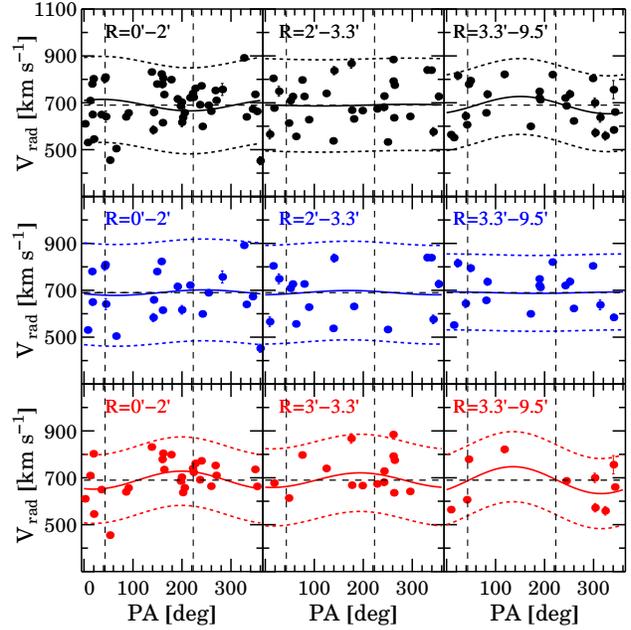} 
\caption{Rotation of NGC~$3377$ GC system with position angle in different radial bins. From the top to the bottom, panels show the rotation for all,
blue and red GC subpopulations respectively. Whereas from the left to the right, panels show the rotation in the radial bin reported on the top left for each panel. 
The photometric major axis and the systemic velocity of the galaxy are both from Table \ref{tab: survey_summary} and represented with dashed 
vertical and dashed horizontal lines respectively. The best fit to eq. \ref{eq:Vobs2} and $\pm 2 \sigma$ envelope are shown as filled and dotted lines respectively.}
\label{fig:NGC3377PA}
\end{figure}

 By way of example, the best fit to eq. \ref{eq:Vobs2} for the GCs in NGC~$3377$ and its kinematic dataset are shown in Figure~\ref{fig:NGC3377PA} and Table \ref{tab: NGC3377}, respectively.

The kinematic method described above was used to investigate the kinematic properties of our surveyed galaxies in two different ways:  
\begin{itemize}
\item First, the kinematic properties of each GC system were calculated as a function of the galactocentric radius. This approach allows us
to compare the kinematics of a given GC system to different kinematic probes (e.g., long-slit spectroscopy and/or PNe kinematics). 
We generally use between $20$ and $30$ GCs per bin depending on the galaxy. These results are shown in Figure~\ref{fig:kinematics}. 

\item Secondly, the confirmed GCs were sorted by their colour to study the GC kinematics as a function of colour. 
Here we use moving colour bins of equal width (with usually $20$ GCs per bin) to investigate the effect of each GC on the final fit. 
This analysis is independent of the photometric dividing colour and it has the advantage of testing whether or not there is a transition
in the kinematics between the two subpopulations. These results are shown in Figure~\ref{fig:CV}.
\end{itemize}

To better appreciate the global kinematic properties of our GC systems, we also construct a 2D smoothed velocity field for our elongated
early-type galaxies (i.e., galaxies with $(b/a)_K <0.6$: NGC~$821$, NGC~$2768$, NGC~$3377$, NGC~$7457$). These galaxies are all part of the ATLAS$^{\rm 3D}$ sample
 \citep{Krajnovic} and are indeed good candidates to compare their moderate--fast stellar rotation in the innermost regions with the kinematics of their GC system.
At every position $(x,y)$ on the sky we compute the local radial velocity $\hat v(x,y)$ via interpolation of the weighted average 
radial velocities of the $N$ nearest neighbours, similar to the technique used in \citet{Coccato}:
\begin{equation}
\hat v(x,y) =\frac{\textstyle \sum_{i=1}^N V_{\rm rad, i} / d_i ^2}{\textstyle \sum_{i=1}^N 1 / d_i ^2} 
\label{eq:2D}
\end{equation}
with the weights being the reciprocal of the square distance $d$ between two neighbour GCs. The 2D velocity field was smoothed using a 
Gaussian filter kernel of variable width for each galaxy on a regularly spaced grid. The kernel size was arbitrarily set between $2$ and $5$ kpc depending
the sampling of the GC system and on the size of the galaxy. For the sake of comparison, we analysed the SAURON 
data-cubes \citep{Emsellem07} of the same four early-type ellipticals in order to reproduce their 2D stellar velocity fields using our technique. These 
results are shown in Figure~\ref{fig:2D}.

We also compute the root-mean-square velocity:
\begin{equation}
V_{\rm rms}^2 =\frac{1}{N} \sum_{i=1}^N (V_{\rm rad, i} - V_{\rm sys})^2 - (\Delta V_{\rm rad, i})^2.
\label{eq:Vrms}
\end{equation}
This quantity coincides with the velocity dispersion in eq. \ref{eq:Vobs2} if the contribution of the rotation is negligible. 
The uncertainty on the $V_{\rm rms}$ is estimated with the formulae provided by \citet{Danese}.

\begin{table*}
\begin{tabular}{@{}l c c c c c c c c c c c}
\hline
ID & RA & Dec & V$_{\rm rad}$ & $\delta$V$_{\rm rad}$ & $g$  & $\delta g$  & $r$    &  $\delta r$    & $i$ & $\delta i$ \\
     & [Degree] & [Degree]  & [\kms]  & [\kms]		                  & [mag] & [mag] & [mag]  &  [mag] & [mag] & [mag]\\
      (1)  & (2)      &  (3)            & (4)            & (5)     & (6)    &  (7)         &  (8)      &       (9)           &       (10)      & (11)   \\
\hline
NGC$3377$\_GC1   &	$161.94713$  & $14.00741$ &	$810$ &	8   & $21.387$ & $ 0.019$ &	$20.855$ &	$0.01$3 &$20.606$ & $ 0.014$\\
NGC$3377$\_GC2  &	$161.90719$ &	 $13.96616$ &	$724$ &	10 & $22.295$ & $ 0.032$ &	$21.588$ &	$0.021$ &	$21.223$ &  $0.021$\\
NGC$3377$\_GC3   & 	$161.89561$  & $13.97022$ &	$729$ &	13 & $22.596$ & $0.032$ &	$21.962$ &	$0.021$ &	$21.671$ &  $0.021$\\
NGC$3377$\_GC4  & 	$161.92600$  & $14.01391$ &	$453$ &	17 & $22.610$ &$0.034$ &	$22.048$ &	$0.023$  &$21.864$ &  $0.025$\\
NGC$3377$\_GC5   & 	$161.89744$  & $13.97086$ &	$681$ &	14 & $22.789$ &$0.037$ &	$22.037$ &	$0.023$ &	$21.69$1 &  $0.022$\\
NGC$3377$\_GC6   & 	$161.93466$  & $13.96388$ &	$779$ &	15 & $22.546$ &$0.034$ &	$21.948$ &	$0.023$ &$21.653$ & $ 0.025$\\
NGC$3377$\_GC7   & 	$161.91688$  & $13.96030$ &	$617$ &	18 & $22.919$ &$0.046$ &	$22.286$ &	$0.031$ &	$22.073$ &  $0.036$\\
NGC$3377$\_GC8   & 	$161.88981$  & $13.95709$ &	$754$ &	5   & $20.002$&$0.009$ &	$19.411$ &	$0.006$ &	$19.116$ &  $0.005$\\
NGC$3377$\_GC9    &     $161.94103$  & $14.03665$ &	$804$ &	6   & $20.150$ &$0.009$ &	$19.632$ &	$0.006$ &	$19.378$ &$ 0.006$ \\
NGC$3377$\_GC10  &     $161.87097$ &	 $13.91145$ &	$820$ &	7   & $20.895$ &$0.013$ &	$20.305$ &	$0.008$ &	$20.094$ & $0.008$ \\
$\cdots$  & $\cdots$  &	$\cdots$ &	$\cdots$ &	$\cdots$   &$\cdots$ &	$\cdots$ &	$\cdots$ &	$\cdots$ &	$\cdots$ & $\cdots$ \\
\hline
NGC$3377$\_star1  & $162.03633$  &	$14.02335$ &	$185$  &	$11$   &	$22.174$ &	$0.026$ &$	21.750$ &$	0.017$ &	$21.58$1 & $0.017$ \\
NGC$3377$\_star2  & $161.86723$  &	$13.90198$ &	$289$  &	$18$   &	$22.547$ &	$0.003$ &$	20.090$ &$	0.020$ &	$21.887$ & $0.002$ \\
$\cdots$  & $\cdots$  &	$\cdots$ &	$\cdots$ &	$\cdots$   &$\cdots$ &	$\cdots$ &	$\cdots$ &	$\cdots$ &	$\cdots$ & $\cdots$ \\
\hline
NGC$3377$\_gal1  & $161.83237$  &	$13.92902$ &	$-$   &	$-$   &	$22.250$ &	$0.025$ &	$21.595$ &	$0.015$ &$	21.370$ &$ 0.015$ \\
NGC$3377$\_gal2  & $161.95037$  &	$14.08945 $&	$-$  &	$-$   &	$24.188$ &	$0.069$ &	$24.214$ &	$0.062$ &	$       24.290$ & $0.087$ \\
$\cdots$  & $\cdots$  &	$\cdots$ &	$\cdots$ &	$\cdots$   &$\cdots$ &	$\cdots$ &	$\cdots$ &	$\cdots$ &	$\cdots$ & $\cdots$ \\
\hline   
\hline
\end{tabular}
\caption{Spectro-photometric catalogue of objects around NGC~$3377$. The horizontal line divides the sample into spectroscopically confirmed GCs, spectroscopically confirmed Galactic
stars and background galaxies. The first column gives the object ID, composed of the galaxy name and the object identification. Columns 2 and 3 give position in right ascension and declination (J$2000$), 
respectively. Columns 4 and 5 are the observed heliocentric radial velocities and uncertainties respectively. The remaining columns are the Subaru photometry in \textit{gri} and respective uncertainties. 
The full table for all the surveyed galaxies is available in the online version.}
\label{tab: NGC3377} 
\end{table*}

\section{Notes on individual galaxies}
\label{sec:individualgalaxies}

In this section we briefly discuss the kinematic results for our new nine GC systems that are presented in Figure~\ref{fig:kinematics}, Figure~\ref{fig:CV} and Figure~\ref{fig:2D}.

\subsection{NGC~821}
\label{sec:NGC821}
NGC~$821$ is an isolated E6 galaxy \citep{Vaucouleurs} with photometric and kinematic signatures of an edge-on stellar disk (\citealt{Emsellem07}; \citealt{Proctor}). 
The dark matter content of this galaxy has been debated in the literature. \citet{Romanowsky03}, and recently \citet{Teodorescu} found that the velocity dispersion
of the PNe decreases with radius. In contrast, \citet{Weijmans}, \citet{Forestell} and \citet{Proctor} found a flat velocity dispersion for the stellar component 
within 100 arcsec.
A photometric study of the GC system of NGC~$821$ was carried out by \citet{Spitler08} using wide-field WIYN and \textit{HST} observations. 
They were able to detect significant colour bimodality only after combining ground-based and space-based observations. 

Our Subaru observations suffer from moderate $g$ band seeing ($\sim 1.2$ arcsec). However, we detect significant GC colour bimodality, although the blue and the red
peaks are not clearly visible. We confirm that the surface density profile of the GC system extends up to $4$ arcmin, as suggested from the WIYN imaging. 

The combination of the poor $g$ band seeing, and the presence of the $10$th magnitude star $2$ arcmin from the galaxy centre, 
resulted in a low return rate of spectroscopically confirmed GCs. In total, we obtained radial velocities for $61$ GCs over $6$ DEIMOS masks. 
We adopt a colour split at $(g-i)=1.0$. This value was used to analyse the kinematics of blue and red GC subpopulations separately.

We detect significant rotation only for the blue GCs and for a small group of red GCs. The blue GCs are found
to rotate at $\sim85 \kms$ along PA$_{\rm kin}=85^{+26} _{-27}$ deg, consistent with the photometric minor axis and in agreement with that
found by \citet{Coccato} using PNe. Such a peculiarity is clearly visible in Figure~\ref{fig:CV}. We will discuss this feature and its implications in more detail later. 
Interestingly, we note that the direction of the GC and PNe rotation coincides with an elongated jet/outflow structure detected in the X-ray \citep{Pellegrini}.
With the adopted colour split, the kinematic position angle of the red GCs is generally unconstrained, except the outer red GCs that counter rotate with respect to the host galaxy stars. 
The velocity dispersion of both the red and the blue GCs declines with radius with a slope similar to that of the PNe and host galaxy stars.
\begin{table*}
\begin{tabular}{@{}l l l l l l l l l l l}
\hline
Galaxy ID & $V_{\rm sys, GC}$& $V_{\rm  rms,A}$ & $V_{\rm rms,B}$ & $V_{\rm rms,R}$ & $(V_{\rm rot} / \sigma)_{\rm A}$  &  $(V_{\rm rot} / \sigma)_{\rm B}$ & $(V_{\rm rot} / \sigma)_{\rm R}$ & PA$_{\rm kin, A}$ & PA$_{\rm kin, B}$ & PA$_{\rm kin, R}$  \\
      &  [\kms]    & [\kms]  &   [\kms]   &  [\kms]     &  &       &      &   [deg]      &   [deg]       &   [deg]               \\

 (1)  &  (2)      &  (3)            & (4)            & (5)     & (6)    &  (7)         &  (8)      &       (9)           &       (10) & (11) \\
\hline
\hline   
NGC~$821$	& $1750$ 	& $158_{-13} ^{+17} $ &	$145_{-18} ^{+28} $ &	$	154_{-17} ^{+26}$ &	$ 0.11_{-0.18} ^{+0.18}$ &	$	0.60_{-0.30} ^{+0.33}$ &	$0.09_{-0.22} ^{+0.27}$ &	$80_{-90} ^{+183}$ &	$	85_{-27} ^{+26}$          &	$	252_{-137} ^{+94}$\\
NGC~$1400$	& $612$ 	& $137_{-11} ^{+14} $ &	$140_{-15} ^{+22} $ &	$	135_{-14} ^{+19}$ &	$ 0.07_{-0.13} ^{+0.17}$ &	$	0.10_{-0.15} ^{+0.25}$ &	$0.54_{-0.26} ^{+0.27}$ &	$39_{-48} ^{+49}$   &	$	200_{-60} ^{+55}$        &	$	30_{-20} ^{+21}$\\
NGC~$1407$	& $1774$ 	& $224_{-8} ^{+9}      $ &	$232_{-12} ^{+14} $ &	$	216_{-11} ^{+12}$ &	$ 0.17_{-0.08} ^{+0.07}$ &	$	0.03_{-0.08} ^{+0.10}$ &	$0.32_{-0.10} ^{+0.10}$ &	$285_{-29} ^{+30}$ &	$	224_{-101} ^{+120}$   &	$       299_{-19} ^{+18}$\\
NGC~$2768$	& $1338$ 	& $165_{-11} ^{+13} $ &	$173_{-17} ^{+23} $ &	$	160_{-13} ^{+17}$ &	$ 0.39_{-0.11} ^{+0.13}$ &	$	0.34_{-0.23} ^{+0.21}$ &	$0.43_{-0.14} ^{+0.15}$ &	$97_{-24} ^{+22}$   &	$	121_{-34} ^{+32}$        &	$	79_{-27} ^{+28}$\\
NGC~$3377$	& $685$ 	& $100_{-6} ^{+7}      $ &	$100_{-9} ^{+12}   $ &	$	85_{-7}   ^{+10} $ &	$ 0.19_{-0.14} ^{+0.13}$ &	$	0.09_{-0.14} ^{+0.15}$ &	$0.58_{-0.20} ^{+0.20}$ &	$197_{-40} ^{+40}$ &	$	275_{-108} ^{+77}$      &	$	181_{-19} ^{+21}$\\
NGC~$4278$	& $637$ 	& $177_{-7} ^{+9}      $ &	$182_{-10} ^{+12} $ &	$	152_{-10} ^{+12}$ &	$ 0.17_{-0.08} ^{+0.08}$ &	$	0.16_{-0.13} ^{+0.11}$ &	$0.25_{-0.14} ^{+0.15}$ &	$200_{-26} ^{+26}$ &	$	194_{-38} ^{+32}$        &	$	207_{-42} ^{+39}$\\
NGC~$4365$	& $1210$	& $248_{-10} ^{+12} $ &	$230_{-16} ^{+20} $ &	$	258_{-13} ^{+16}$ &	$ 0.11_{-0.10} ^{+0.09}$ &	$	0.17_{-0.14} ^{+0.18}$ &	$0.26_{-0.12} ^{+0.11}$ &	$129_{-42} ^{+37}$ &	$	307_{-40} ^{+29}$        &	$	121_{-26} ^{+26}$\\
NGC~$5846$	& $1706$ 	& $238_{-11} ^{+13} $ &	$266_{-18} ^{+22} $ &	$	207_{-14} ^{+17}$ &	$ 0.02_{-0.12} ^{+0.15}$ &	$	0.19_{-0.10} ^{+0.14}$ &	$0.12_{-0.15} ^{+0.14}$ &	$157_{-79} ^{+60}$ &	$	259_{-71} ^{+60}$        &	$	107_{-56} ^{+43}$\\
NGC~$7457$	& $847$ 	& $68_{-9} ^{+12}      $ &	$-$ 				&	$-$				      &	$ 1.68_{-0.40} ^{+0.37}$ &	$-$					    &	$-$ 			                &	$324_{-10} ^{+10}$ &	$-$				      &					$-$\\
\hline
NGC~$3115$	& $710$  &  $164_{-8} ^{+9}   $ &	$166_{-11} ^{+13} $ &	$	161_{-10} ^{+13}$ &	$ 0.66_{-0.09} ^{+0.09}$ &	$	0.74_{-0.13} ^{+0.12}$ &	$0.61_{-0.13} ^{+0.14}$ &	$35_{-12} ^{+11} $ &	$	47_{-13} ^{+13}$ &	$	20_{-17} ^{+16}$\\
NGC~$4486$	& $1336$  &  $328_{-12} ^{+14} $ &	$296_{-22} ^{+28} $ &	$	337_{-14} ^{+16}$ &	$ 0.07_{-0.04} ^{+0.05}$ &	$	0.09_{-0.08} ^{+0.07}$ &	$0.29_{-0.17} ^{+0.17}$ &	$105_{-38} ^{+40}$ &	$	92_{-48} ^{+49}$ &	$	123_{-43} ^{+39}$\\
NGC~$4494$	& $1338$ & $99_{-12} ^{+14}    $ &	$98_{-7} ^{+9}        $ &	$	99_{-11}  ^{+15}$ &	$ 0.60_{-0.11} ^{+0.12}$ &	$	0.65_{-0.15} ^{+0.15}$ &	$0.43_{-0.23} ^{+0.21}$ &	$174_{-17} ^{+16}$ &	$	167_{-1} ^{+16}$ &	$	195_{-31} ^{+33}$\\\hline
\hline
\end{tabular}
\caption{GC kinematic results for our galaxy sample. Column (2) shows the systemic velocity of the host galaxy obtained from the weighted average of the GC radial velocities of the azimuthal-complete innermost region.
The remaining columns are the overall root-mean-square velocity $V_{\rm rms}$, the rotational dominance parameter $(V_{\rm rot} / \sigma)$ and kinematic position angle PA for all, blue and red GC subpopulations respectively.}
\label{tab:KinematicsTable} 
\end{table*}

\subsection{NGC~1400}

NGC~$1400$ has been classified both as a face-on S$0$ \citep{Jarrett} and as an E0 (e.g., \citealt{daCosta}). It is the second brightest galaxy in the 
Eridanus group after NGC~$1407$. We have assumed that NGC~$1407$ and NGC~$1400$ lie at the same 
distance of $26.8$ Mpc. NGC~$1400$ is characterised by a uniformly old stellar age up to $\sim1.3\Reff\ $\citep{Spolaor} and by an unusually 
low systemic velocity (V$_{\rm sys} = 558 \kms$) for its distance. \citet{Forbes06A} studied the GC system of this galaxy using Keck/LRIS in imaging mode, detecting significant bimodality. 

The photometric GC selection was performed within $8$ arcmin from the galaxy centre to minimise the contamination from NGC~$1407$ GC system, resulting in a clear
bimodal distribution, with the colour separation occurring at $(g-i)=0.98$. The GC surface density was corrected for the local NGC~$1407$ contribution.
This galaxy shows the steepest red surface density profile among our galaxies, similar to the slope of the galaxy surface brightness. In contrast,
the profile of the blue GC subpopulation is more radially extended and also requires a power-law fit, because a S\`ersic function does not return a satisfactory solution.

We present here radial velocities for $34$ blue and $37$ red GCs respectively, for a total of $71$ spectroscopically confirmed GCs. Despite their small angular
separation $(\sim10$ arcmin), NGC~$1407$ and NGC~$1400$ also have a large peculiar velocity difference ($\sim1200$ \kms), assuring a reliable 
separation of their spectroscopically confirmed GCs. The red GCs mimic the rotation of the stars in the inner regions, whereas the rotation of the blue GCs is consistent with zero
with a marginal signature of counter-rotation in the inner radial bin. The velocity dispersion of the red GCs is in agreement with long--slit data in the region of overlap, with a slightly 
increasing trend towards the outer regions. Conversely, the velocity dispersion of the blue GCs decreased with the radius.

\subsection{NGC~1407}
\label{sec:NGC1407}
NGC~$1407$ is a massive E$0$ galaxy at the centre of the dwarf galaxy dominated Eridanus A group \citep{Brough}. It shows moderate rotation along the photometric 
major axis and has a weak central AGN \citep{Zhang}. The stellar population analysis of \citet{SpolaorB} found the galaxy to possess a uniformly old age within $\sim0.6\Reff$. 
A dynamical analysis of NGC~$1407$ was given by \citet{Romanowsky09}. They used Suprime-Cam imaging and DEIMOS spectra for $172$ GCs and
found a massive dark halo. 

In this work, we use the photometric results presented in \citet{Romanowsky09} in which the Suprime-Cam imaging was reduced and 
analysed with the same methodology described in this paper. The colour distribution shows clear bimodality with the colour separation occurring at $(g - i)=0.98$. 
We supplement the spectroscopic sample of \citet{Romanowsky09} with $6$ additional DEIMOS masks, that make NGC~$1407$, with a total of $369$ 
spectroscopically confirmed GCs, the most populous spectro-photometric dataset in our galaxy sample. 

We detect rotation along the photometric major axis for both GC subpopulations. The blue GCs rotate in the innermost radial bin along the photometric minor axis and between $3$ and $5$ \Reff\ along the photometric major axis. 
Similarly, the rotation of the red GCs occurs along the photometric major axis directions, but in this case the rotation signal is generally larger and better constrained than that of the blue GCs. 
Both GC subpopulations appear to rotate in the outermost regions of the galaxy along the major axis of the galaxy.
The velocity dispersion of the red GC subpopulation in the inner regions is consistent with stellar data from \citet{Spolaor} and \citet{Proctor} showing a decreasing profile up to 10 \Reff. 
Conversely, the velocity dispersion of the blue GCs increases with radius.

\subsection{NGC~2768}
\label{sec:NGC2768}
NGC~$2768$ is classified as E$6$ in \citet{Vaucouleurs} and as S$0_{1/2}$ in \citet{Sandage}. Spectroscopic studies of
the very central regions \citep{McDermid} have also suggested a young ($2.5$ Gyr) stellar population associated with the disk, as supported 
by the recent supernova SN2000ds \citep{Filippenko}. \citet{Kundu} studied the \textit{HST}/WFPC2 photometry for 113 GC candidates in this galaxy, finding a
statistically significant probability of it having a bimodal colour distribution.

In this work, NGC~$2768$ was imaged with $R_{\rm C}iz$ filters, in good seeing conditions but with the lowest exposure time among our sample of galaxies. 
Using KMM we found the colour distribution to be bimodal, with the blue and red peaks at $(R_{\rm C}~-~z)~=~0.41 \pm 0.01$ and $0.70 \pm 0.02$ respectively. 
Bimodality was also found in the \textit{HST}/ACS West pointing used to design the DEIMOS masks in the central region. 
The colour peaks occur at (F435W~$-$~F814W)~$=1.77 \pm 0.03$ and $2.17 \pm 0.01$ respectively. 

We find 109 spectroscopically confirmed GCs over 5 DEIMOS masks observed in sub-arcsec seeing conditions. 
Using a colour split at $(R_{\rm C} - z)=0.57$ and excluding the marginal GCs, we investigate the kinematics of the resulting $60$ blue and $42$ red GCs respectively. 

We find significant rotation only for the red GCs, which rotate roughly along the photometric major axis in agreement with the host galaxy stars. 
The $2$--D velocity field of the red GCs is akin to that of ATLAS$^{\rm 3D}$ in the inner regions (Figure \ref{fig:2D}). 
The blue GCs have a marginally higher velocity dispersion profile than the red GCs.

\begin{figure*}
\centering
\includegraphics[scale=.43]{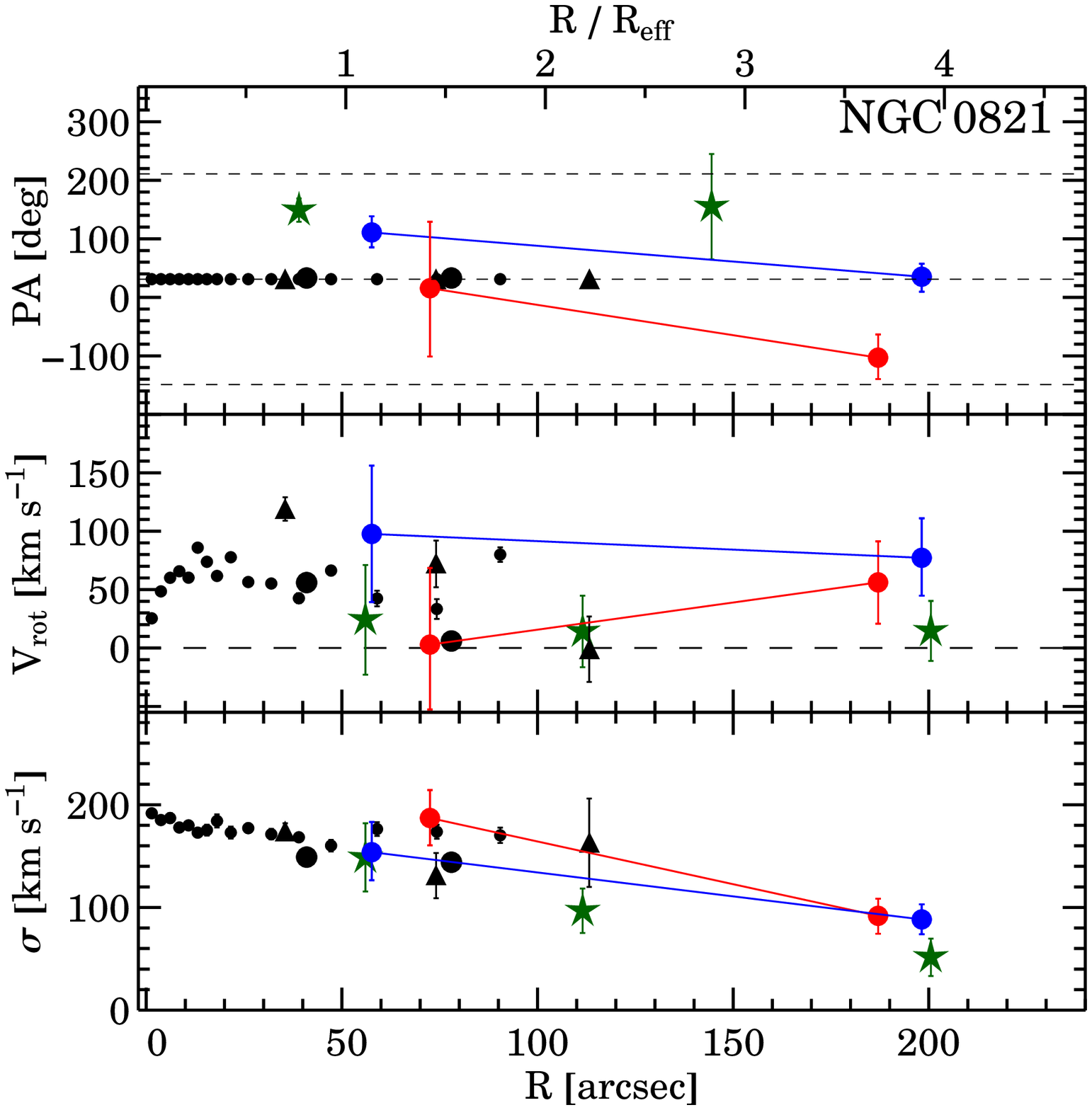} 
\hspace{1 cm} \includegraphics[scale=.43]{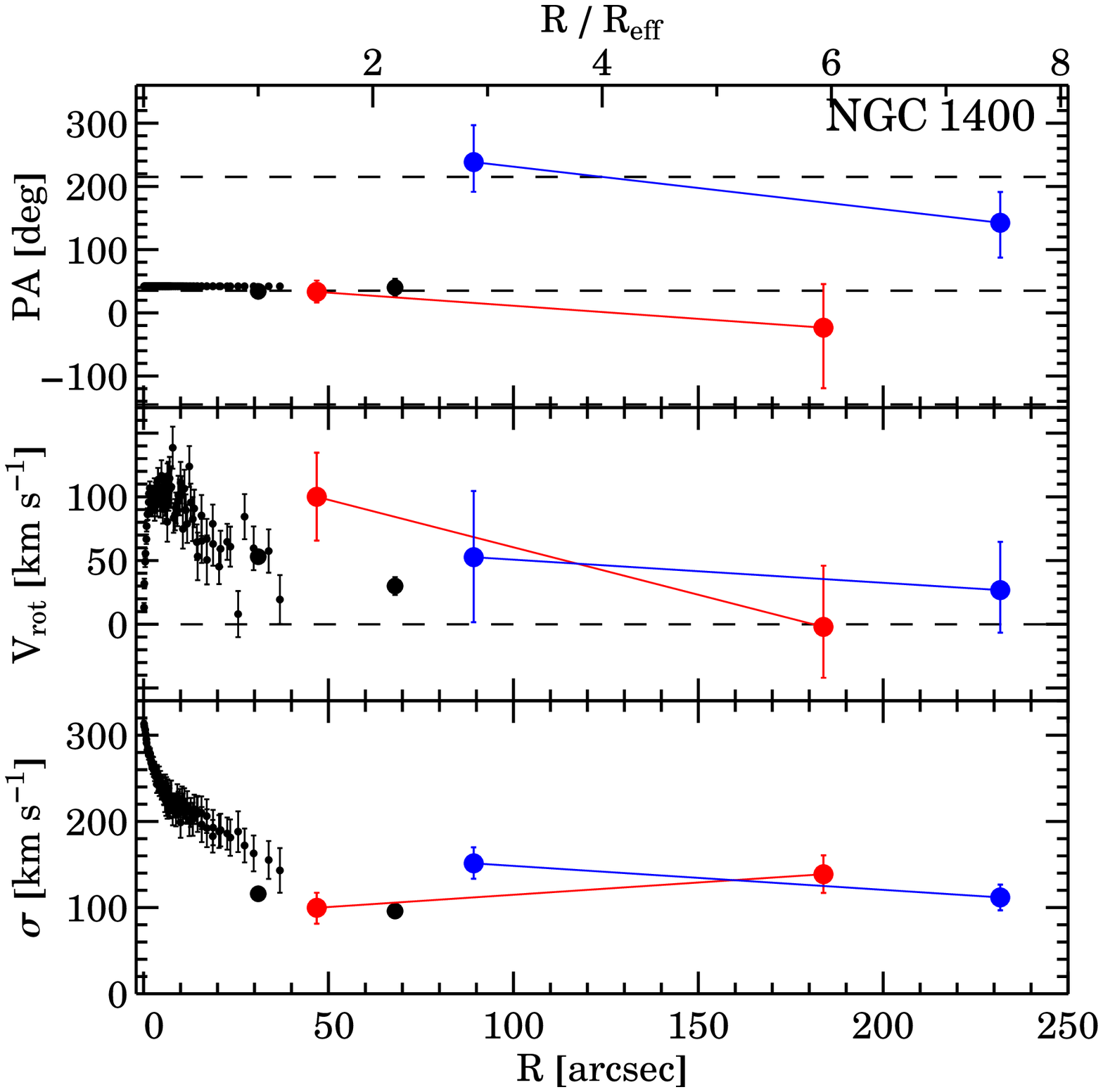} 

\includegraphics[scale=.43]{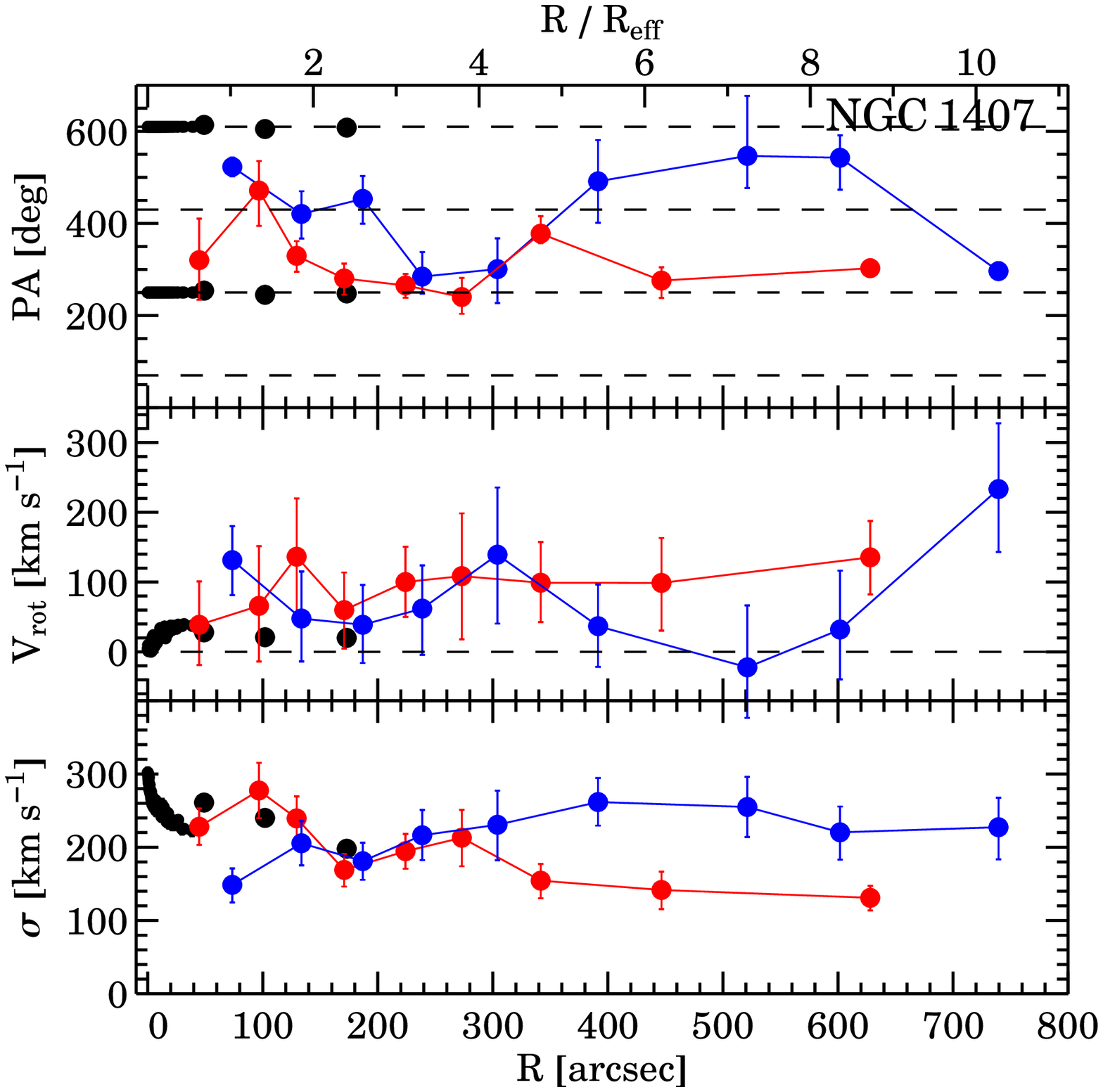} 
\hspace{1 cm} \includegraphics[scale=.43]{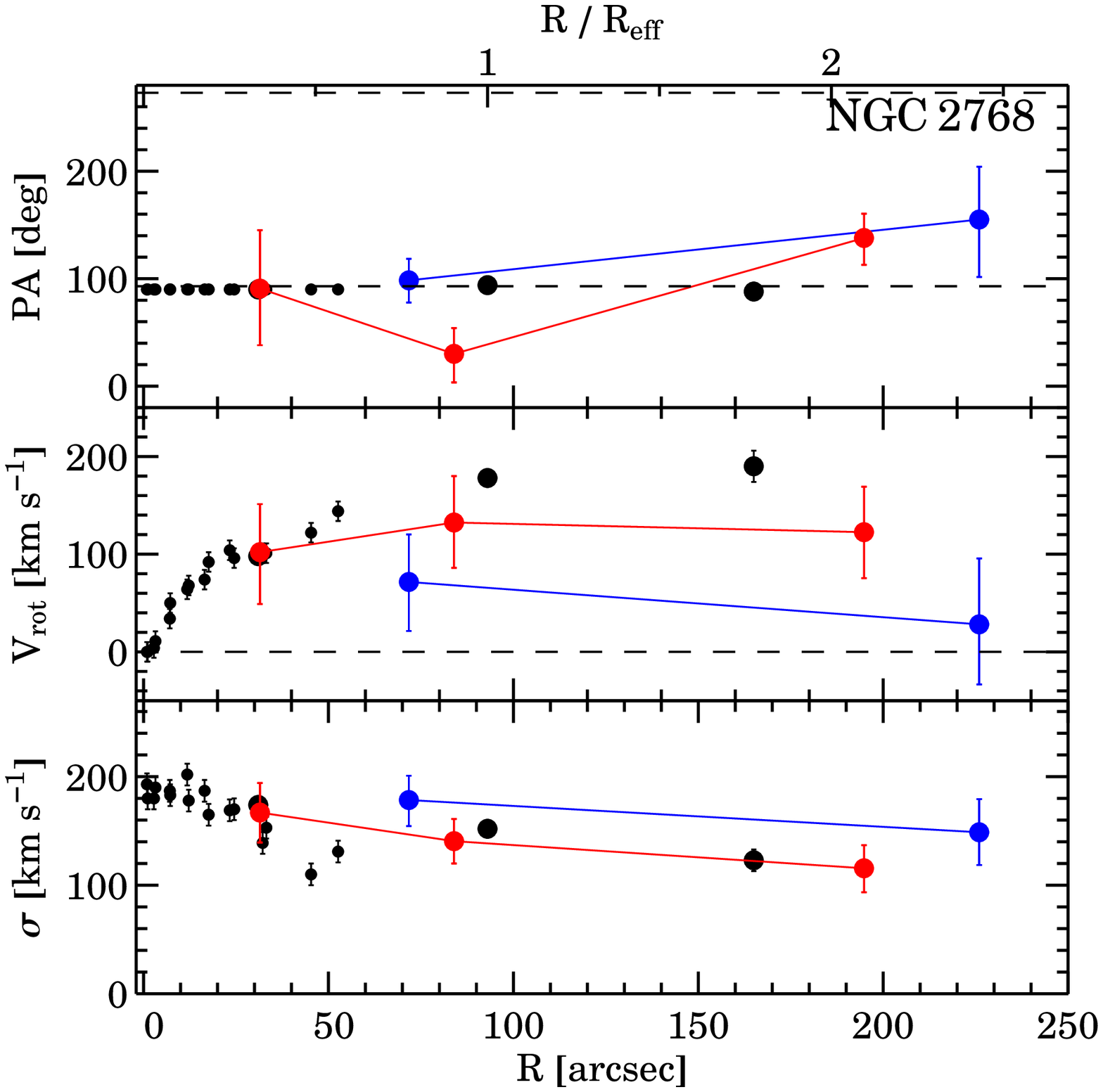} 

\caption{GC system kinematics compared to stellar data. For each galaxy (labeled on the top right of each panel) the plots represent the tern of parameters $(PA_{\rm kin}, V_{\rm rot}, \sigma)$ that minimise
eq. \ref{eq:GC_LR} as a function of the radius. Data points are joined for clarity, with the error bars representing the $68$ per cent confidence intervals. 
The dashed horizontal lines in top and middle panels represent the photometric $PA \pm 180$ deg from Table \ref{tab: survey_summary} and $V_{\rm rot} = 0$, respectively.
All the PNe data (green stars) are from \citet{Coccato}. Integrated stellar light data are shown as small black points (if from long--slit) and large black points (if from \citealt{Proctor}). References for
long--slit data are: \citet{Forestell} for NGC~$821$; \citet{Proctor} for NGC~$1400$ and NGC~$1407$; \citet{Fried} for NGC~$2768$; \citet{Coccato} for NGC~$3377$; \citet{vanderMarel93} for NGC~$4278$;
\citet{Bender94} for NGC~$4365$; \citet{Kronawitter} for NGC~$5846$; \citet{Simien} for NGC~$7457$. Black trangles in NGC~$821$ are SAURON-IFU data from \citet{Weijmans}.
For NGC~$7457$, because of the low number statistics, we only show the total GC kinematic profile in black.}
\label{fig:kinematics}
\end{figure*}

\begin{figure*}
\centering
\includegraphics[scale=.43]{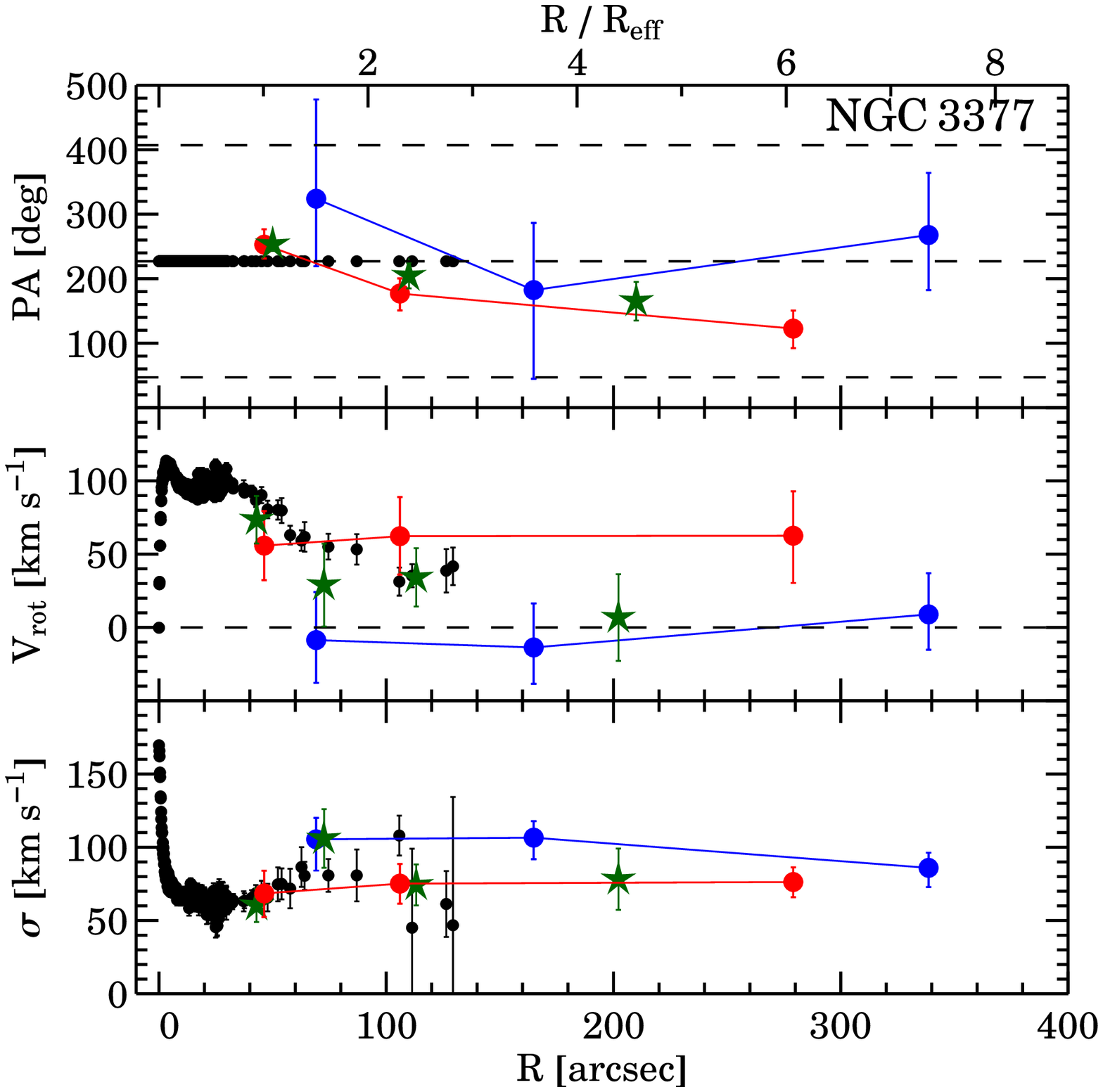} 
\hspace{1 cm}  \includegraphics[scale=.43]{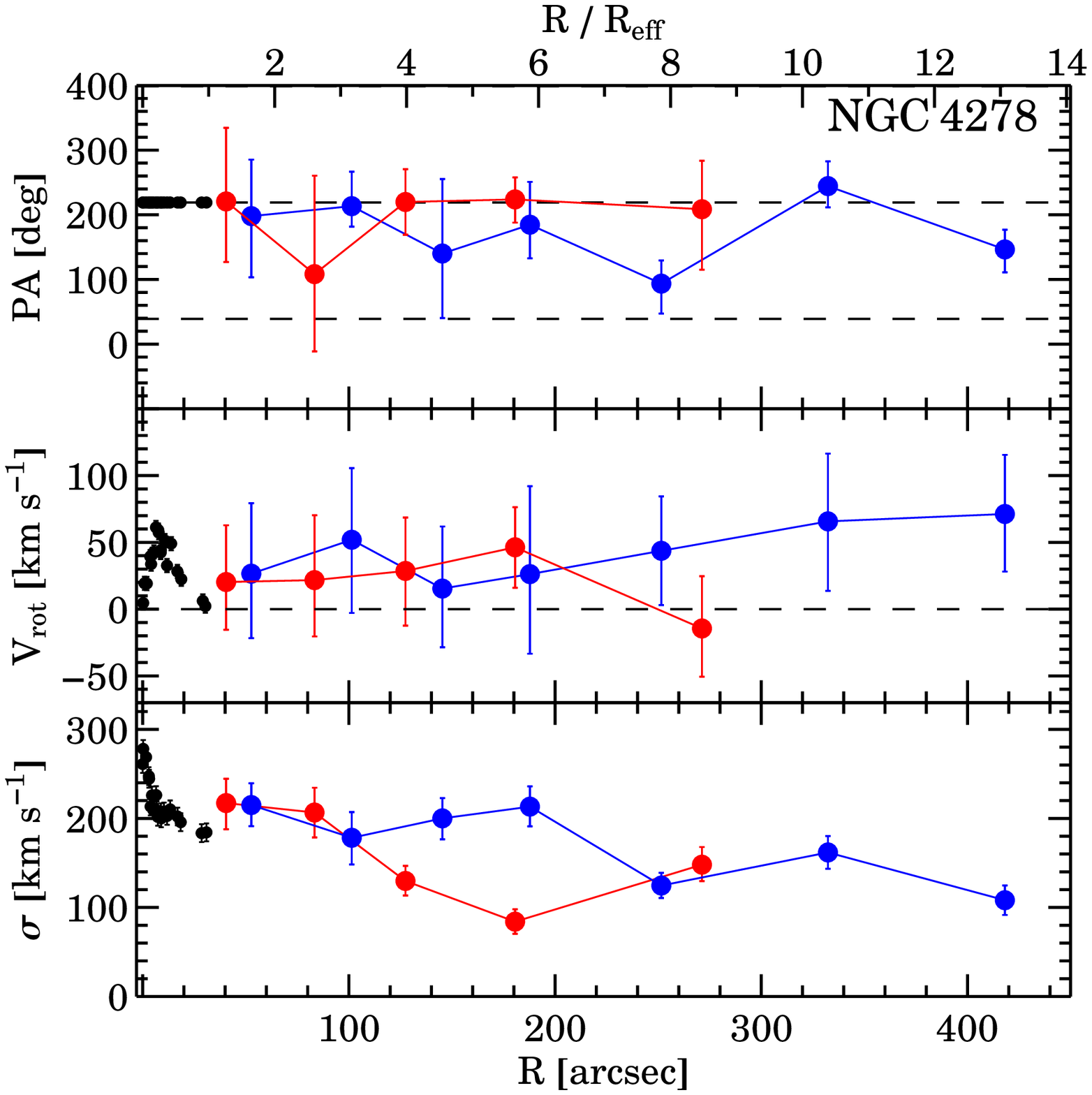}  \\

\includegraphics[scale=.43]{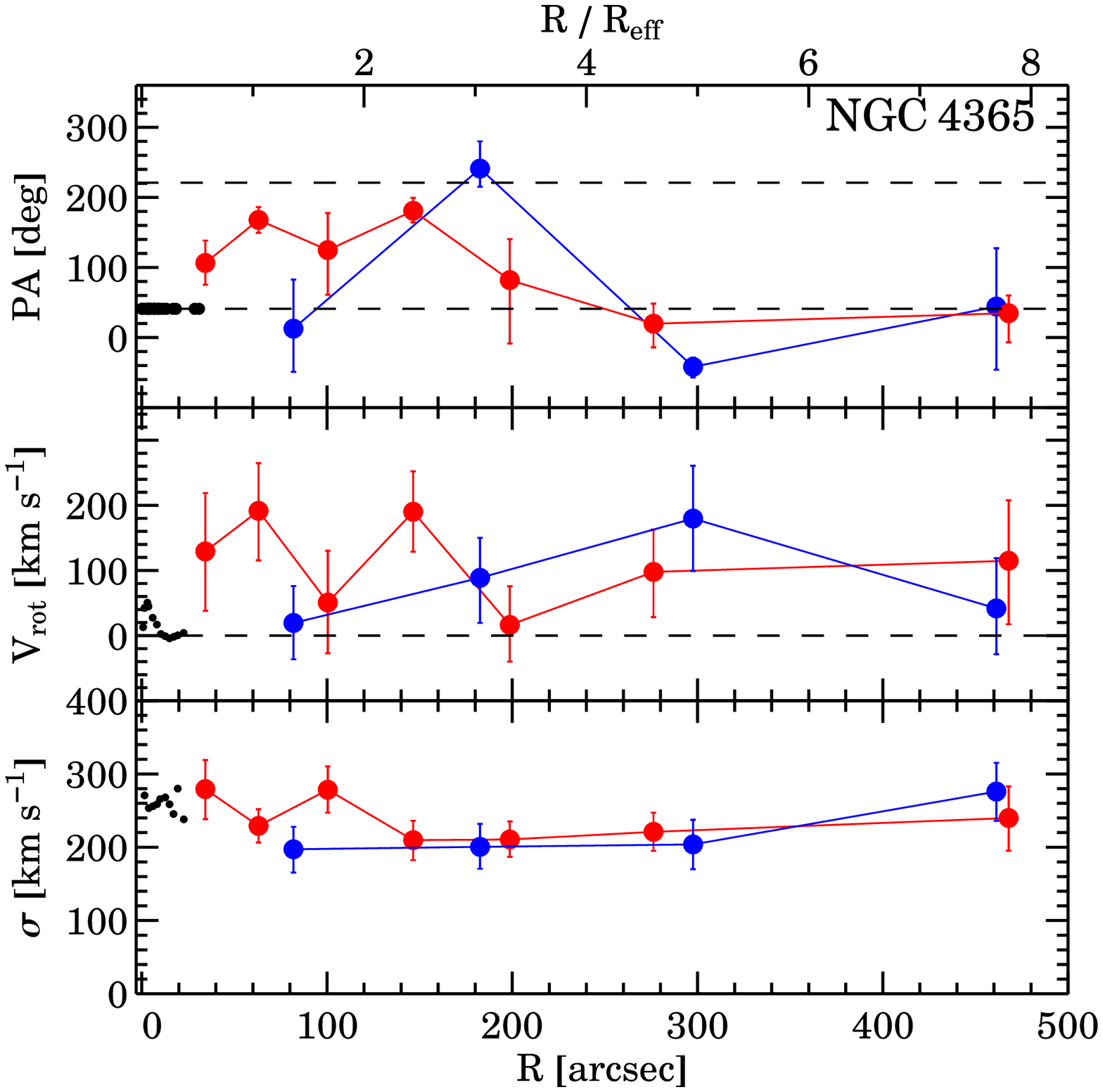} 
\hspace{1 cm} \includegraphics[scale=.43]{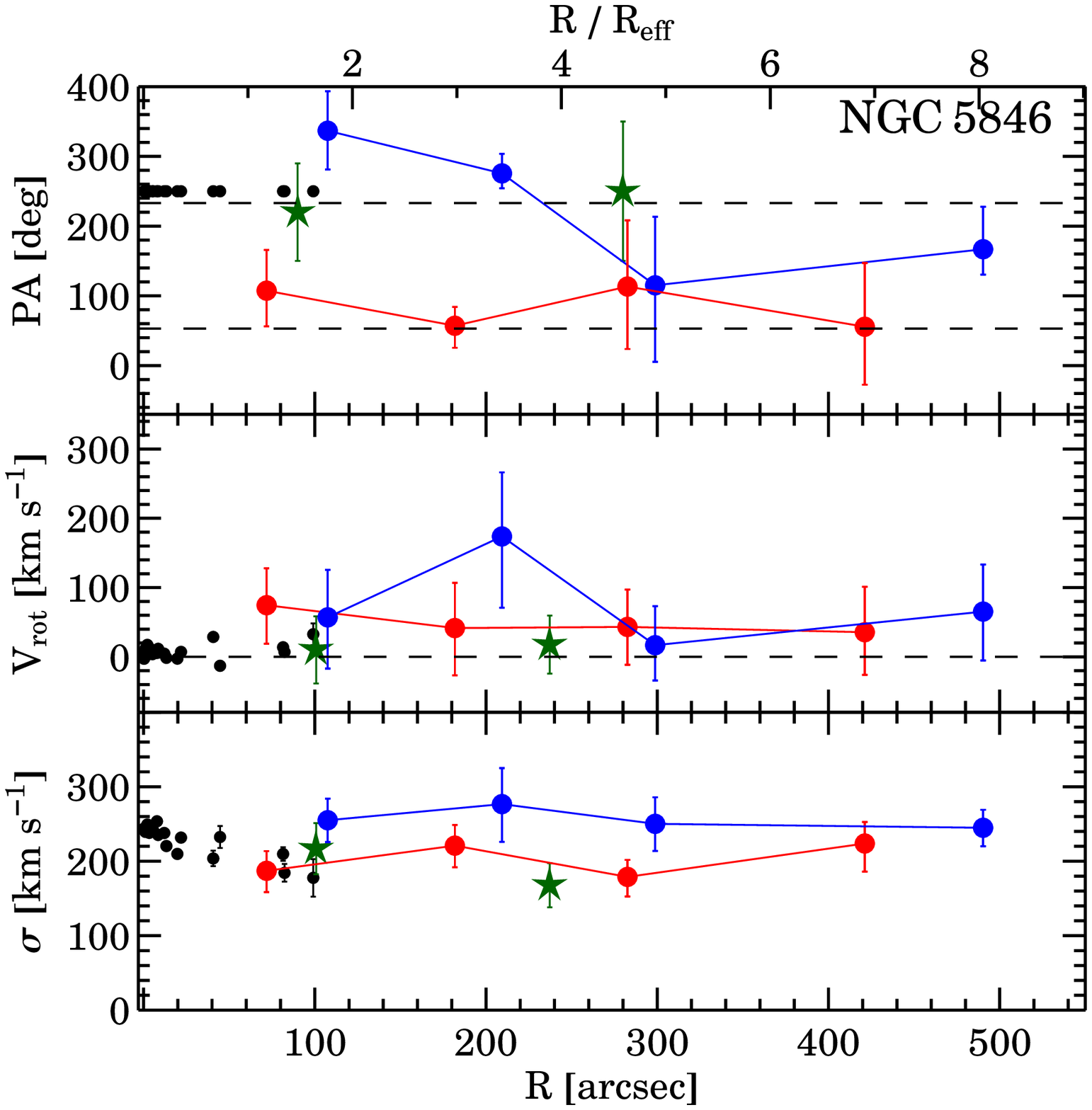}  \\
\includegraphics[scale=.43]{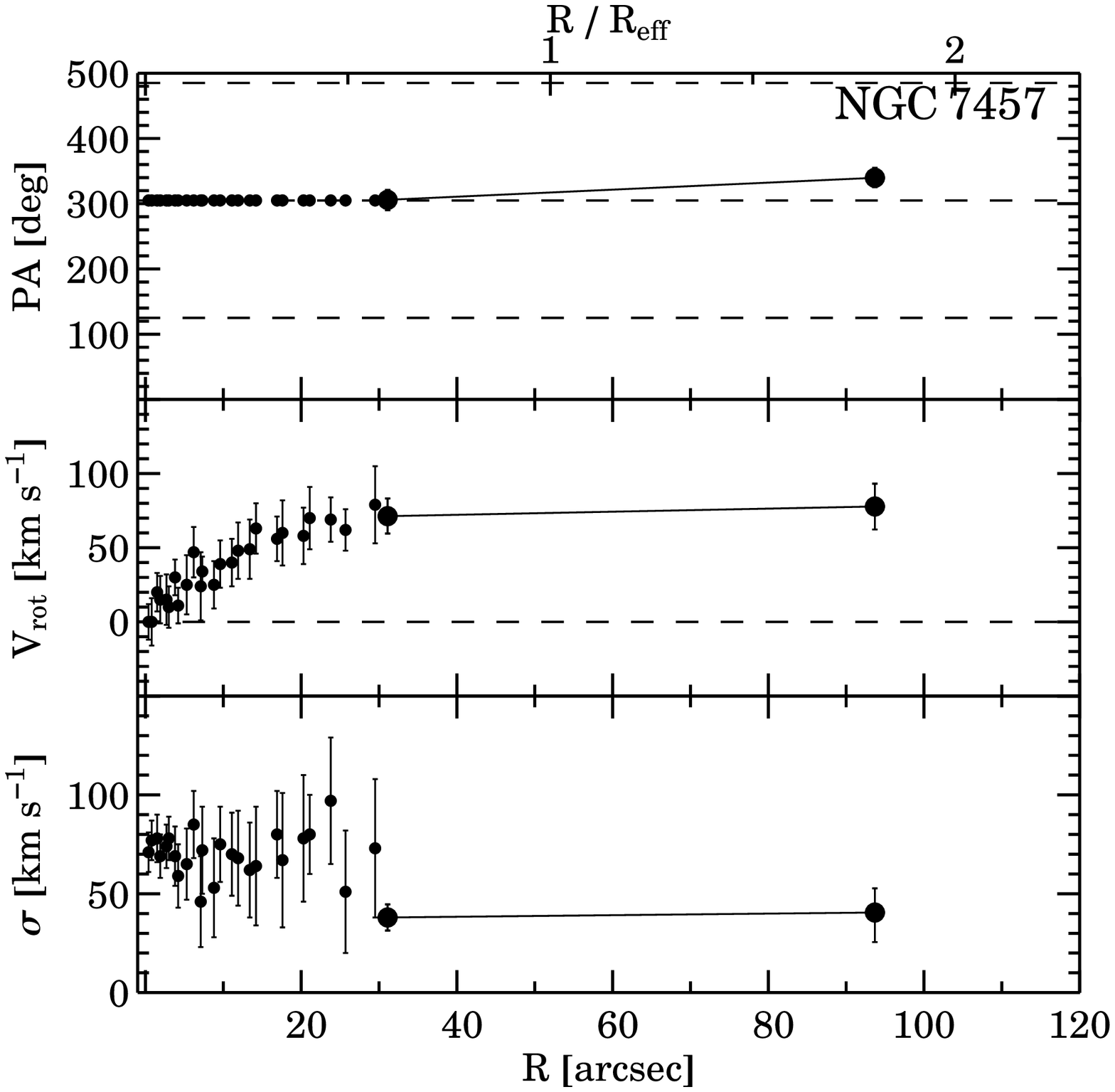} 

\contcaption{}
\end{figure*}

\subsection{NGC~3377}
\label{sec:NGC3377}

%%%%%%%%%%%%%%%%%%%%%%%%%%%%%%COLOUR%%%%%%%%%%%

\begin{figure*}
\centering
\includegraphics[scale=.45]{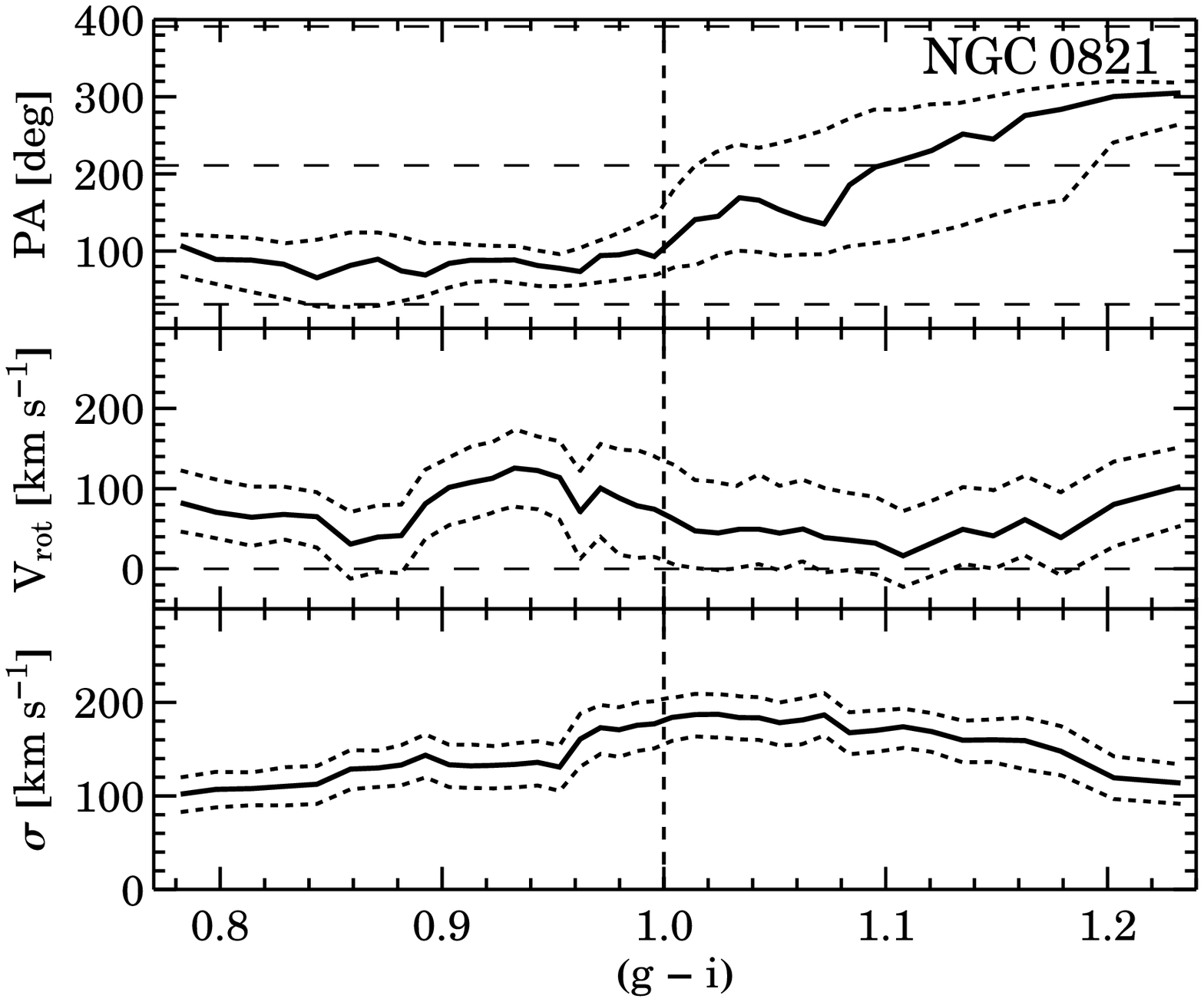} 
\hspace{1 cm} \includegraphics[scale=.45]{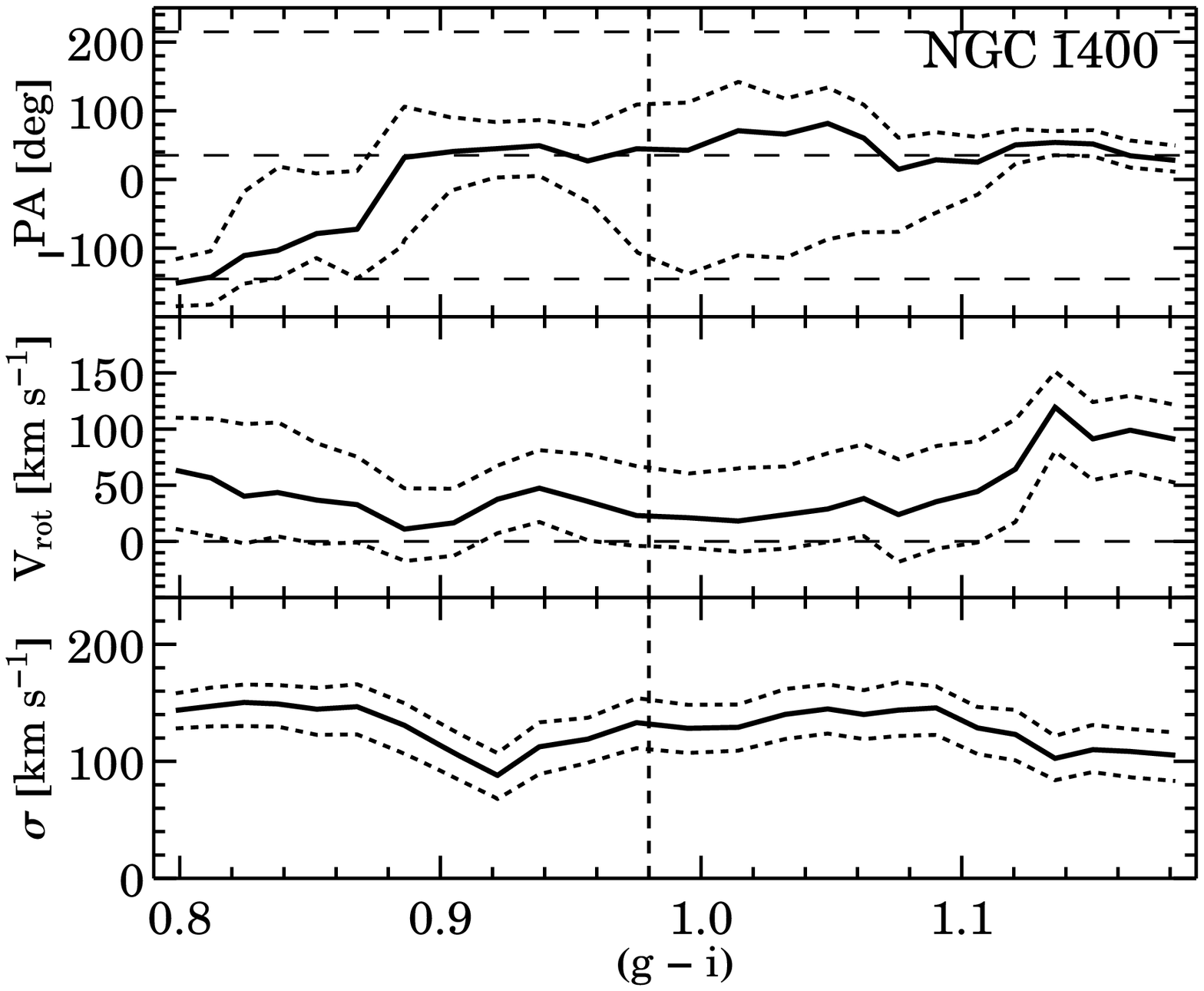} 

\includegraphics[scale=.45]{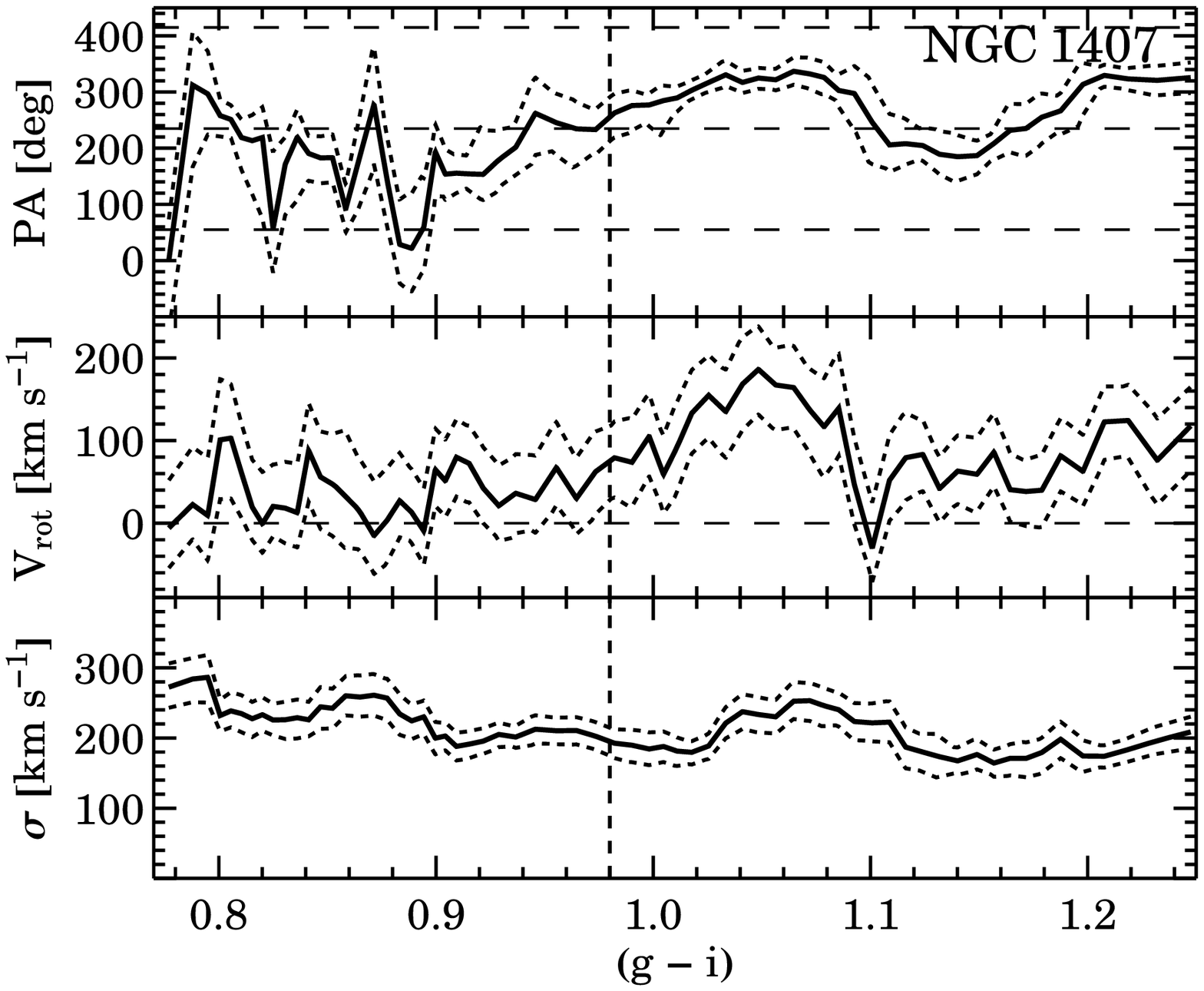} 
\hspace{1 cm} \includegraphics[scale=.45]{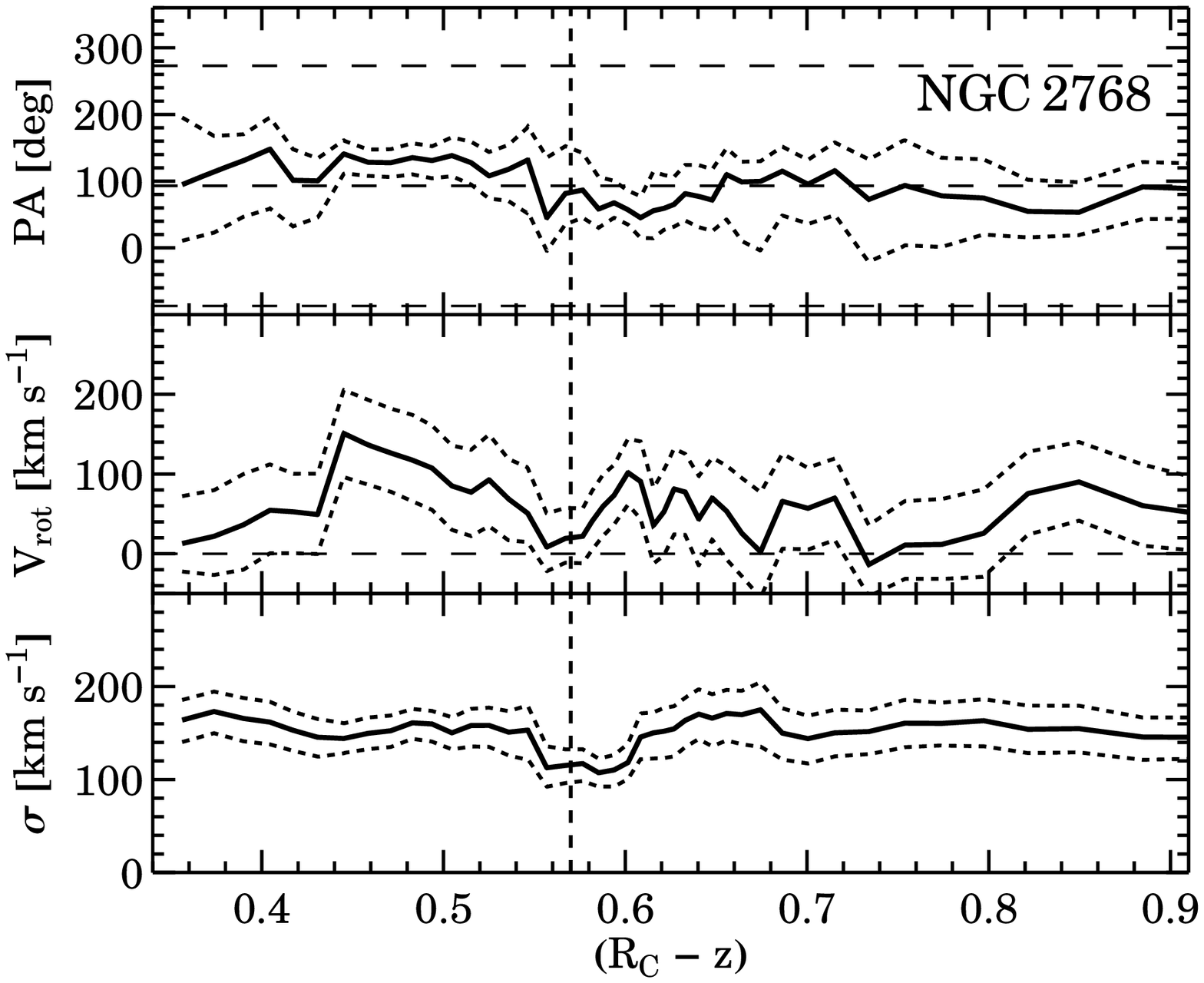} 

\caption{GC kinematics as a function of colour. 
Kinematic position angle, rotation velocity and velocity dispersion are shown in the top, central and bottom panel respectively. Dotted lines represent $68$ per cent confidence intervals. 
Also shown are the photometric major axis (horizontal dotted lines) from Table \ref{tab: survey_summary} and the colour split between blue and red GC subpopulations 
(vertical dashed line) as derived from the KMM analysis.}
\label{fig:CV}
\end{figure*}

\begin{figure*}
\centering
\includegraphics[scale=.43]{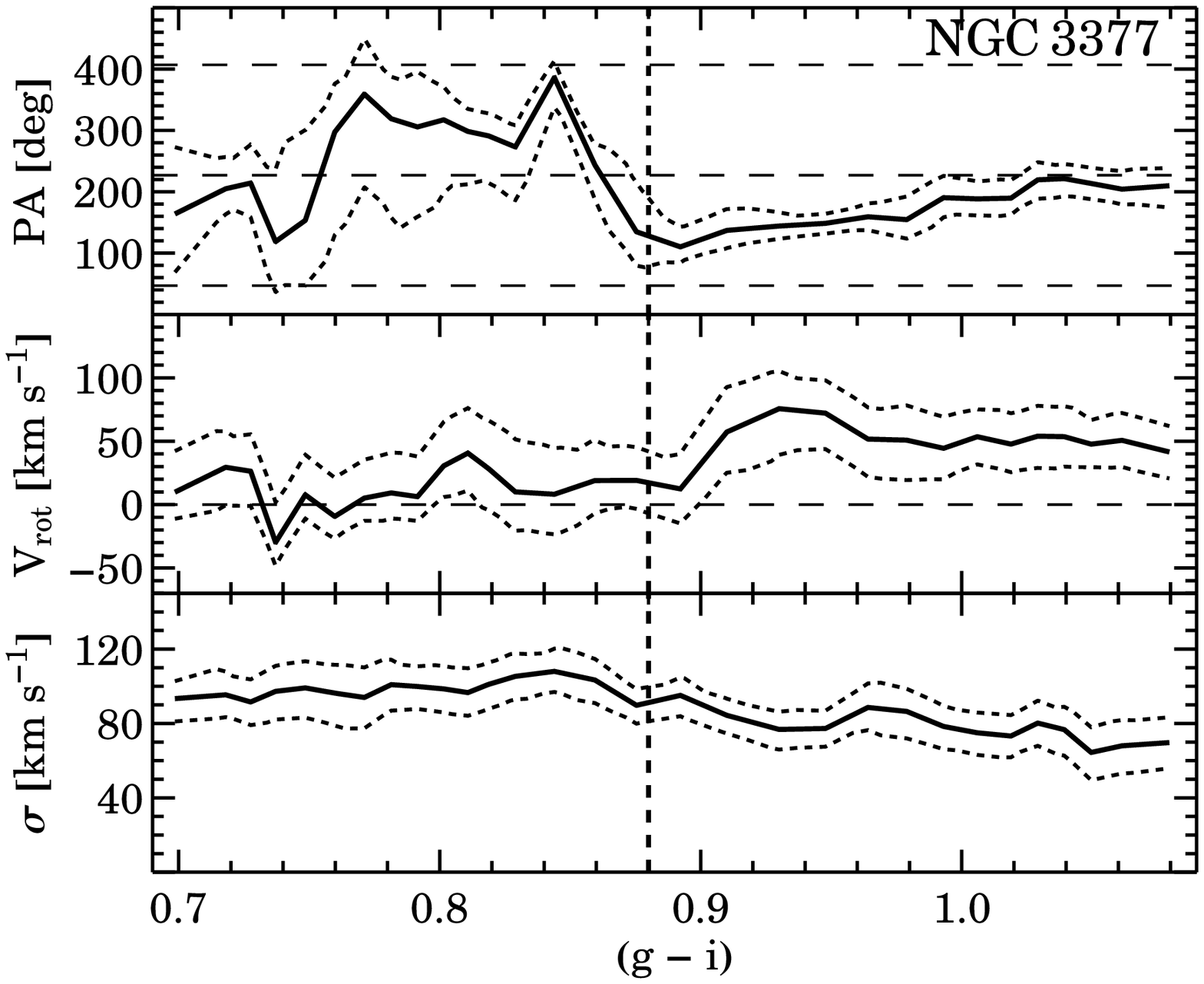} 
\hspace{1 cm}  \includegraphics[scale=.43]{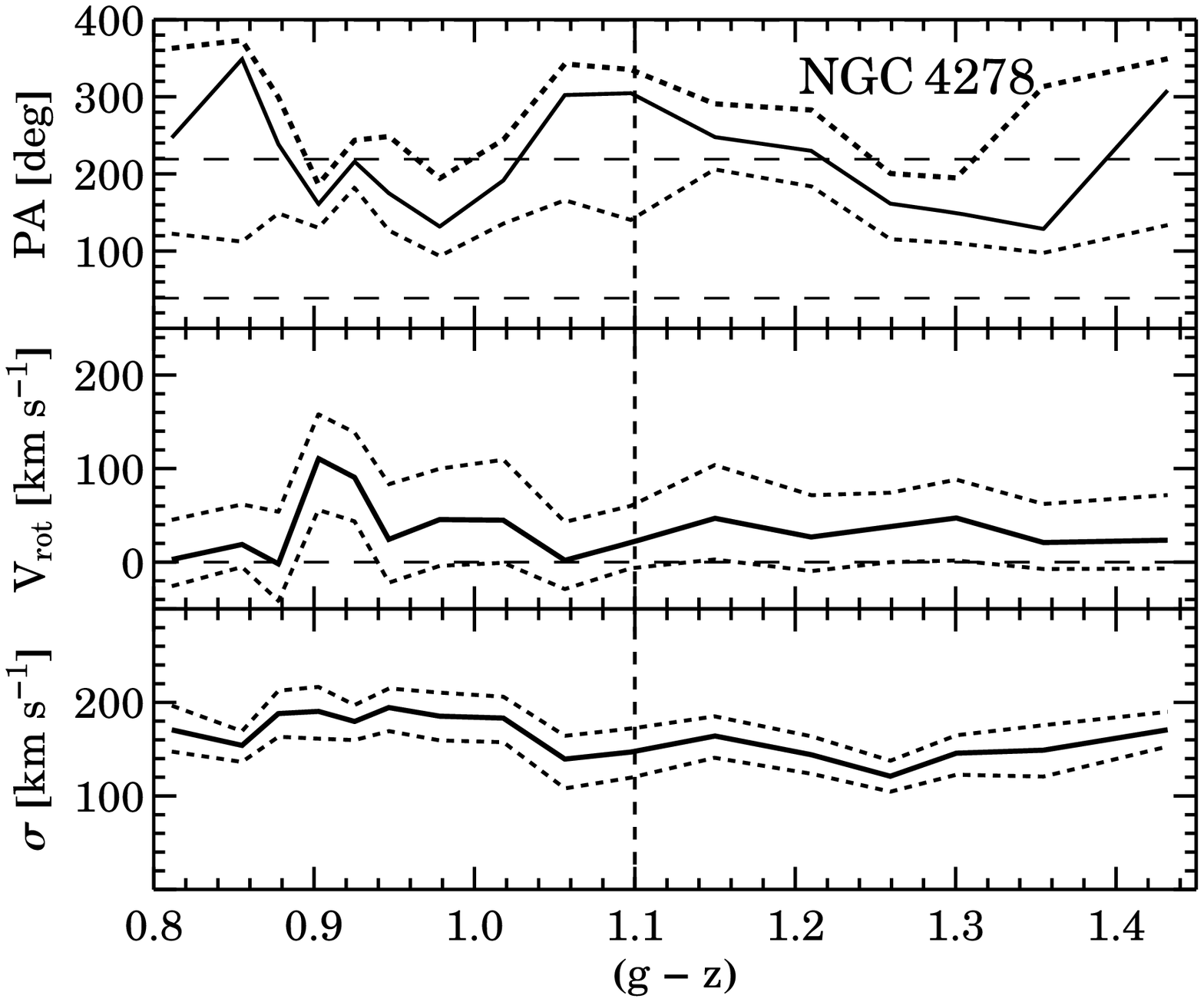}  \\

\includegraphics[scale=.43]{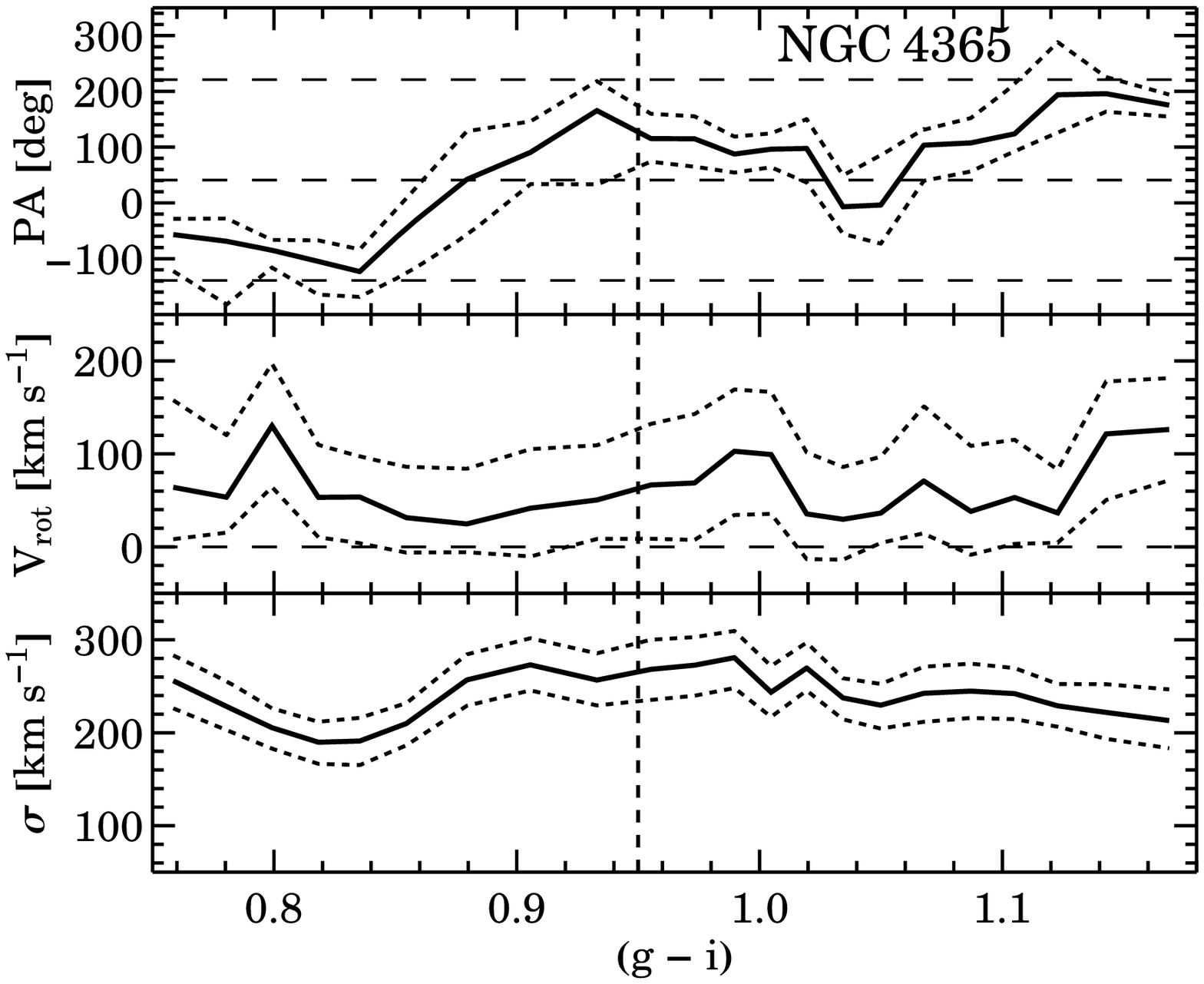} 
\hspace{1 cm} \includegraphics[scale=.43]{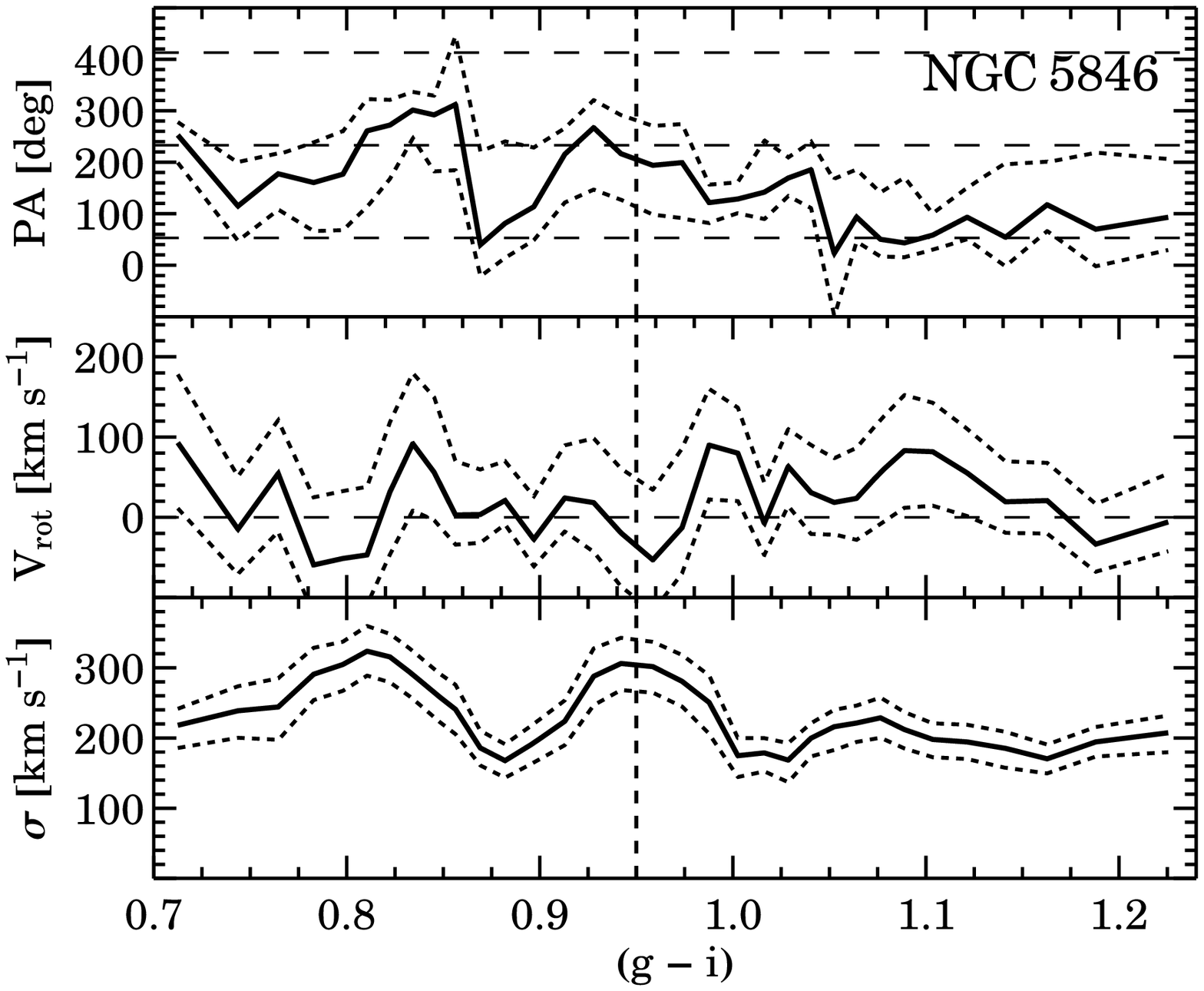}  \\
\includegraphics[scale=.43]{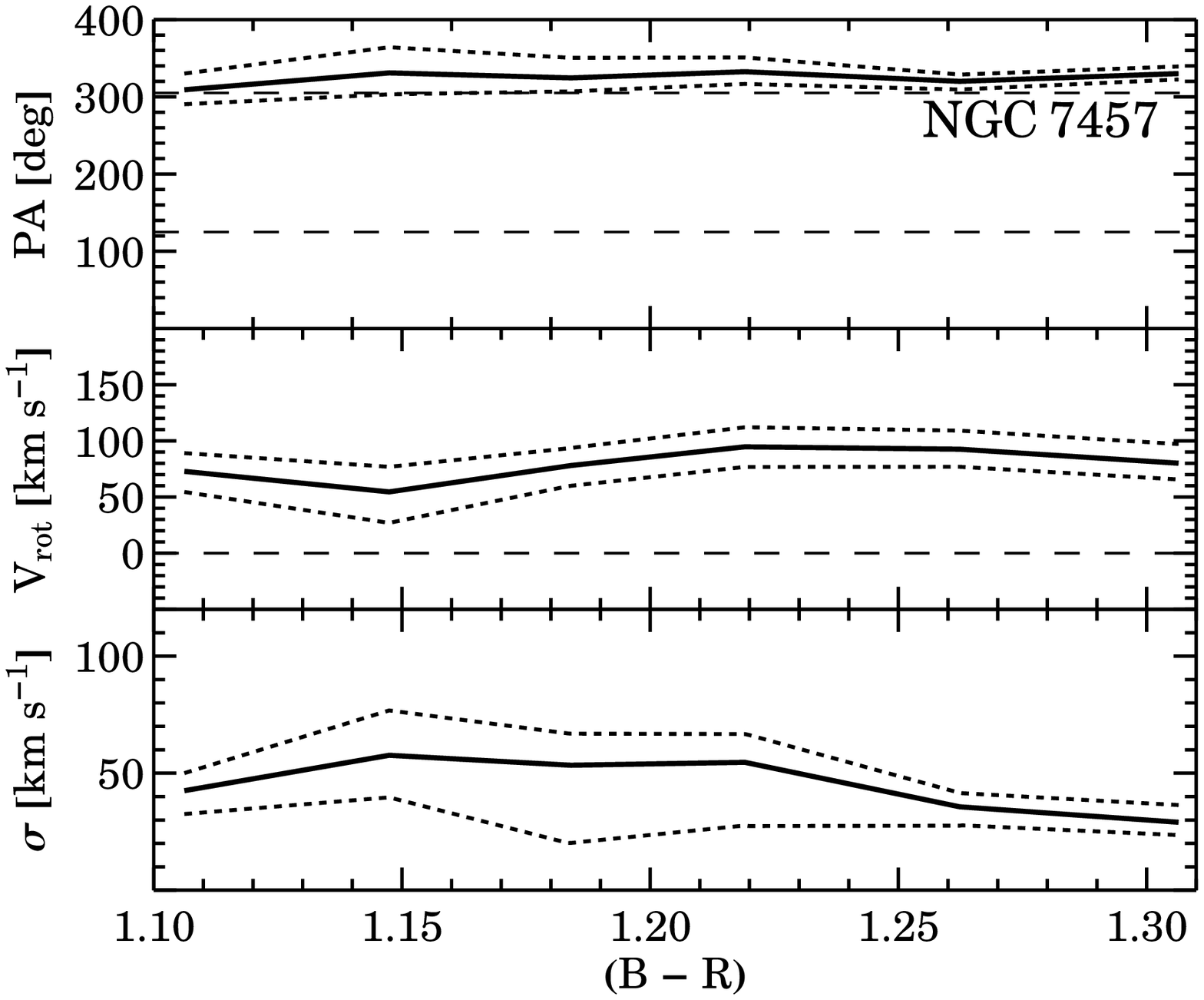} 

\contcaption{}
\end{figure*}

%%%%%%%%%%%%%%%%%%%%%%%%%%%%%%COLOUR ENDS%%%%%%%%%%%

NGC~$3377$ is a classical E6 elliptical in the Leo I group and the closest galaxy in our sample. 
The resolved stellar properties of NGC~$3377$ have been studied by \citep{Harris07A} that found no evidence for any young ($< 3$ Gyr) stellar population.  
\citet{Coccato} studied $159$ PNe in this galaxy, detecting significant rotation within $2$ arcmin and a kinematic major axis twist of $\sim 80$ deg from the 
major towards the minor axis. \citet{Kundu}, \citet{Chies-Santos} and \citet{Cho}
studied the GC system of this galaxy using \textit{HST} imaging, finding the colour distribution to be likely bimodal. 

NGC~$3377$ was imaged with $gri$ filters and \textit{gz} filters for ground based and space based observations respectively.
We found a high probability for the colour distribution to be bimodal, both in our Suprime-Cam and in the supplementary
ACS archive imaging, as already found in \citet{Chies-Santos}. 

In this work, we present radial velocities for 126 GCs, observed in 4 DEIMOS masks. It is worth noting that we extend the kinematics
of this galaxy to $8~\Reff\ \sim 25$ kpc, two times further than the PNe studies. 
We rule out two GCs that are likely to belong to the spiral galaxy NGC~$3377$A (V$_{\rm sys} \sim$ 573 \kms) that lies $\sim 8$ arcmin North-West from NGC~$3377$. 
With a colour split at $(g-i)=0.88$ we study the kinematics of $57$ blue and $60$ red GCs, respectively.

We detect significant rotation for the red GC subpopulation along the photometric major axis in agreement with PNe and long--slit data \citep{Coccato}. 
As found for the PNe, we discover that the kinematic position angle of the red GC subpopulation twists with radius from $250$ deg to $150$ deg (see also Figure \ref{fig:NGC3377PA}). 
The $2$--D velocity field of the red GCs is consistent with ATLAS$^{\rm 3D}$ in the inner regions (Figure \ref{fig:2D}). 
The position angle of the blue GC subpopulation is unconstrained, implying an overall null rotation with radius.
The velocity dispersion of the red GC subpopulation is very flat and it matches the stellar and the PNe. The blue GCs
have an overall higher velocity dispersion profile than the red GC subpopulation.

\subsection{NGC~4278}
\label{sec:NGC4278}
NGC~$4278$ is an elliptical (E$1$--$2$) member of the Coma I cloud. It has been extensively studied at different wavelengths: in radio, \citet{Nagar} detected two sub-parsec jets;  
in X-rays, a long (months) time scale variability \citep{Ho} and a dominant nuclear source \citep{Younes} were detected. It has long been known for  its massive 
($\sim 10^8$~M$_{\sun}$) H{\,\small I} disc extending beyond $10$ \Reff\ \citep{Gallagher} and used to infer the dark matter content at large radii \citep{Bertola93}.
The bimodality of the GC system of NGC~$4278$ has been under debate. Both \citet{Forbes96b} and \citet{Kundu} investigated the GC system of NGC~$4278$
with \textit{HST}/WFPC2, finding a marginal or ``likely'' probability for the $(V- I)$ distribution to be bimodal. Recently, \citet{Chies-Santos} combined \textit{HST}/ACS and
WHT/LRIS imaging, finding evidence for bimodality in space-based but not ground-based datasets. 

In this work we present new wide field $BVI$ Suprime-Cam imaging (with the $B$ band in poor seeing conditions), combined with four $g$ and $z$ 
\textit{HST}/ACS pointings downloaded from the Hubble Legacy Archive (see Section \ref{sec:subarudata}). We find statistically significant bimodality both 
in our ground-based and space-based observations, with the colour separation occurring at $(B-I)~\sim~1.9$ and $(g~-~z)~=~1.1$, respectively. 

We spectroscopically confirm $256$ GCs over four DEIMOS masks observed in good seeing conditions. Given the wide field coverage
of the four \textit{HST} pointings, we decided to use \textit{HST} magnitudes to divide the kinematics of blue and red GC subpopulations, converting $(V-I)$ colours into $(g-z)$ if only 
Subaru photometry was available. As for other group members, we exclude from the kinematics analysis GCs that might be associated with other galaxies.
We select and remove three GCs whose position and radial velocity are consistent with the galaxy NGC~$4283$ $(V_{\rm sys} = 984 \kms)$ that lies $3.6$ ($\sim 15$ kpc) 
arcmin from NGC~$4278$. 

We find no clear evidence for red GC rotation, but there is a hint that the blue GCs rotate in the outer regions along a direction 
intermediate between the major and the minor axis. The velocity dispersion of the red and blue GCs seem to be in good agreement with long--slit data 
in the inner regions although there is no direct overlap between the two datasets. The blue GCs have an overall higher velocity dispersion at intermediate radii. 

\subsection{NGC~4365}
\label{sec:NGC4365}
NGC~$4365$ is a massive elliptical (E3) behind the Virgo cluster with signs of an ongoing merger (Mihos et al. 2012, in preparation). 
It has a kinematically distinct core \citep{Krajnovic} and an unusual stellar rotation along the photometric minor axis \citep{Bender94}. 
Its GC system is mainly known for an odd GC colour distribution that seems to suggest the presence of three, instead of the classic
two, subpopulations \citep{Larsen03,Larsen05,Brodie05,Blom}. 

We make use of the photometric results presented in \citet{Blom} based on $gz$ \textit{HST}/ACS and $gri$ Suprime-Cam imaging.
In this work, we will treat the GC system of NGC~$4365$ as made up of two classic subpopulations and we refer to Blom et al. (2012, in preparation) for a detailed study
of the kinematics of the three subpopulations. We use a colour split at $(g-i) \sim 0.91$ that includes the ``green'' GCs as part of the red subpopulation. 
We represent here the surface density profile published in \citet{Blom} that shows that red GCs is more centrally concentrated than the blue GC subpopulation, with a 
trend similar to that of the galaxy stellar light (Figure \ref{fig:SD}). 

We spectroscopically confirm $269$ GCs over 6 DEIMOS masks. We study the kinematics of $87$ blue and $164$ red GCs respectively. 
We also add 9 GCs from the combined dataset of \cite{Larsen03} and \citet{Brodie05} as described in \S \ref{sec:Repeated}.
We find that the red GCs rotate within $R<200$ arcsec along the photometric minor axis, mimicking the kinematics of the stars. 
The rotation of the blue GCs is only significant at intermediate radii. The velocity dispersion profile is identical for the blue and the red GCs.
\begin{figure}
\includegraphics[scale=.20]{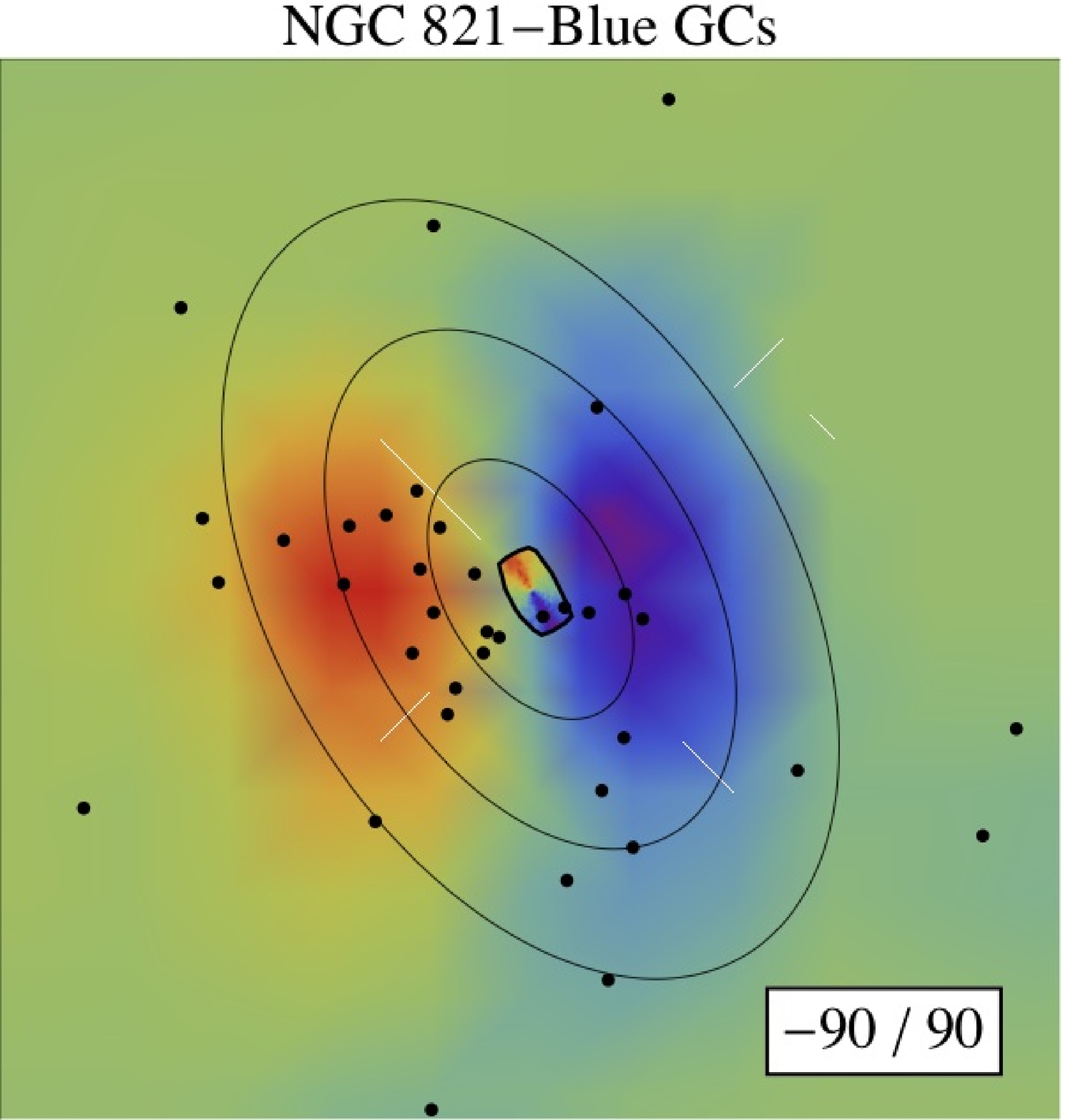} 
\includegraphics[scale=.20]{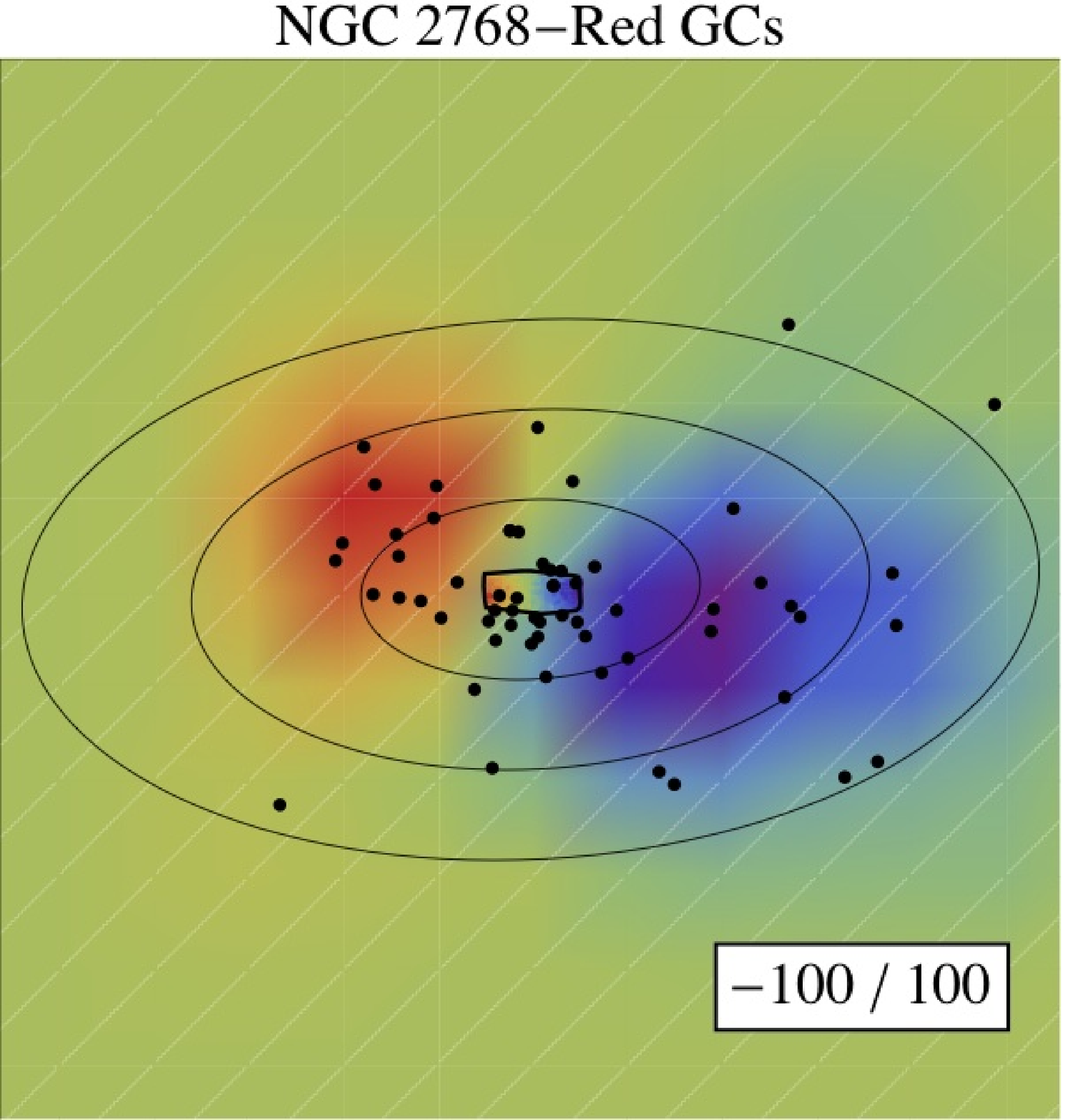} \\
\includegraphics[scale=.20]{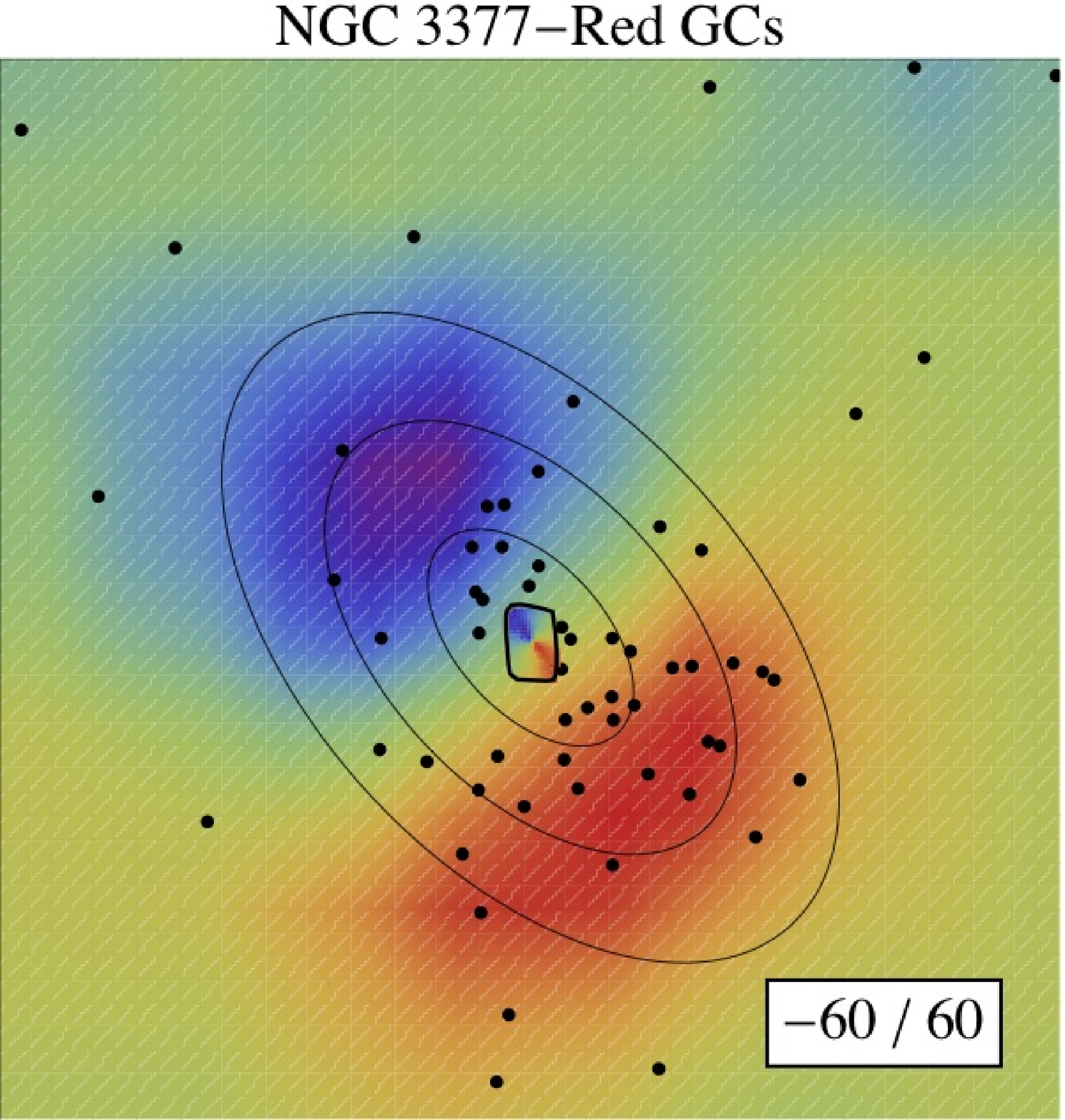} 
\includegraphics[scale=.20]{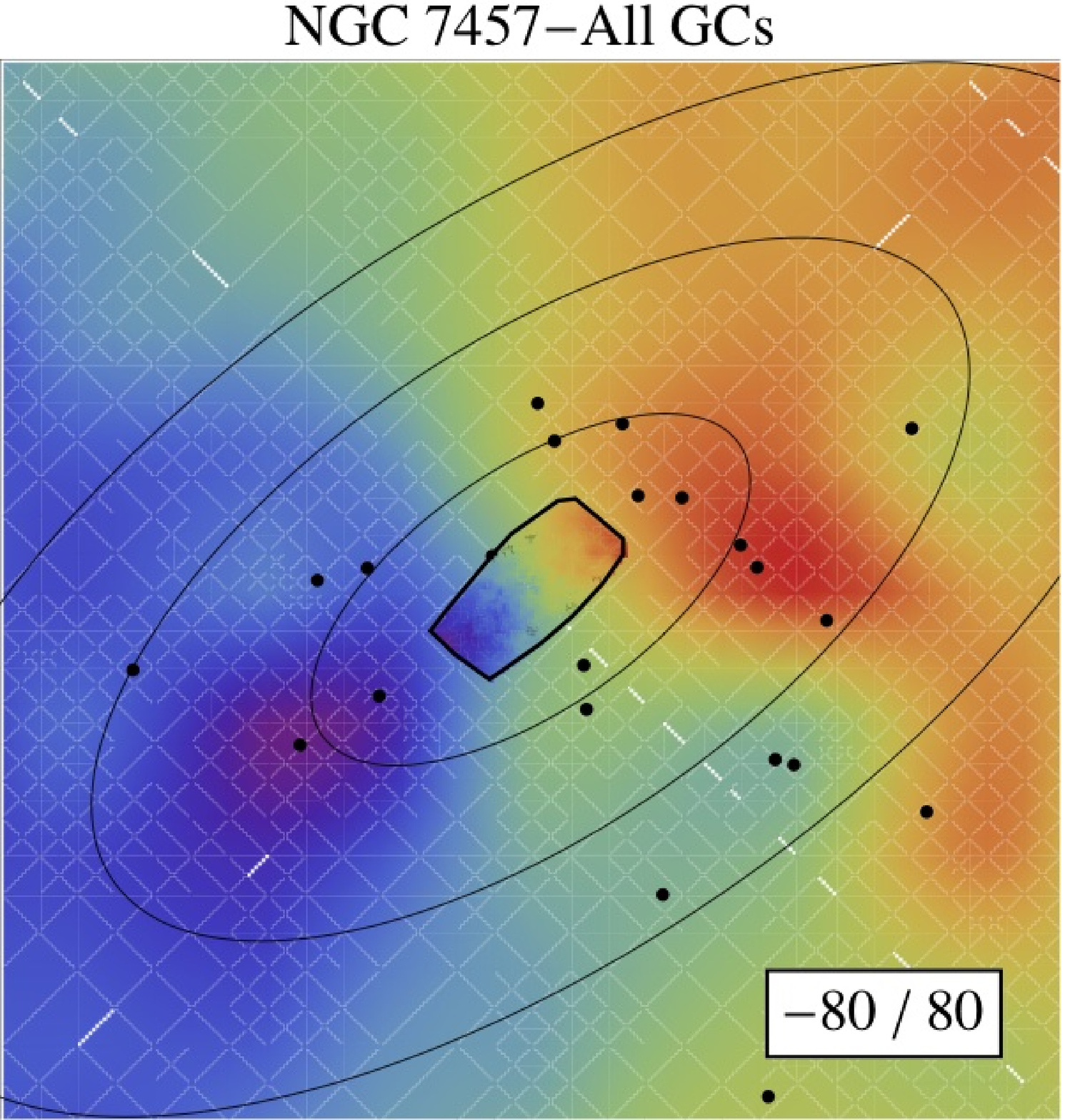} 
\caption{Two-dimensional smoothed velocity fields of our disky early-type galaxies. 
Black points and ellipses represent the locations of the spectroscopically confirmed GCs for a given population and the $2, 4, 6$ \Reff\ isophotes
corrected for the galaxy ellipticity respectively. The white boxes are the maximum red-shifted and minimum blue-shifted velocities in 
the galaxy rest frame in unit of \kms. The SAURON stellar velocity maps are also shown in the innermost region of the galaxies. The 
agreement between the two datasets (except for NGC~$821$) is extremely good.}
\label{fig:2D}
\end{figure}

\subsection{NGC~5846}
\label{sec:NGC5846}
NGC~$5846$ is the brightest member of a galaxy group. The dynamics of NGC~$5846$ has been modelled within 1\Reff\ (\citealt{Cappellari07}; \citealt{Kronawitter}). 
Recently, \citet{Das} derived the mass of this galaxy up to 11\Reff\ exploiting X-ray and PNe observations.
An \textit{HST}/WFPC2 analysis of the NGC~$5846$ GC system was given in \citet{Forbes97B} and then revisited by \citet{Chies-Santos06}. In addition to the classic bimodality, they
also discovered that the GC system is better aligned with the galaxy's minor axis than its major axis and that this galaxy has an unusually low specific frequency compared to 
similar dominant ellipticals in groups or clusters.

In this work we present new Subaru photometry in $gri$ filters, with the $g$ band in moderate seeing conditions ($\sim 1$ arcsec). Although the red peak is
not clearly visible (Figure~\ref{SpectroPhoto_Matrix}), KMM returned a high probability for the colour distribution to be bimodal with a colour split at $(g-i)=0.95$. 

We spectroscopically confirm $195$ GCs over $6$ DEIMOS masks. We also note that some radial velocities might be associated with other bright members of the group as NGC~$5846$A 
({$V_{\rm sys}=2200 \kms$) and NGC~$5845$ ({$V_{\rm sys}=1450 \kms$). However the similar systemic velocities of these galaxies with NGC~$5846$ ($V_{\rm sys}=1712 \kms$)
makes any attempt to distinguish their GC populations problematic. We also supplement our DEIMOS catalogue with 22 GCs from the dataset of \citet{Puzia04} as described 
in \S \ref{sec:Repeated}. In summary, we study the kinematics of $104$ blue and $108$ red GCs. 

We detect rotation only for the blue GCs, which rotate between $150$ and $300$ arcsec close to photometric major axis. 
This feature might be caused by the GCs of NGC~$5846$A, which contaminates our sample within $300$ arcsec. 
The velocity dispersion of the red GCs is flat with radius and consistent with other studies, although the PNe seem to suggest a slightly decreasing slope. 
The blue GCs have a systematically higher velocity dispersion than the red GC subpopulation. 

\subsection{NGC~7457}
\label{sec:NGC7457}
NGC~$7457$ is an isolated S$0$ with a pseudo-bulge detected both photometrically \citep{Tomita} and kinematically \citep{Pinkney} that shows
an unusually low central velocity dispersion for its luminosity. \citet{Emsellem07} revealed that this galaxy has a small counter 
rotating core that might be the result of a merger. \citet{Chomiuk} gave an overview of the GC system of NGC~$7457$ using
\textit{HST} observations and Keck/LRIS spectra for $13$ GCs. They find evidence for a third intermediate population of GCs sharing 
the same age ($2$ --$2.5$ Gyr) as the young nuclear (radius of 1$.5$ arcsec) stellar population discovered by \citet{Sil'chenko}. 

NGC~$7457$ is the only galaxy in our sample not observed with Suprime-Cam. We have instead used the photometric results of \citet{Hargis} to 
design our DEIMOS masks, and we refer to their paper for a detailed description of the data reduction and analysis. They provide wide field 
WIYN/Minimosaic photometry of the GC system of this galaxy, finding that the colour distribution is not bimodal and showing the total population to be
made up of $210 \pm 30$ GCs with a radial extent of $12 \pm 2$ kpc. 

We find 21 spectroscopically confirmed GCs in two DEIMOS masks observed in average $0.9$ arcsec seeing conditions. We also obtained spectra
for $7$ of the $13$ GCs already observed by \citet{Chomiuk}, finding their LRIS and our DEIMOS radial velocities to be consistent within the errors, 
but with a mean difference of $\sim 12\kms$ (see \S \ref{sec:Repeated}). Therefore, we add the remaining $6$ Chomiuk GCs corrected for the velocity 
offset to our GC sample. The final GC catalogue consists of 27 GC radial velocities.

Our confirmed GCs appear  to be distributed within 2 arcmin~$\sim 7.7$ kpc from the galaxy centre, reflecting the expected poor
extent of the underlying GC subpopulation. The mean velocity of the system ($V_{\rm sys} = 847 \pm 20 \kms$) is in good agreement with both \citet{Chomiuk} 
and literature data. As this galaxy appears to be unimodal in colour, we fit the kinematics of all the GCs without any split in colour and we compare the results 
with the long-slit data of \citet{Simien}.
We find that the GC system of NGC~$7457$ is rapidly rotating along the photometric major axis, with a flat rotation curve at $\sim 80 \kms$ up to $2\Reff$. 
In contrast, the velocity dispersion is low $(< 50 \kms)$ at all radii. The agreement with the ATLAS$^{\rm 2D}$ velocity map is also very good (Figure \ref{fig:2D}).
We will discuss the implication of these results in more detail in \S \ref{sec:a case study}.

\section{GC formation models}
\label{sec:GC formation models}

Having determined the photometric and kinematic properties of our GC sample, in the following sections we will discuss these results as a whole. 
We look for common kinematic features that might retain key information about the formation history of the galaxies themselves and then 
we compare these results to theoretical models. In order to do so, it is first worth giving a brief summary of the main GC formation scenarios
proposed in the literature, focusing on their kinematic predictions. 

To date, the formation of GC bimodality has been investigated both in a cosmological context \citep{Kravtsov,Diemand,Moore}
and at smaller scales, with models fine tuned to reproduce the properties of specific galaxies (see \citealt{Vesperini} for NGC~$4486$ or \citealt{Deason} for the
Milky Way and M31). The challenge of producing simulations of GC bimodality formation is intertwined with our poor understanding of how GC systems formed in the 
first place \citep[e.g.,][]{Beasley02,Elmegreen12}.
As a consequence, the three classic formation scenarios proposed in the literature and summarized below have few, if any, theoretical predictions directly comparable with the 
observations. In particular, there is a general dearth of GC kinematics predictions. 

In the \textit{major merger scenario} \citep{Ashman92} two or more gas-rich disk galaxies with pre-existing blue and red GC subpopulations merge to form an elliptical galaxy. 
New red GCs may form from the star formation induced by the merger \citep{Bekki2002}. 

This scenario has been simulated by \citet{Bekki2005} (hereafter B$+$05). In this simulation the pre-existing blue and red GCs are assumed to have a Milky Way-like spatial 
distribution, but they are both pressure-supported. This is a reasonable assumption for the blue but not for the red GCs of the Milky Way that are known to be rotation-supported \citep{Cote99,Deason}.  
B$+05$ predicted that a merger with even mass-to-mass ratio produces strong rotation for both GC subpopualtions in the outer regions of the remnant. This result is independent of the orbital configuration of the merger. 
The velocity dispersion profile is predicted to decrease with radius, but it would be flatter in case of multiple mergers. Also, the blue GCs are expected to show a larger central velocity dispersion than red GC subpopulation.
Finally, the ratio of the maximum rotational velocity $V_{\rm m}$ to the central velocity dispersion $\sigma_0$ of the GC systems ranges from
0.2 to 1.0 within 6\Reff\ for both the blue and red GC subpopulations, but most GC systems, viewed from various angles, have $(V_{\rm m} / \sigma_0)<0.5$. 
\citet{BekkiPeng} carried out a simulation similar to that of B$+05$ (with  the rotation of the disk in the spiral progenitors included), fine tuned to study the dynamics of the planetary nebulae (PNe) in elliptical galaxies.
They found that the effect of the residual spin disk (additional to the initial orbital angular momentum) on the final PNe kinematics enhances the rotation at all radii, making the merger remnant rotation-supported.
 
In the \textit{multiphase dissipational collapse scenario} \citep{Forbes97}, GCs are the result of an early two phase collapse, 
and hence two main star formation episodes, that the galaxy undergoes. The blue GCs form during the first star formation episode in a metal-poor cloud, 
whereas red GC subpopulation form in a second phase after the gas in the galaxy is self-enriched. In this scenario, a fraction of blue GCs might also come from
accreted satellite galaxies, similar to the most recent two-phase galaxy formation model \citep{Oser}. No numerical simulations have been performed so far
to test this model. Since the red GCs are coeval with the galaxy stellar component, \citet{Forbes97} infer that they should share the same kinematic 
properties as the galaxy stars, and hence also the same spatial distribution. No significant rotation is expected for the blue GC subpopulation.

In the \textit{dissipationless accretion scenario} \citep{Cote98}, the red GCs form from the monolithic collapse of a giant protogalactic cloud, 
whereas the blue GCs are accreted from low mass galaxies due to mergers or tidal stripping. To date, simulations based on similar scenarios have been performed 
\citep[e.g.,][]{Oser} but they do not include GC particles, and they have not made kinematic predictions. In addition to the observation of infalling co-moving groups, \citet{Cote98} also infer that we
should expect to observe blue GCs with radially biased orbits without overall rotation. 

\citet{Bekki2008} (hereafter B$+$08) performed a high resolution N-body cosmological simulation combined 
with semi-analytic models. They found that almost half of the simulated $10^5$ galaxies show clear bimodality in their GC metallicity distribution. 
The majority of GCs form in low-mass galaxies at redshift greater than 3 with the blue GCs being slightly older ($<1$ Gyr) than the red GC subpopulation. 
B$+$08 also made predictions for the overall kinematics of each GC system, whereas current spectroscopic observations only sample a smaller fraction of it. 
They find that the velocity dispersion of blue and red GC subpopulations increases with the total luminosity of the host galaxy and that the ratio of the velocity dispersion of the blue to red GC subpopulation is 
$(\sigma_{\rm B}/\sigma_{\rm R}) \sim 1$ for a wide range of luminosity. Finally, they also predicted that the GC systems of most galaxies
are mainly pressure supported with ($V_{\rm rot} / \sigma) < 0.3$. This is in contrast with the wider range of $V_{\rm rot} / \sigma$ found by B$+05$.
\begin{figure*}
\includegraphics[scale=.5]{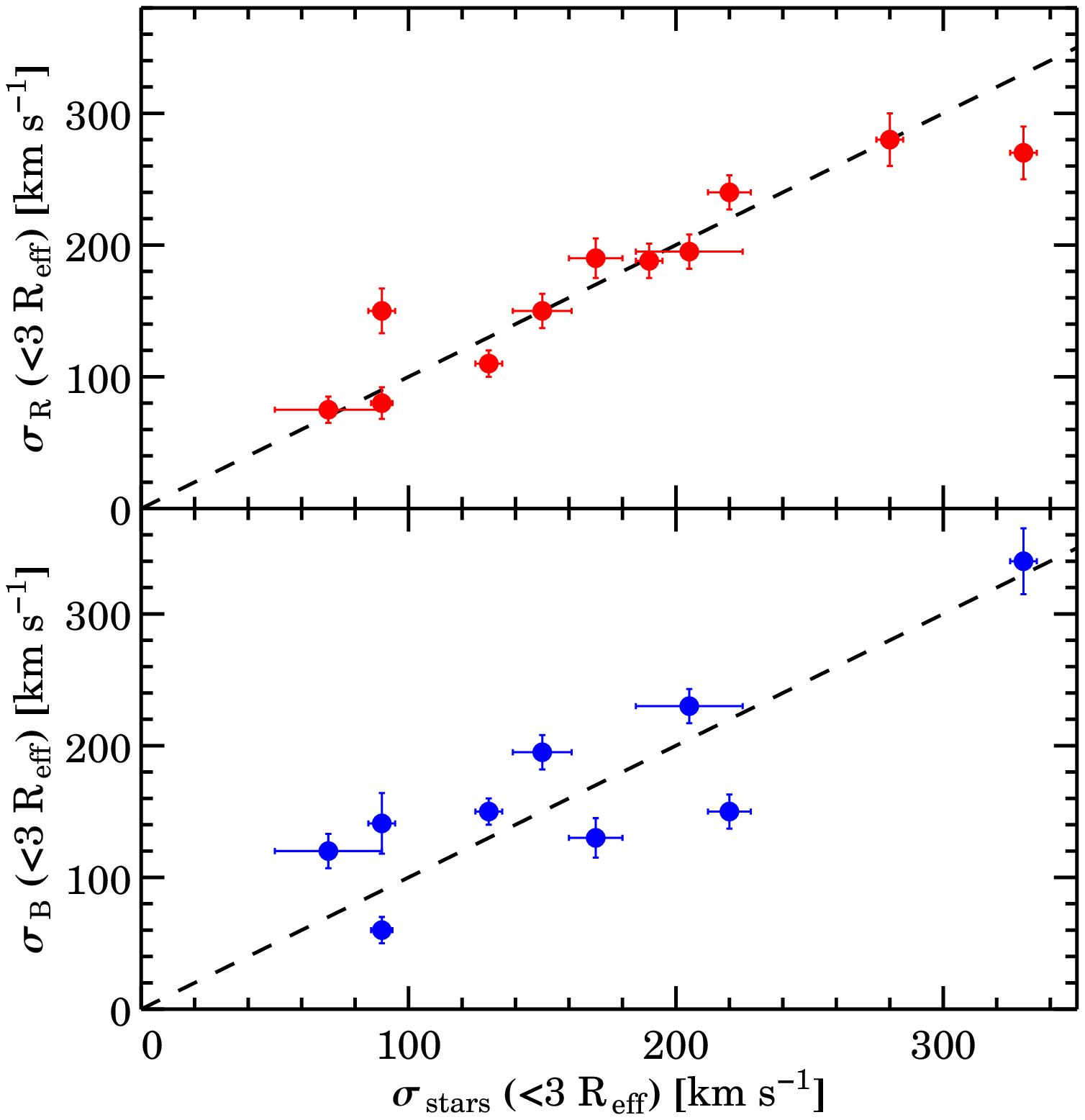} 
\includegraphics[scale=.5]{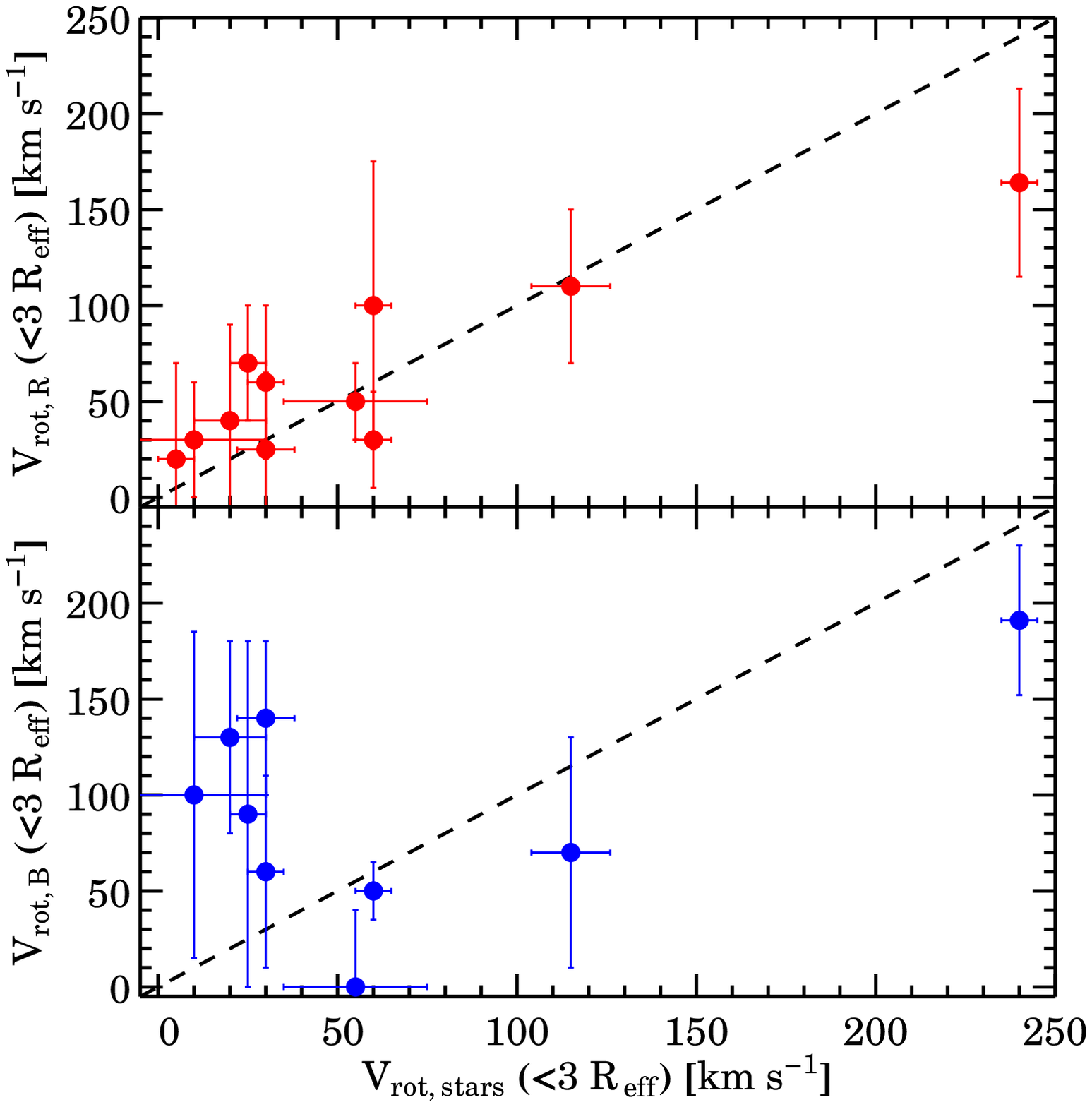} 
\caption{GC rotation velocities and velocity dispersions compared to the galaxy stars/PNe. Rotation velocities (\textit{left panel}) and 
velocity dispersions (\textit{right panel}) are extracted within 3\Reff. The exact extraction radius is usually set to 2\Reff\ but this threshold is arbitrarily adjusted by $\pm 1\Reff$ in case the GCs do not
overlap with the stellar data. The blue and the red GC subpopulations are shown in the \textit{bottom} and \textit{upper panel} respectively. Dashed lines represent a one-to-one relationship.  
Blue GC data for NGC~$4365$ and NGC~$4278$ are not shown due to limited radial range. NGC~$7457$ is not shown. 
The red GC subpopulation shows better agreement with stars than the blue~GCs.}
\label{fig:SigmaStar}
\end{figure*}

\section{Global results from our GC sample}
\label{sec:Summary of generic results}

In this section we summarize the results for the GC systems presented in this paper, including the three galaxies previously published by us (i.e., NGC~$3115$, NGC~$4486$, NGC~$4494$). 
Section \ref{sec:Merging} will incorporate other literature data, and tackle topics not covered here.

\subsection{Spatial Distribution}
\label{sec:Summary Photometry}
\begin{figure}
\includegraphics[scale=.51]{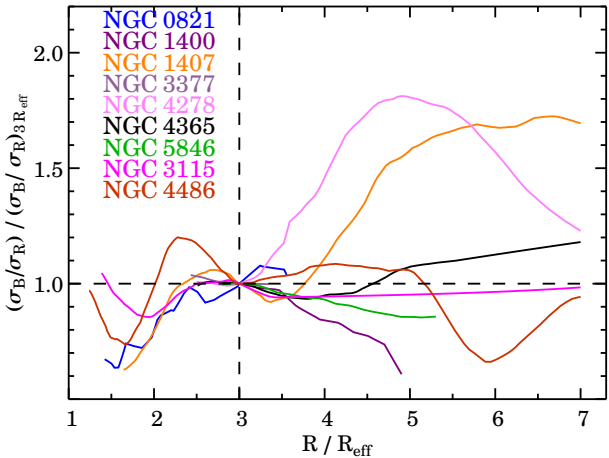} 
\caption{Velocity dispersion profiles. For each galaxy, we show the ratio of the velocity dispersion of blue to red GC subpopulations normalised to $3$ \Reff\
(vertical dashed line) as a function of the effective radius. Different colours correspond to different galaxies. The horizontal and vertical lines show 
$(\sigma_{\rm B} / \sigma_{\rm R}) = (\sigma_{\rm B} / \sigma_{\rm R})_{3 \, \Reff}$ and $R=3$\Reff, respectively. NGC~$7457$ and NGC~$4494$ are not shown. } 
\label{fig:SigmaReff}
\end{figure}

Significant colour bimodality was detected in all the GC systems, except for NGC~$7457$. All GC system formation scenarios, as well as numerical simulations, 
predict that this bimodality should be reflected in different spatial distributions of the two subpopulations around the parent galaxy. A glance at Figure~\ref{fig:SD} 
reveals that this is also the case for our galaxy sample. Generally speaking, the red GCs are more centrally concentrated than the blue GCs, as already found in 
other studies \citep[e.g.,][]{Schuberth,Faifer11,Strader11}. 

The advent of wide field imaging has shown that the radial distribution of the red GC subpopulation matches that of the host galaxy light, suggesting that they might 
have shared a similar formation history \citep{Bassino}. Here we have exploited the $34 \times 22$ arcmin$^2$ field of view of our Suprime-Cam imaging to compare 
the GC surface density to the surface brightness of the respective host galaxy (shifted by an arbitrary constant). Qualitatively speaking, the galaxy starlight has a similar
slope to that of the red GC subpopulation.  The discrepancies between the surface brightness and the GC spatial distribution in innermost regions (e.g., see NGC~$4365$
and NGC~$1407$ in Figure \ref{fig:SD}) is due to the core-like distribution of the GCs, that makes the GC density profile flatter.
This feature is interpreted as the effect of the GC disruption, stronger in the central regions \citep{Baumgardt98,Baumgardt}.

\subsection{Rotation}
\label{sec:Summary Rotation}

We find a large variety of GC rotation profiles. Both the blue and red GCs show some degree of rotation, but there does not seem to be a clear correlation between the rotation patterns and
the property of the host galaxy. A glance at Figure~\ref{fig:kinematics} reveals that the red GCs rotate more consistently with the photometric 
major axis than the blue GC subpopulation does. For our galaxy sample the photometric position angle coincides with the photometric major axis 
of the galaxy stars (except for NGC~$4365$), which means that the rotation velocity of the 
red GC subpopulation is similar to that of the galaxy stars. This is also true for the overall GC system of NGC~$7457$. 

To quantify this phenomenon, in Figure~\ref{fig:SigmaStar} we compare the major axis rotation velocity of the blue and the red GC subpopulations within 3\Reff\ to the rotation velocity  
of the stars and PNe (if available) at the same galactocentric distance. The exact extraction radius was set to 2\Reff, but in some cases this was relaxed by $\pm 1\Reff$ to maximise 
the overlap between the GCs and the stars. The rotation velocity of the host galaxy stars was extracted along the photometric major axis for all the galaxies, except for NGC~$4365$ 
in which the bulk of the stellar rotation is occurring along the photometric minor axis. The red GC rotation for our galaxies is similar to those of the host galaxy stars, at least for galaxies with a 
conspicuous amount of stellar rotation. On the other hand, the star-GC system coupling is also evident for only blue GC subpopulations, such as in NGC~$4494$, NGC~$821$ and NGC~$3115$.

The rotation of the blue GCs is more puzzling than the red GCs. Overall the rotation of the blue GCs is lower than the red GCs and often consistent with zero, 
yet we also detect minor axis rotation in NGC~$821$, also seen in PNe, in contrast with integrated starlight. 

We detect GC rotation at large galactocentric radii for some galaxies. This feature is only significant in NGC~$1407$, NGC~$4486$ and only marginal in NGC~$4278$. 
The outer rotation occurs in the same direction for both the blue and the red GC subpopulations and it usually coincides with the photometric major axis.
\begin{figure*}
\centering
\includegraphics[scale=.52]{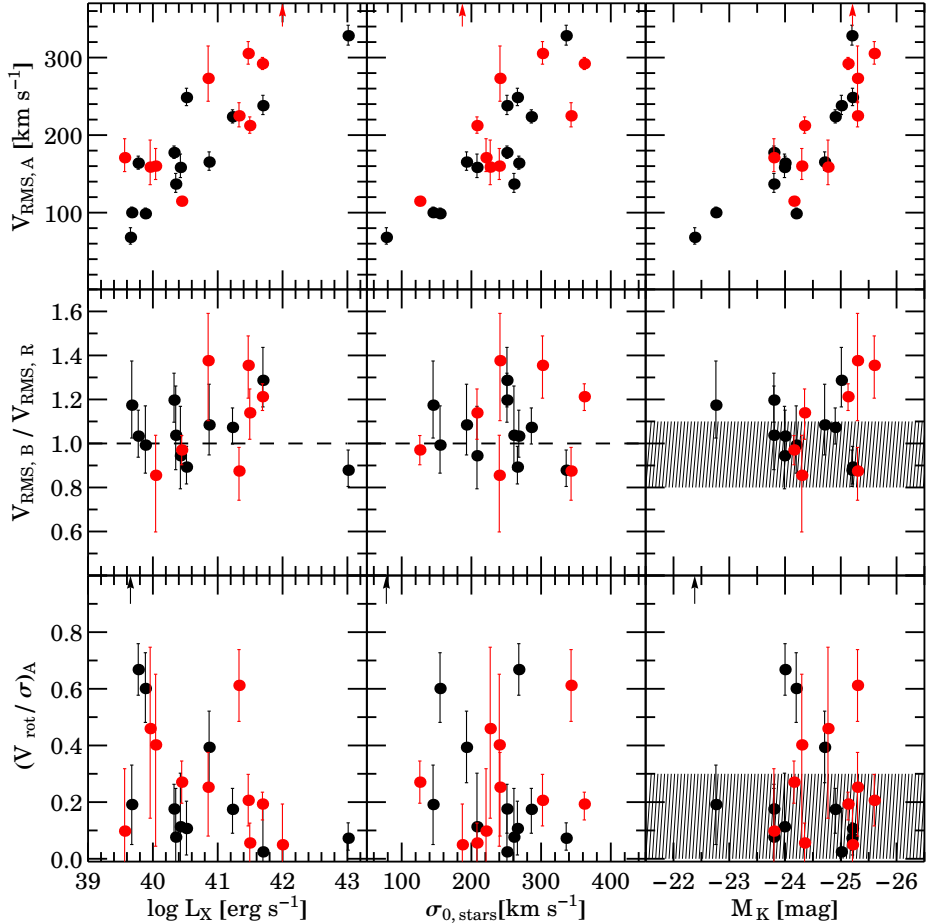} 
\caption{Summary of GC system kinematics as a function of the host galaxy properties. Literature data and our sample are shown as red and black points respectively.
From the top to the bottom: root-mean-square velocity for the overall GC system ($V_{\rm rms, A}$), ratio of the $V_{\rm rms}$ of the blue to the red GC subpopulation
($V_{\rm rms, B}/V_{\rm rms, R}$), ratio of the rotation velocity to the velocity dispersion for the overall GC system $(V_{\rm rot} / \sigma)_{\rm A}$. 
From the left to the right: X-ray luminosity (log $L_X$) from \citet{O'Sullivan01} at the distance given in Table \ref{tab: survey_summary} and Table \ref{tab: survey_literature}, 
central stellar velocity dispersion ($\sigma_{0, \rm stars}$) from \citet{Prugniel} and total K-band magnitude (M$_{\rm K}$) from 2MASS. 
The black and red arrows indicate the position of NGC~$7457$ at $(V_{\rm rot} / \sigma)_{\rm A} = 1.68$ and of NGC~$3311$ at $V_{\rm rms, A}=653 \kms$, respectively.
The horizontal dashed line marks $V_{\rm rms, B}/V_{\rm rms, R} = 1$ to guide the eye.
The grey shaded shaded region represents the theoretical predictions from the cosmological simulation of \citet{Bekki2008}.}
\label{fig:SuperPlot}
\end{figure*}

\subsection{Velocity Dispersion}
\label{sec:Summary Dispersion}
\begin{table*}
\begin{tabular}{@{}c c c c c c c c c c c c}
\hline
Galaxy ID & Hubble &$M_{\rm K}$ & $D$  & N$_{\rm GCs}$&$V_{\rm  rms,A}$ & $V_{\rm rms,B}$ & $V_{\rm rms,R}$ & $(V_{\rm rot} / \sigma)_{\rm A}$  &  $(V_{\rm rot} / \sigma)_{\rm B}$ &  $(V_{\rm rot} / \sigma)_{\rm R}$    & Ref. \\
      & Type     &  [mag]  & [Mpc]     &  &   [\kms]       & [\kms]  &  [\kms]       &          &           &                 \\
 (1)  & (2)      &  (3)            & (4)            & (5)     & (6)    &  (7)         &  (8)      &       (9)           &       (10) &  (11)    & (12)      \\
\hline
\hline   
NGC~$1380$ & SA0 &$-24.3$ &$17.1$ & $42$      &	$160_{-17} ^{+23} $ &	$	142_{-25} ^{+38} $ &	$	170_{-22} ^{+32} $ &	$	0.39_{-0.35} ^{+0.25} $ &	$0.32_{-0.54} ^{+0.79}$ &	$	0.05_{-0.22} ^{+0.30}$& P$+$04 \\	 
NGC~$1399$ & E1    &$-25.1$ &$19.4$ & $738$   &	$292_{-8} ^{+8}     $  &	$	325_{-12} ^{+13} $ &	$	266_{-9} ^{+10}  $  & 	$	0.19_{-0.05} ^{+0.04} $ &	$0.28_{-0.06} ^{+0.06}$ &	$	0.07_{-0.08} ^{+0.06}$  & S$+10$ \\	
NGC~$3311$ & cD    &$-25.2$ &$46.7$  & $116$  &        $653_{-40} ^{+48} $ &	$-$ &	$-$ &	$	0.03_{-0.12} ^{+0.14} $ &	$-$ &         $-$ & M$+11$\\    
NGC~$3379$ & E1    &$-23.7$ &$10.2$   & $39$     & 	$171_{-18} ^{+24} $ &	$-$ &	$-$ &	$	0.08_{-0.15} ^{+0.22} $ &	$-$ &	$-$ &  P$+06$; B$+06$\\ 
NGC~$3585$ & E6    &$-24.7$ &$19.5$  & $20$    &	$159_{-23} ^{+35} $ &	$-$ &	$-$ &	$	0.43_{-0.31} ^{+0.28} $ & $-$  &	$-$ &  P$+$04 \\
NGC~$3923$ & E4    &$-25.3$ &$22.3$  & $79$    &	$273_{-29} ^{+42} $ &	$	273_{-29} ^{+42} $ &	$	200_{-19} ^{+26} $ &	$	0.24_{-0.16} ^{+0.12} $ &	$0.31_{-0.27} ^{+0.26}$ &	$	0.00_{-0.22} ^{+0.14}$& N$+12$\\
NGC~$4472$ & E2    &$-25.6$ &$16.7$  &  $225$ & 	$305_{-14} ^{+15}  $ &	$	334_{-19} ^{+22} $ &	$	256_{-19} ^{+22} $ &	$	0.19_{-0.10} ^{+0.09} $ &	$0.27_{-0.10} ^{+0.10}$ &	$	0.08_{-0.07} ^{+0.14}$ & C$+03$ \\
NGC~$4636$ & E2    &$-24.3$ &$14.2$  &  $259$ &	$212_{-10} ^{+11}      $ &	$	226_{-15} ^{+18} $ &	$	200_{-12} ^{+15} $           &	$	0.05_{-0.11} ^{+0.07} $ &	$0.09_{-0.12} ^{+0.14}$ &	$	0.18_{-0.10} ^{+0.13}$ & L$+08$ \\
NGC~$4649$ & E2    &$-25.3$ &$16.3$ & $121$ &		$225_{-14} ^{+17} $ &	$	209_{-16} ^{+20} $ &	$	257_{-26} ^{+37} $           &	$	0.61_{-0.12} ^{+0.12} $ &	$0.59_{-0.15} ^{+0.16}$ &	$	0.70_{-0.25} ^{+0.24}$& H$+08$ \\
NGC~$5128$ & S$0$ &$-24.1$&$4.1$  & $449$ &		$115_{-5} ^{+5}   $ &	$	113_{-6}  ^{+7} $ 	&		$	117_{-7} ^{+7}   $             &	$	0.26_{-0.05} ^{+0.08} $ &	$0.25_{-0.11} ^{+0.10}$ &	$	0.28_{-0.11} ^{+0.09}$ & W$+10$ \\  
\hline
\end{tabular}
\caption{Kinematical properties of literature GC systems. The host galaxy name (1) and Hubble Type (2) are from NED database. 
K~band absolute magnitude (3) is from 2MASS apparent magnitude at the distances given in column 4 and corrected for the foreground Galactic 
extinction from NED database. The distance in Megaparsec (4) was obtained by subtracting 0.06 mag \citep{Mei} from the distance modulus from \citet{Tonry01}, 
except for NGC~3311 for which we adopt the distance of the Hydra I cluster of $(m - M)=33.37$ mag \citep{Mieske}. The number of spectroscopically confirmed
GC system (5) is after the outliers filtering as described in the text. Columns (6), (7) and (8) are the root-mean-square velocity $V_{\rm rms}$ 
for all, blue and red GC subpopulations respectively. Columns (9), (10) and (11) are the $(V / \sigma)$ value for all, blue and red GC subpopulations respectively. The data references are:
P$+$04 \citep{Puzia04}; S$+10$ \citep{Schuberth}; M$+11$ \citep{Misgeld}; P$+06$ \citep{Pierce06}; B$+06$ \citep{Bergond06}; N$+12$ \citep{Norris12}; C$+03$ 
\citep{Cote03}; H$+08$ \citep{Hwang}; L$+08$ \citep{Lee08}; W$+10$ \citep{Woodley10}.}
\label{tab: survey_literature} 
\end{table*}
Our results show that the velocity dispersion profiles are quite flat for most of the galaxies, 
both as a function of radius (Figure~\ref{fig:kinematics}) and as a function of the colour (Figure~\ref{fig:CV}).
The slope of the velocity dispersion profiles (including the contribution of the rotation) will be discussed below.

As for the rotation, we study how the velocity dispersion of the two GC subpopulations compares to the velocity dispersion of the host galaxy stars. 
We extract the velocity dispersion at the same galactocentric distances as done with the rotation. Figure~\ref{fig:SigmaStar} 
shows that the velocity dispersion of the red GC is very similar to that of the host galaxy stars in the region of overlap. Conversely, 
the blue GC subpopulations seem to avoid the one-to-one line. The standard deviation from the one-to-one line is $22\kms$ 
and $50\kms$ for the red and the blue GC subpopulations respectively. 

To quantify the differences between the velocity dispersion profiles of the two GC subpopulations, in Figure~\ref{fig:SigmaReff} we
plot the ratio of the velocity dispersion of the blue GCs to the velocity dispersion of the red GCs ($\sigma_{\rm B} / \sigma_{\rm R}$) normalised 
to 3 \Reff\ as a function of the effective radius. In this case we show the velocity dispersion using rolling radial fits to appreciate the details of the profiles. 
Figure~\ref{fig:SigmaReff} shows that the ratio ratio $(\sigma_{\rm B} / \sigma_{\rm R})/(\sigma_{\rm B} / \sigma_{\rm R})_{3 \, \Reff}$ is generally consistent with $1$, 
but it increases towards the outer regions for some galaxies, as already observed for more massive galaxies \citep{Lee}. 

The overall velocity dispersion profiles are in rough agreement with their merger remnants assembled via multiple mergers, with the velocity dispersion of the 
blue GCs being larger or equal to that of the red GCs. However, the model fails in reproducing the irregularity of some dispersion profiles (e.g., NGC~$1407$ and NGC~$4278$).

\section{Including literature data}
\label{sec:Merging}
\begin{figure}
\centering
\includegraphics[scale=.33]{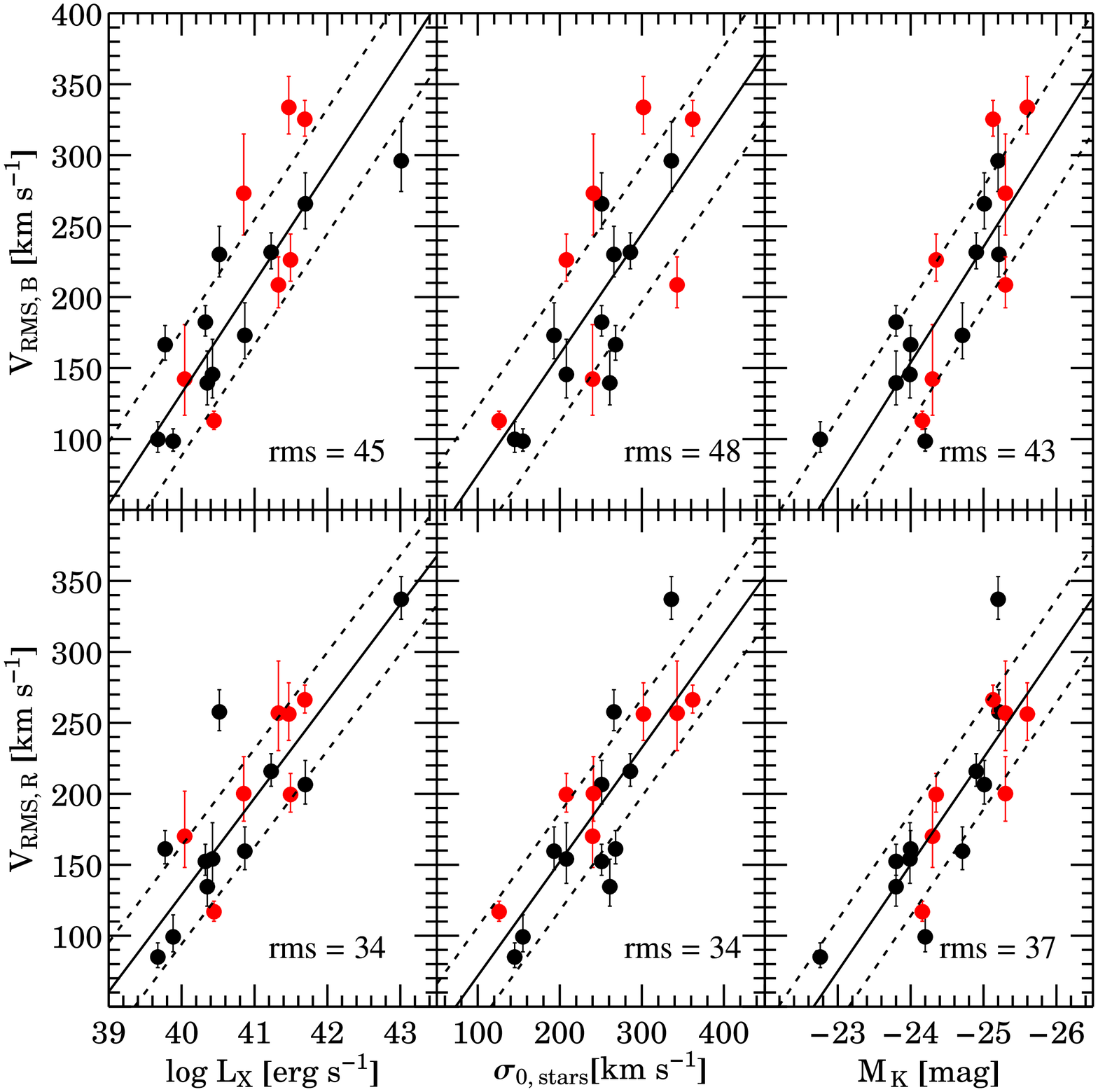} 
\caption{GC system \Vrms\ as a function of the host galaxy properties. Literature data and our sample are shown as red and black points, respectively.
The top and the bottom panels represent the velocity dispersion of the red and blue GC subpopulations as a function of the X-ray luminosity (log $L_X$), central stellar velocity
dispersion ($\sigma_{0, \rm stars}$) and total $K$-band magnitude (M$_{\rm K}$), respectively. The solid and dashed lines represent the weighted linear fit to the
data and its rms in \kms\ quoted on the bottom right of each panel, respectively. The dependence of red GC subpopulation on the galaxy properties is tighter than for the blue GCs.}
\label{fig:SuperPlot_BlueRed}
\end{figure}

We examined the literature for galaxies with a reasonable number of GC radial velocity measurements. We require at least 20 spectroscopically 
confirmed GCs for a galaxy to be included in the literature sample. We use the same friendless algorithm as described in \S \ref{sec:spectroscopic_selection} 
to clip outliers in radius-velocity space. This selection returned ten GC systems, which are summarized in Table \ref{tab: survey_literature}. Most of
these galaxies are dominant group/cluster galaxies discussed in \citet{Lee}, whereas the less massive galaxies are part of the survey of \citet{Puzia04}. 
We also include NGC~$3923$ from \citet{Norris12} which shows strong signatures a recent interaction.
A first comparison between our GC dataset and external datasets was already given in Figure~\ref{fig:magNGC} in which we showed that we cover 
a wider range of mass with three times better velocity accuracy than previous studies. 

We ran the kinematic likelihood code on the literature galaxies to get the \textit{overall} best fit values to eq. \ref{eq:Vobs2}, but we do not show 
the kinematic radial profiles for the literature sample. We investigate the kinematics of blue and red GC subpopulations for seven out of ten galaxies
using the dividing colour quoted in the respective papers. For consistency with the kinematic analysis of our galaxies, we do not exclude UCD 
candidates from the literature data. We find that our likelihood-method tends to reduce both the velocity dispersion (by $\sim 10 \kms$) 
and the rotation amplitude (up to $\sim 30\kms$) if the contribution of the rotation is low ($V_{\rm rot}/ \sigma < 0.4$). We attribute this discrepancy
to the fact that some previous GC kinematics studies have employed error-weighted least-square rotation fittings that are less appropriate for a 
system with an intrinsic dispersion. Another explanation might be that, to our knowledge, previous studies have never explicitly taken into 
account the bias introduced by an unconstrained position angle (see \S \ref{sec:kinematics_analysis}), which has an important effect at lower 
velocities. It is important to note that the maximum-likelihood approach does not alter the general results of previous works.

A final caveat to bear in mind concerns the different kinematic axes convention. Based on the fit of the GC radial velocities with the 
position angle (Figure~\ref{fig:NGC3377PA}), we define the kinematic position angle as the angle between the direction of 
maximum rotation amplitude and North, that is 90 degrees away from the angular momentum vector. Some previous studies have instead
defined the kinematic position angle as the direction around which the rotation is occurring. In other words, our kinematic major axis would 
correspond to the kinematic minor axes quoted by \citet{Lee}. Our convention is in line with that used in other galaxy kinematic studies 
(e.g. ATLAS$^{\rm 3D}$). For the galaxy properties of the literature sample we use the same sources quoted in Table \ref{tab: survey_summary}.

\section{Results}
\label{sec:LiteratureResults}

In this section we analyze the extended sample of 22 GC systems (literature plus our own galaxies) , 18 of which have kinematics for both GC subpopulations. 
We carry out a number of tests to compare the kinematic properties of the two subpopulatinos with each other and with host galaxy properties.

\subsection{Correlations with host galaxy properties}
\label{sec:Discussion}

In Figure~\ref{fig:SuperPlot} we show the GC system kinematics as a function of the general properties of the host galaxies.
We now have the chance to better constrain some relations examined by  \citet{Lee} that were uncertain because of their limited sample size of six galaxies (see their figure 12).

The upper panels of Figure~\ref{fig:SuperPlot} show that, as found from previous studies, a correlation exists between \Vrms\ of the GC systems and their respective host 
galaxy X-ray luminosity $L_X$,  central velocity dispersion of the stars $\sigma_0$ and absolute $K$-band magnitude $M_K$. We find that these correlations also hold for less massive galaxies. 
The correlations with galaxy properties are always tighter for our GC dataset as consequence of the improved reliability of our velocity measurements. 
\begin{figure}
\centering
\includegraphics[scale=.5]{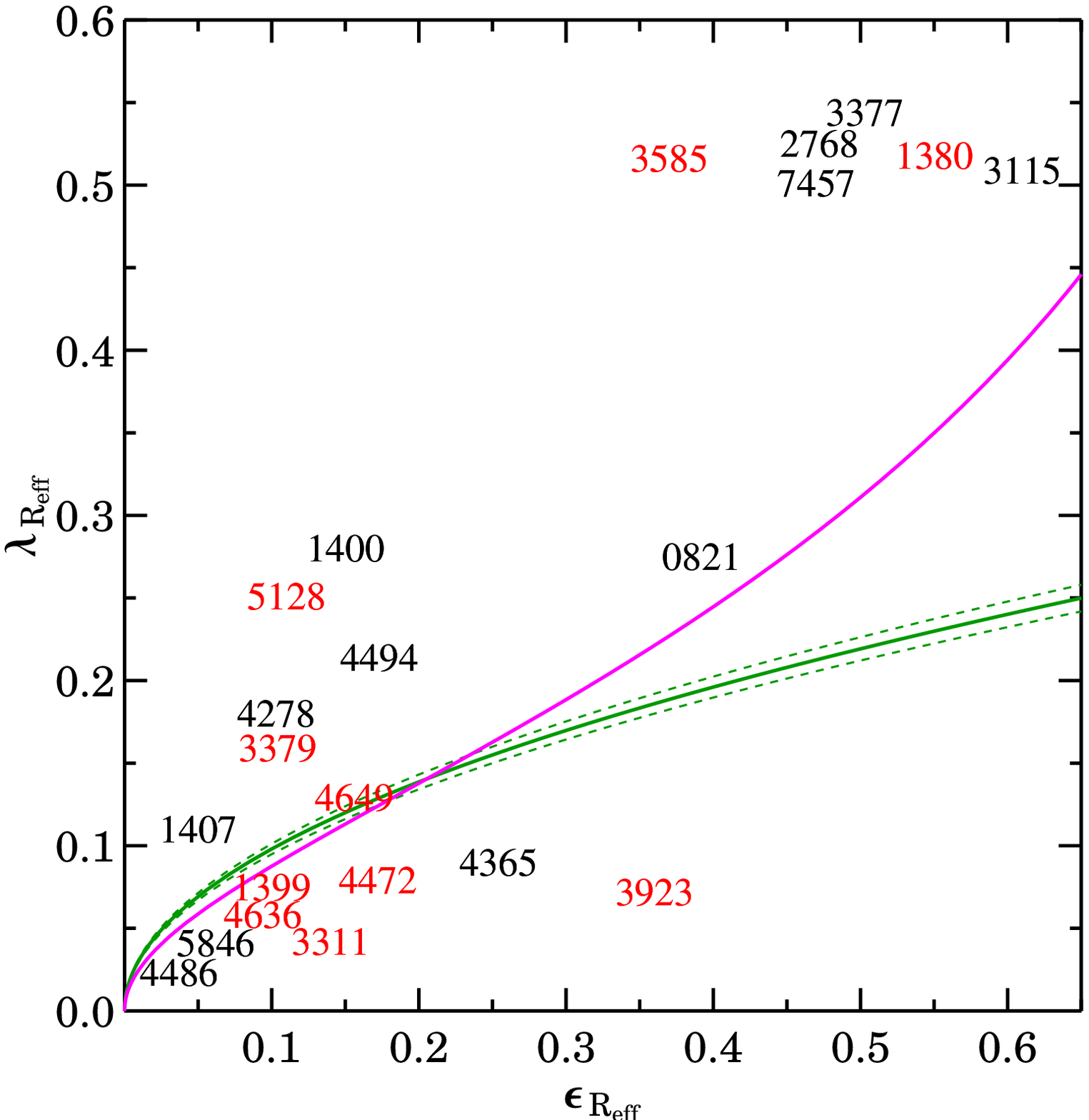} 
\caption{Rotational dominance parameter of the stellar component $\lambda_{\Reff}$ versus the galaxy ellipticity $\epsilon_{\Reff}$. 
Both quantities are measured within 1\Reff. Data points are represented by the NGC number of our
galaxy sample (black) and literature data (red). Measurements are from \citet{Emsellem11} if the galaxy is in ATLAS$^{\rm 3D}$. If not they are estimated from literature data (see text). 
The magenta curve is the inferred edge-on average relation for nearby fast-rotators. The solid green curve is $(0.31 \pm 0.01) \times \sqrt{\epsilon_{\Reff}}$, with the dashed lines representing its uncertainties. 
Galaxies are divided into fast-rotators and slow-rotators if they are above or below the green curve respectively.}
\label{fig:Lambda}
\end{figure}

The correlation of \Vrms\ with galaxy properties holds when we plot the \Vrms\ of the blue and red GC subpopulations separately (see Figure~\ref{fig:SuperPlot_BlueRed}).
The main difference here is that the two subpopulations scatter differently with respect to the best fit lines. The rms of this difference for the blue GCs is systematically larger for the red 
GCs. 
\begin{figure}
\centering
\includegraphics[scale=.5]{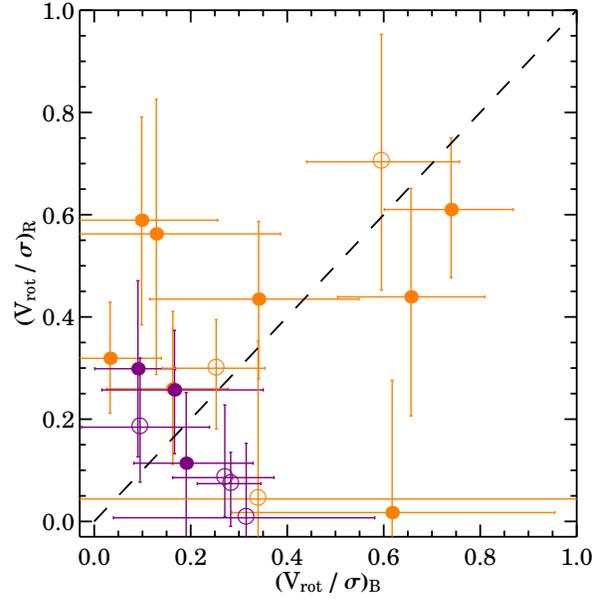} 
\caption{The rotational dominance parameter of GC systems. The plot compares the $(V_{\rm rot}/ \sigma)$ parameter of the blue 
GCs $(V_{\rm rot}/ \sigma)_{\rm B}$ to that of the red GCs $(V_{\rm rot}/ \sigma)_{\rm R}$. The dashed line marks a one-to-one line. 
GC systems are colour coded according to the kinematics of the host galaxy: purple if slow-rotators and orange if fast-rotators.
Filled and open circles represent our GC systems and the data from the literature, respectively.}
\label{fig:VsigmaBR}
\end{figure}

\begin{figure*}
\centering
\includegraphics[scale=.55]{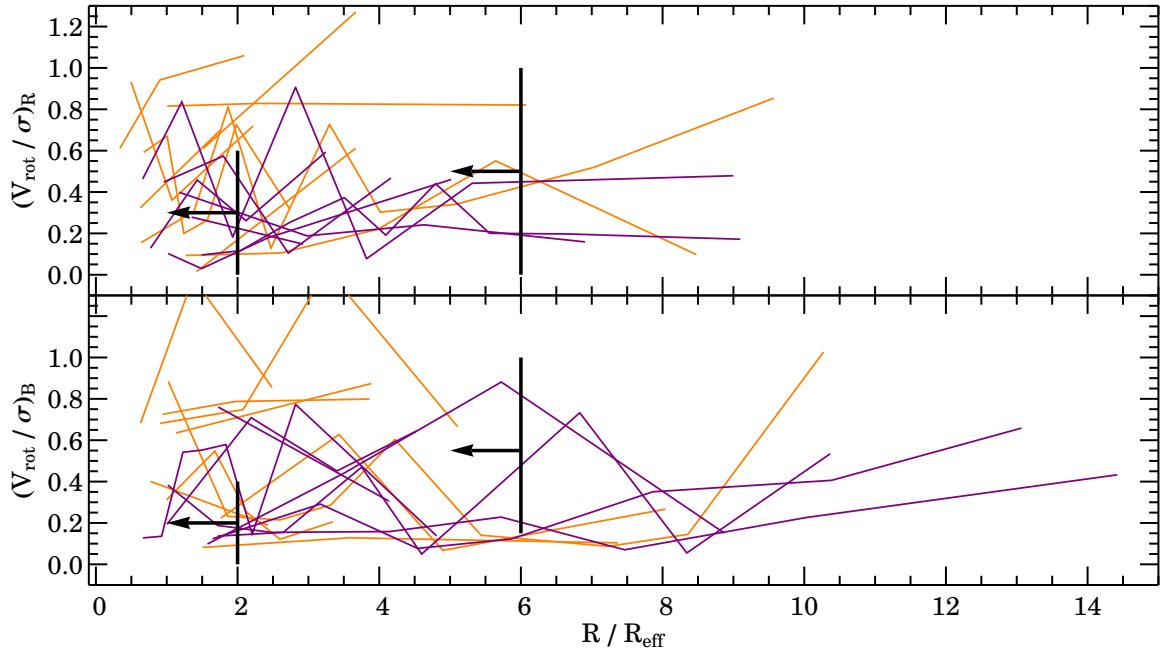} 
\caption{$(V_{\rm rot}/ \sigma)$ radial profiles for our galaxy sample. Colours are as in Figure~\ref{fig:VsigmaBR}. The top and bottom panels show the rotational dominance parameter for the red and blue GC systems respectively.
The black lines and arrows are predictions from \citet{Bekki2005} and represent the range of $(V_{\rm max} / \sigma_0)$ within $2$ and $6$ \Reff\ respectively. There is no clear trend of $(V_{\rm rot}/ \sigma)$ with radius as predicted from B$+05$.}
\label{fig:Vsigmainout}
\end{figure*}

In Figure~\ref{fig:SuperPlot}, the middle panel shows that most of the GC systems have $V_{\rm rms, B}/V_{\rm rms, R} \ge 1$. This is seen more clearly 
as a function of the X-ray luminosity which is larger in galaxies with an extended massive halo to which most of the blue GCs are thought to belong \citep{Forbes12}.
These results are in general agreement with the cosmological simulation of B$+08$ that predicted the $V_{\rm rms, B}/V_{\rm rms, R}$ 
ratio to be slightly $\ge 1$ for a wide range of host galaxy magnitudes. We also tested the significance of the apparent correlation between $V_{\rm rms, B}/V_{\rm rms, R}$ 
and $M_K$ finding a Spearman's rank correlation coefficient of $r_s=0.35$, implying that a correlation is unlikely. 

In the bottom panel of Figure~\ref{fig:SuperPlot} we quantify the GC system rotation by studying the overall ($V_{\rm rot}/ \sigma$), 
which estimates the contribution of the GC rotation over its random motions. We find that galaxies are spread between $0<(V_{\rm rot}/ \sigma)_{\rm A}<0.8$ 
except for NGC~$7457$ with $(V_{\rm rot}/ \sigma)_{\rm A}=1.68$. It also appears that galaxies with lower $L_X$ and lower $\sigma_0$ are more rotation-supported than other galaxies. 
The bulk of GC systems with ($V_{\rm rot}/ \sigma) \ge 0.3$ is consistent with cosmological simulation of B$+08$. However, as previously discussed, trends of rotation with radius 
and colour are also important to examine.

\subsection{Rotation}
\label{sec:discussion_Rotation}

To make things clearer we divide the host galaxies into two kinematic groups that will be used throughout this section. 
After considering various possibilities, we finally decided to divide the GC systems according to the rotation dominance parameter of their host galaxy. We use the 
$\lambda_{\Reff}$ parameter within 1\Reff\ as defined in \citet{Cappellari07} with a threshold of $0.31 \times \sqrt{\epsilon}$ to discriminate 
between slow and fast-rotators, where $\epsilon_{\Reff} = 1- (b/a)$ is the galaxy ellipticity.  For our sample, we use $\lambda_{\Reff}$ from \citet{Emsellem11} if the galaxy is in ATLAS$^{\rm 3D}$, 
whereas we use the conversion formulae provided by the same authors to convert $V_{\rm max} / \sigma_0$ into $\lambda_{\Reff}$, where $V_{\rm max}$ and $\sigma_0$
are the maximum rotation amplitude and the central velocity dispersion of the host galaxy respectively (Figure \ref{fig:Lambda}). 
This separation has the advantage of being independent of the galaxy inclination. It also allows us to test if the slow-fast rotator separation in the inner regions persists in the outer regions. 
The first group of galaxies is composed of slow-rotator galaxies and includes mostly round massive galaxies at the centre of groups or clusters. 
The second group includes fast-rotators but it covers a wide range of galaxy masses and morphologies, from E$0$ to S$0$. Fast-rotator galaxies 
with $\lambda_{\Reff} > 0.25$ have stellar disky-like structures (photometric and/or kinematic) in the inner regions.

In Figure~\ref{fig:VsigmaBR} we compare the \textit{overall} rotational dominance parameter ($V_{\rm rot}/ \sigma$) of the blue and red GC subpopulations colour-coded according to their fast/slow rotator
classification.  
We find that a dichotomy exists between the GC systems in slow and fast rotators. 
Most of the GC systems in the slow-rotator galaxies have $0<(V_{\rm rot}/ \sigma)<0.3$. Again, we want to emphasise that this does not necessarily imply that these systems lack rotation because if 
their rotation occurs only in a limited radial range, it may have been smeared out in the overall ($V_{\rm rot}/ \sigma$). 
The GC systems in fast-rotator galaxies are more rotation-supported and in general at least one GC subpopulation has a conspicuous amount of rotation. There is a group of 
galaxies with $(V_{\rm rot}/ \sigma)\sim0.6$ which includes the fast-rotator galaxies with large $\lambda_{\Reff}$, but it also includes NGC~$4649$ with $\lambda_{\Reff}\sim0.12$. 

In Figure~\ref{fig:Vsigmainout} we compare the radial $(V_{\rm rot} / \sigma)$ profiles for the blue and red GC systems with the numerical simulations of B$+05$.  We note that
B$+05$ quote the ratio of the maximum rotation velocity to the central velocity dispersion $(V_{\rm max} / \sigma_0)$ within $2$ and $6$ \Reff, whereas we plot the overall $(V_{\rm rot} / \sigma)$ in each radial bin. 
Therefore, the predictions plotted in Figure~\ref{fig:Vsigmainout} are typically upper limits. Another caveat is that B$+05$ simulate dry mergers, which leads to a systematic mismatch with the observed rotation 
in the galaxy centers. However, this effect is small for $R\gtrsim2\Reff$ \citep{Hoffman}, which is the radial range with better GC coverage and the most relevant for this study. 

We find that for galaxies in the fast-rotator group, at least one GC subpopulation has $(V_{\rm rot} / \sigma) \ne 0$. Slow-rotator galaxies have generally slower rotating GC systems. 
Although the comparison with B$+05$ suggests an overall agreement of our results with their predictions, the scatter in both $(V_{\rm rot} / \sigma)$ and in the simulations is too large to draw any strong conclusions.
In more detail, the simulations predict $(V_{\rm rot} / \sigma)$ increasing monotonically with radius but we see no evidence for this feature in our data. Moreover,
a direct comparison with  B$+05$ simulations is not possible because they did not publish the $(V_{\rm rot} / \sigma)$ profiles to $6\Reff$.
Figure~\ref{fig:Vsigmainout} also shows that many GC systems are characterized by significant rotation spikes. These features might be partly caused by spatial incompleteness, outliers or projection effects. 
Alternatively, one speculative possibility is that these rotation spikes are the imprints of minor mergers. These events are expected \citep[e.g.,][]{Vitvitska,Hetznecker}, and observed
\citep[e.g.,][]{Romanowsky11}, to perturb the halo kinematics of massive galaxies. Detailed studies of GC system phase-space for each galaxy are needed to quantify this effect.
\citet{Norris12} also suggested that the missing outer rotation could be explained if the reservoir of angular momentum was located beyond the radii mapped by current GC system studies. 
This idea would be supported by the large rotation $(V_{\rm rot} / \sigma\sim1.0)$ we observe at $\sim 10\Reff$ for both GC subpopulations of NGC~$1407$. However, at such galactocentric distances, results are uncertain 
because of spatial incompleteness and low number statistics. The same concerns were addressed by \citet{Strader11} on the suspicious outer rotation $(V_{\rm rot} / \sigma\sim0.6)$ observed in NGC~$4486$.

\subsection{Kinematic misalignment}
\label{sec:discussion_misa}
\begin{figure}
\centering
\includegraphics[scale=.5]{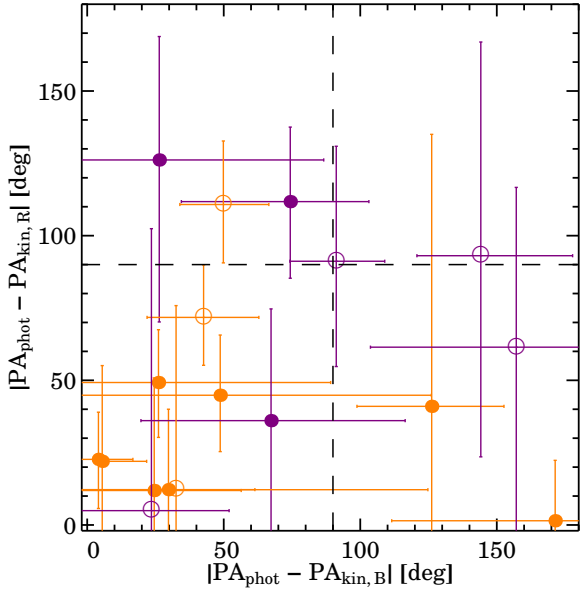} 
\caption{GC system position angle misalignments. The horizontal and the vertical axes show the misalignment between the photometric and the 
kinematic major axes for the blue and the red GC subpopulations respectively.  Dashed lines mark a constant 
$90$ degree angle. Rotation along the photometric major axis occurs at $0$ deg, rotation along the minor axis at 90 deg and counter rotation 
at 180 deg.  Galaxies with both a blue and red GC kinematic position angle unconstrained are not included for clarity. 
The rotation of the red GCs is more consistent with the photometric major axis.}
\label{fig:PA_misa}
\end{figure}

It is also important to study how the rotation of the GC systems aligns with respect to the host galaxy isophotes. An eventual 
position angle misalignment might contain information about the triaxiality of the system \citep{Krajnovic}. 

In Figure~\ref{fig:PA_misa} we compare the PA misalignment of the blue and red GC subpopulations. 
For clarity, we have removed from the plot galaxies for which neither the $PA_{\rm kin}$ of the blue nor of the red GC subpopulation was constrained.
A caveat here is that we study \textit{overall} $PA_{\rm kin}$ that will be biased by kinematic twists with radius.
Broadly speaking, we note that most of the galaxies are located in the bottom-left quadrant of Figure~\ref{fig:PA_misa} towards $|\Delta PA|_{\rm} = 0$, suggesting that 
the red GC subpopulation rotates more consistently with the photometric major axis than the blue GC subpopulation. 
We do not see a sharp separation between the GC systems in fast and slow-rotator galaxies. The GC rotation in fast-rotator galaxies seems to be equally consistent with the photometric major axis
for both GC subpopulations. 

The B$+05$ simulations also predicted kinematic misalignments with respect to the galaxy isophotes, albeit with an amplitude smaller than our findings. 
Depending on the orbital configuration, these authors can also reproduce the minor-axis rotation for the blue and red GC subpopulations, but they do not 
quantify this effect. Unfortunately, B$+05$ do not provide radial kinematic misalignments profiles, but they find that the rotation of both GC subpopulations in the outer regions of a merger
remnant should be aligned along the photometric major axis, which is in rough agreement with our findings.
\begin{figure}
\centering
\includegraphics[scale=.5]{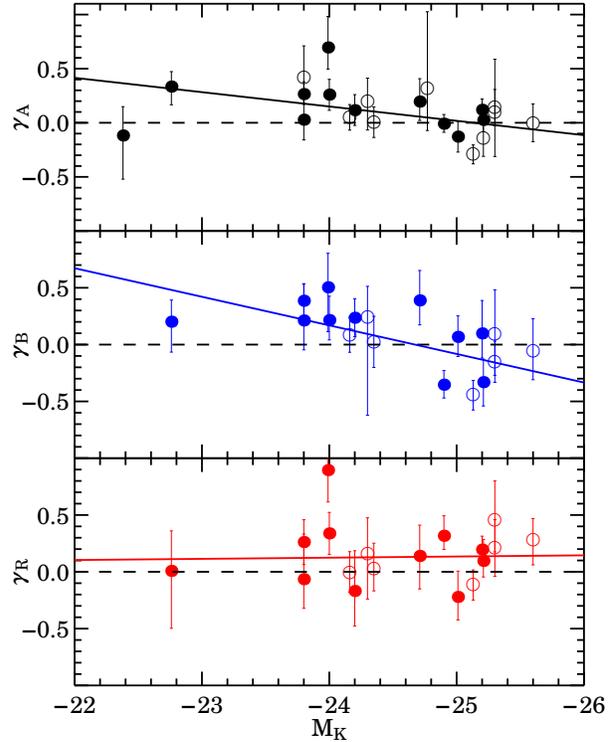} 
\caption{Log-slopes of the GC system $V_{\rm rms}^2$ profiles. Symbols are as in Figure~\ref{fig:VsigmaBR}. 
For each GC system the plot shows the best fit $\gamma$ parameter to eq. \ref{eq:power-low} for all (top panel), 
blue (middle panel) and red GC subpopulation (bottom panel) as a function of the absolute magnitude $M_K$ of the respective host galaxy. 
Lines are weighted linear fits to the data, whereas the dashed lines show an ideal flat $V_{\rm rms}$ profile $(\gamma=0)$.
The blue GC subpopulation tends to have a decreasing $V_{\rm rms}$ slope $(\gamma<0)$ in more massive galaxies, 
while the red GC subpopulation has steady $\gamma \ge 0$ over the whole magnitude range.}
\label{fig:gamma}
\end{figure}

\subsection{The slope of the \Vrms\ profile}
\label{sec:discussion_slope}
The \Vrms, sometimes freely named as velocity dispersion if the contribution of the rotation is low, is related the line-of-sight kinetic energy of a galaxy per unit of mass. 
Its profile has been classically used to model dark matter halos. Here, instead of showing the detailed \Vrms\ profile 
for our GC system sample, we think it is more convenient to parametrize these profiles and to study their overall slopes. We use a maximum-likelihood 
approach to fit a power-law function to the non-binned \Vrms\ profile, similarly to that done in eq. \ref{eq:Vobs2} for the GC system kinematics. 
In this case the $\chi^2$ function to be minimized is \citep{Bergond06}:
\begin{equation}
\chi^2= \sum^{i=N}_{i=1} \left[\frac{(V_{{\rm rad},i}-V_{\rm sys})^2}{V_{\rm rms}^2+(\Delta V_{{\rm rad},i})^2}+\ln (V_{\rm rms}^2+(\Delta V_{{\rm rad},i})^2)\right],
\label{eq:power-low-like}
\end{equation}
and the function to be modelled is:
\begin{equation}
V_{\rm rms}^2(R)=V_{\rm rms,0}^2 \times (R/R_0)^{-\gamma},
\label{eq:power-low}
\end{equation}
where $R_0$ is a scale parameter, set here to the median galactocentric distance of each GC system. $V_{\rm rms,0}$ and $\gamma$ are two parameters free to vary, with the latter 
representing the slope of the power-law: increasing if $\gamma<0$ and decreasing if $\gamma>0$. Uncertainties on this method were obtained by bootstrapping
the sample 2000 times in order to obtain the $68$ per cent confidence levels.

Bearing in mind that the shape of the profile might change with radius (see the velocity dispersion profile of NGC~$1407$), in Figure~\ref{fig:gamma} we show the best 
fit $\gamma$ for each GC subpopulation (including the combination of the two) as a function of the host galaxy magnitude. The overall slope of the  $V_{\rm rms}^2$ profile ($\gamma$) 
of the the GC systems is generally positive, but it becomes $\le 0$ for brighter galaxies, implying a constant or even increasing \Vrms\ profile.
If we now consider the blue and red GC subpopulations separately, it is clear that this effect is caused \textit{only} by the blue GC subpopulations in galaxies with $M_K < -25$ mag. 
The red GC subpopulations are always consistent with having $\gamma \ge 0$ and a linear fit to the data suggest no significant trend with the magnitude.
It is not clear why we do not observe $\gamma \le 0$ also for the red GC subpopulation. 
The two GC subpopulations might have different orbital configurations as a consequence of different formation histories and this would alter 
the shape of the \Vrms\ profiles. Therefore, it is important also to study the orbits of the GC systems.
\begin{figure*}
\centering
\includegraphics[scale=.5]{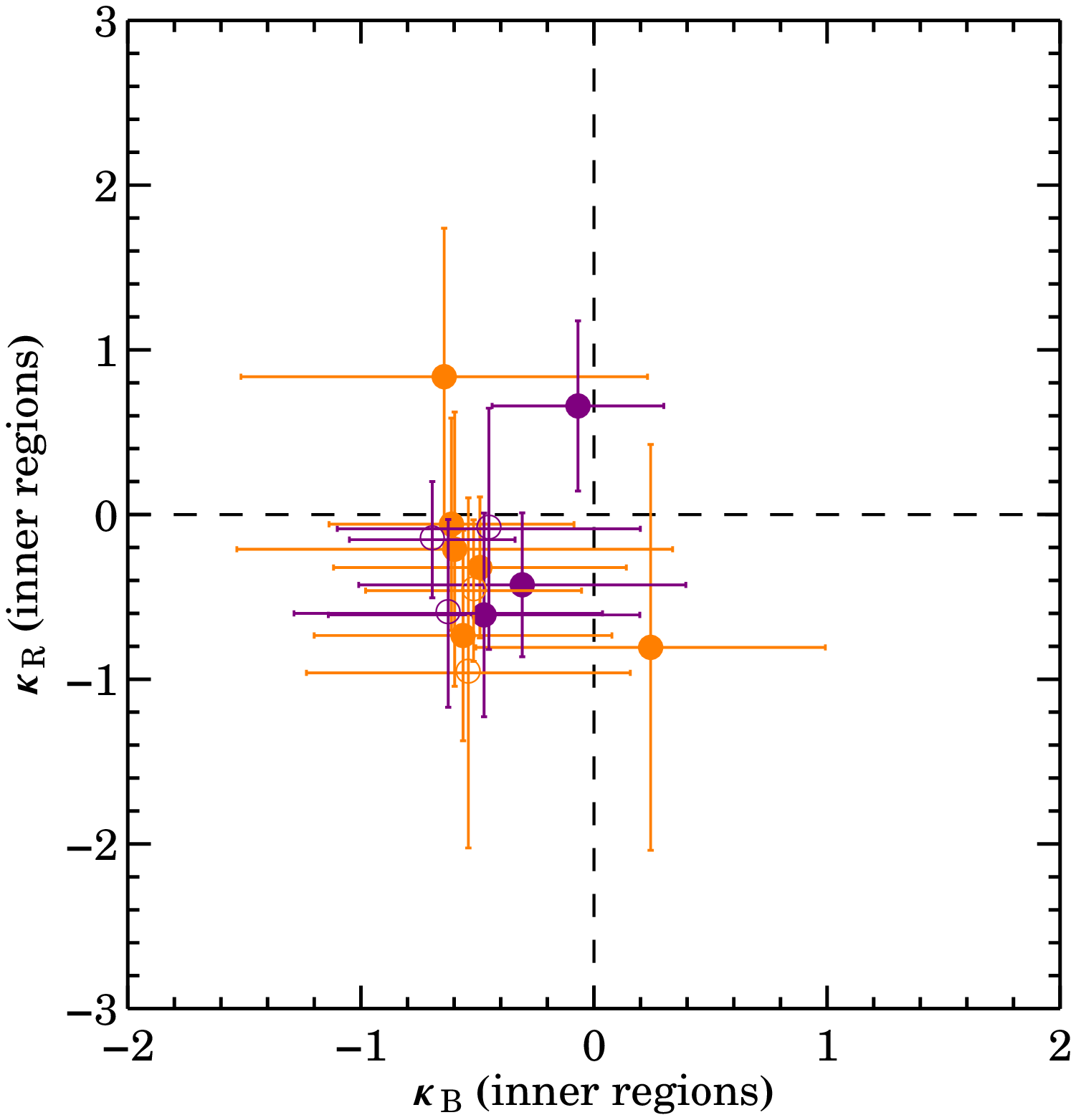} 
\hspace{1cm}
\includegraphics[scale=.5]{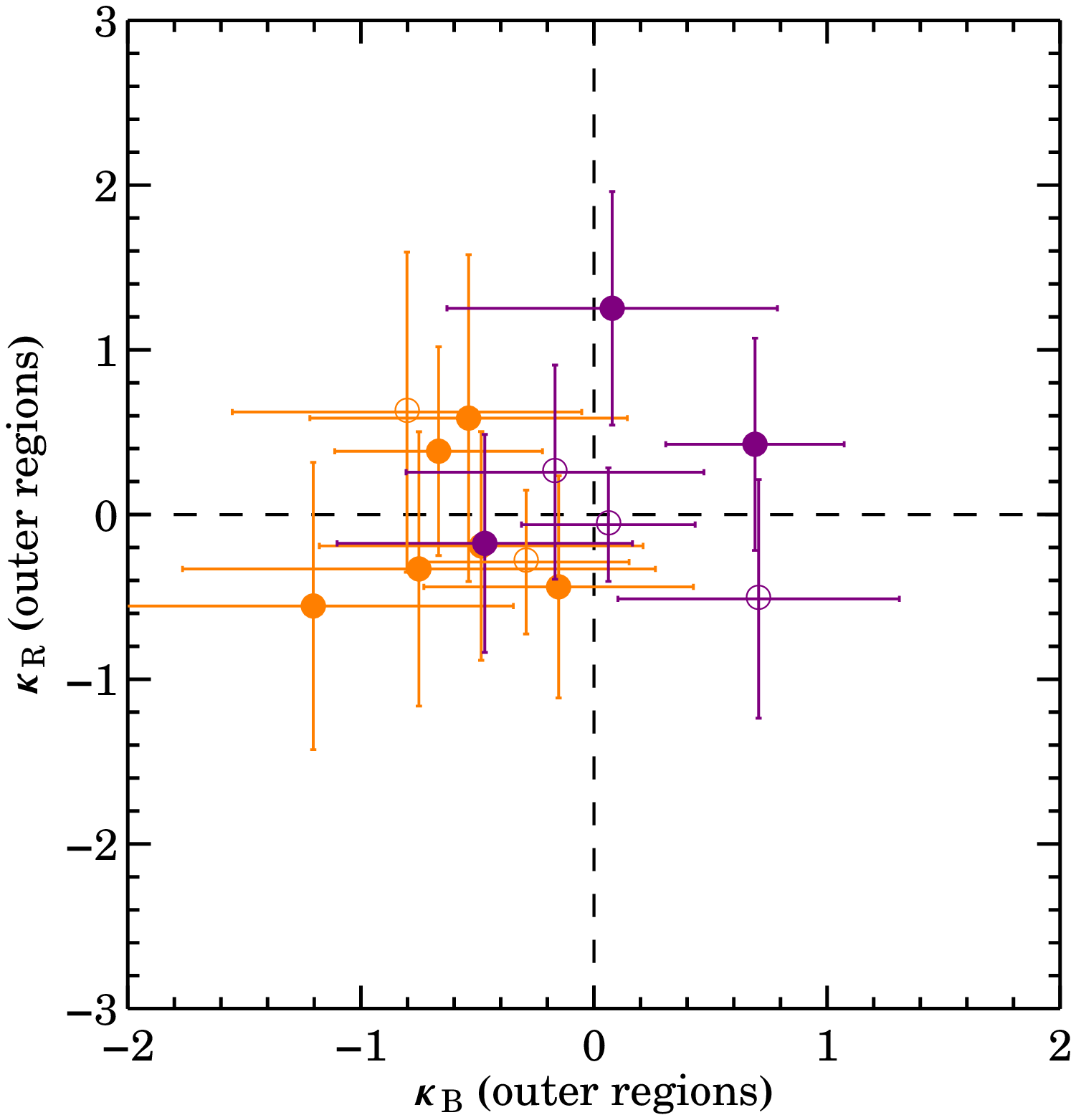} 
\caption{GC system velocity kurtosis. Symbols are as in Figure~\ref{fig:VsigmaBR}. The plot compares the kurtosis $\kappa$ of the blue and red the GC systems in the inner 
(left panel) and outer radial bins (right panel). Dashed lines represent Gaussian distributions at $\kappa = 0$. GC system orbits are tangential and radial 
if $\kappa < 0$ and $\kappa > 0$ respectively.}
\label{fig:kurtosis}
\end{figure*}

\subsection{Kurtosis}
\label{sec:discussion_kurto}
The galaxy anisotropy has classically been measured through the parameter $\beta$ that gives the relative ratio of the tangential to the radial velocity dispersion for a given 
dynamical tracer \citep{Binneybook}. The calculus of $\beta$ for our GC systems is not a topic of this paper and we will instead keep things simple by
calculating the kurtosis $\kappa$ that measures the deviation of a velocity distribution from a Gaussian. In the regime of a constant velocity dispersion the kurtosis 
provides a rough estimate of the GC orbits. These will be will be radial $(\kappa>0)$, isotropic $(\kappa = 0)$ or tangential $(\kappa<0)$, respectively. 
Broadly speaking we measure the fourth-order velocity moment:
\begin{equation}
\kappa=\frac{1}{N} \sum_{i=1}^N \frac{(V_{\rm rad, i} - V_{\rm sys})^2}{V_{\rm rms}^4} - 3 \pm \sqrt{\frac{24}{N}},
\label{eq:kurtosis}
\end{equation}
but in practice we will use a more complicated bias-corrected formula from \citet{Joanes}.

We divide the radial kurtosis profiles in two radial bins in order to have roughly the same number of GCs in each bin. 
In this case the required number of GC radial velocities per bin is larger because the uncertainties are $\propto N^{1/2}$. We require at least $30$ GCs 
per bin and this excludes smaller galaxies like NGC~$1400$ or NGC~$821$. As the calculation of $\kappa$ does not take the GC radial velocity uncertainties 
into account, a bias might be present due to the different velocity resolution of the literature and of our GC samples. Results are presented in Figure~\ref{fig:kurtosis}. 

We find that in the innermost bin most of the GC systems have $\kappa_B \le 0$ and $\kappa_R \le 0$, suggesting isotropic or even tangential orbits. 
This is also true for kurtosis in the outer bin, but some galaxies (such as NGC~$4365$, NGC~$4486$ and NGC~$4472$) have the tendency of having more radially-biased orbits. In all cases the kurtosis of the blue 
and red GCs are consistent with isotropic orbits.

\section{Discussion}
\label{sec:discussion}
\subsection{GC kinematic bimodality}

The origin of GC colour bimodality is still an area of debate. On the theoretical side, there is a lack of consensus regarding 
the leading processes that drive the formation of GC systems (i.e. merging, collapse, accretion) and whether one \citep[see][]{Muratov} 
or more of these processes (e.g., \citealt{Lee}) are needed to explain the observed colour bimodality. The contrasting numerical predictions
are in part due to the diverse recipes with which simulations were built \citep{Bekki2005,Bekki2008,Moore,Kravtsov,Diemand}, but they might
also reflect a real diversity of formation mechanisms that built up the GC systems in the first place. 

On the observational side, the GC colour bimodality has been generally accepted because the two colour subpopulations differ in 
mean size and spatial distribution, and they depend differently on host galaxy properties \citep{Forbes97,Peng06}. 
Recently, the colour bimodality itself has been claimed to be an artifact of a strongly non-linear colour-metallicity relation \citep{Yoon06,YoonSecond}, 
which would make the metallicity distribution of a GC system unimodal, skewed towards metal-poor metallicities. The resulting metallicity 
distribution function would be more metal-rich than the one classically obtained with linear, or broken linear, colour-metallicity relations 
\citep{Peng06,Faifer11,Sinnott,Alves-Brito}. 

We have found that GC systems around most of the surveyed galaxies have different kinematics between the blue and red sides of their GC colour distribution. 
This includes different kinematic position angle, rotation amplitudes and velocity dispersions. The kinematics generally change smoothly with the GC colour, 
but we also observe cases in which there is a sharp kinematic transition corresponding to the blue-red dividing colour (see NGC~$821$, 
NGC~$1407$ and NGC~$3377$ in Figure~\ref{fig:CV}). \textit{Our kinematic results are therefore strong evidence against colour bimodality simply being a consequence of 
a strong non-linear colour-metallicity relation}.

Most of the surveyed galaxies have small, yet significant, blue/red kinematic diversity (especially in their 
velocity dispersion profiles). The fact that we do not see any sharp kinematic transition with colour in all galaxies is not surprising because these systems lack intrinsic 
rotation of both GC subpopulations. Therefore, the rotation amplitude of GC systems can be insensitive for testing kinematic bimodality, especially 
for massive galaxies, for which one has instead to rely on other proxies such as the slope of the \Vrms\ profile (Figure~\ref{fig:gamma}).

\subsection{The star-GC system connection}

Although the link between the red GCs and the host galaxy was somewhat expected from their similar spatial distributions, 
its kinematic confirmation was limited to a handful of galaxies \citep{Schuberth,Norris12,Romanowsky09,Strader11}. The results that we found for
our galaxy sample are diverse in this regard.

On one side we find excellent agreement between the kinematics of the stars (and/or PNe) and that of the red GCs, indicating that the coupling between 
stars and GC systems holds also for less massive galaxies. On the other hand, we have also found galaxies in which the blue GCs behave 
as the galaxy stars or, more interestingly, as the PNe. NGC~$3377$ and NGC~$821$ are two examples. These are very similar field disk galaxies  
with a falling stellar rotation curve. However, the blue GCs in NGC~$3377$ do not rotate and are decoupled from the PNe. Conversely, 
in NGC~$821$ blue GCs mimic the minor axis rotation of the PNe, but not of the stars. 
What is most surprising in the latter case is indeed the fact that the PNe are decoupled from the stars as well. 
Based on stellar population inferences, \citet{Proctor05} concluded that NGC~$821$ has undergone a minor merger or 
tidal interaction in the last $4$ Gyr. Our results would support this scenario. 
However, it remains to be explained why the PNe are not akin to the underlying stars and how this feature can influence the dark matter estimate of this galaxy \citep{Romanowsky03}.

How the star-GC system connection arose in the early times is still an open debate.
Galaxy mergers may contribute to building the GC system of the remnant. \citet{Bekki2002} have shown that tidal shocks induced by galaxy merging 
can compress giant molecular clouds to form both new metal-rich GCs and stars. However, based on stellar population studies, \citet{Forbes07} have shown 
that most of the red GCs are consistent with being uniformly old, and few, if any, formed in late epoch gaseous mergers.

A possible consensus  scenario may arise from the idea that part of the red GCs might have formed during the early turbulent phases of galaxy formation at $z\sim2$ \citep{Kravtsov,Shapiro}. 
At this redshift, galaxies are characterised by fast rotating thick discs \citep{Elmegreen09} fragmented into super-star forming clumps. 
Observations have shown that the origin of these clumps is related to disc instability processes \citep{Wisnioski}, but they may be also caused 
by cold flows \citep{Dekel09}. It is thought that these gas-rich clumps migrate inwards to feed the nascent bulge (dragging along newly formed 
metal-rich GCs) and losing their rotational motion to dynamical friction, finally acquiring a typical $V_{\rm rot} / \sigma < 1$ for the bulge. During
this migration, $\sim 50\%$ of the clump mass will be stripped off and it will end up forming a rotationally supported disk \citep{Bournaud07}. 

This formation mechanism seems to qualitatively explain the spatial and kinematic similarity between stars and red GCs. However,
as far as we know, no numerical simulations have to date been produced to directly compare observations with the kinematics of the red GCs 
forming during the `gas-rich clump-driven phase'. Also, the study of dwarf ellipticals in the Virgo cluster \citep{Beasley09} and in the 
Sculptor group \citep{Olsen} suggested that disks may have been significant sites of GC formation at early times. 

The poor radial overlap between studies of galaxy stars and GC systems complicates the kinematic comparison between these two. Extended stellar kinematic
studies, a la \citet{Proctor}, will help to test the galaxy star-GC system connection. Also, it will be interesting to see what is the role of the PNe in this scenario 
and if the PNe-GCs connection in NGC~$821$ is the rule rather than the exception.

\subsection{Orbital anisotropy}
\label{sec:Orbital anisotropy}

The GC orbital anisotropy, in this context represented by the kurtosis~$\kappa$, can provide other clues about the formation of GC bimodality. 
Theoretical studies predict that dark matter haloes, including the baryonic tracers within them, have radially-biased orbits in their outer regions \citep{McMillan,Prieto,Dekel05}. 
However, the small number of studies in this field, which have usually employed deep integrated stellar light and PNe, seems to suggest that early-type galaxies harbour 
quasi-isotropic orbits \citep{Gerhard98,Douglas07,deLorenzi09,Napolitano,Napolitano11,Deason}. How does this compare to the GC anisotropy? 

Our analysis suggest that in the inner regions both GC subpopulations have isotropic or tangential orbits, whereas in the outer regions there is a hint that some red GC systems 
might have radial orbits. We recognise that in both cases the GC orbits are still consistent with being isotropic. 

As already pointed out by \citet{Romanowsky09}, the lack of radially-biased orbits for the blue GC subpopulation might be connected to the evolutionary history
of the GC system within the galaxy. Processes such as evaporation, dynamical friction and the consequent disruption
of GCs may change the anisotropy. The disruption of GCs will be more efficient for those on radial orbits because 
they plunge deepest into the galaxy centre \citep{Baumgardt98,Baumgardt}. Accordingly, they will become less luminous and eventually vanish altogether, making 
GCs on tangential and isotropic orbits the most likely ones to survive and be observed today. 

The GC orbital anisotropy will need further in-depth investigation but the overall pattern suggests that, if the blue GCs formed from the accretion or minor merger of satellite galaxies, 
some disruption effects have to be taken into account to explain the general absence of blue GC on radially-biased orbits. 
The scenario discussed above supports the idea that present day blue GCs were brought into the host galaxy via accretion and/or minor mergers. 

The key point here is that the different GC orbits mean that the resulting \Vrms\ profiles will be shaped by the orbits themselves (the mass-anisotropy degeneracy).
Tangential anisotropy, for example, could mimic the presence of dark matter, whereas radial anisotropy could deplete it. Interestingly, we find that the blue GC subpopulations with
tangentially-biased orbits tend to have increasing \Vrms\ profiles (suggesting a more dominant dark-matter halo), whereas the red GC subpopulation is consistent with having 
a decreasing \Vrms\ profile and more radial orbits. 
It will be interesting to test how a classic Jeans analysis that makes use of the different GC anisotropies found in this paper
compares with recent studies that do not require assumptions on the dark matter halo \citep{Walker}. 

% % % % % % % % % % % % % % % % % % % % % % % % % % % % % % % % % % % % % % % % % % % % % % % % % % % % % % % % % % % % % % % % 

\subsection{NGC~7457: a case study for the formation of S0s}
\label{sec:a case study}
GC systems have been exploited to study the unsolved conundrum of the formation of lenticular galaxies \citep{Barr,Arnold}.
In fact, it is still not clear whether S$0$s are quenched spirals: the remnant of a gentle gas shut off process that would preserve the kinematic signature of the progenitor spiral 
\citep{Byrd90,Williams}, or very disky ellipticals: the products of violent merging events \citep{Bekki98,Cretton} that would diminish the original disk-rotation. 
The observational differences between these two scenarios are minimal and it also depends upon the galaxy environment \citep{Kormendy12}. 
S$0$s might still be rotationally supported if they form in uneven-mass galaxy mergers, with mass-ratio ratio between $1$:$4.5$ and $1$:$10$ \citep{Bournaud}.
\begin{figure}
\centering
\includegraphics[scale=.5]{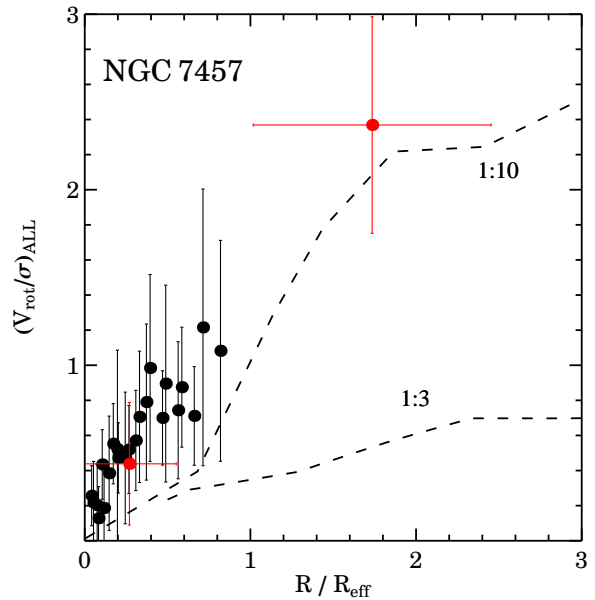} 
\caption{GC rotation rate in NGC~$7457$. GC data from this work (supplemented with \citealt{Chomiuk}) and long--slit data from \citet{Simien} 
are shown as red and black points respectively. Also shown are the simulated merger remnant with a $1$:$10$ (top dashed line) 
and $1$:$3$ (bottom dashed line) mass ratio from \citet{Bournaud} and \citet{Cretton}, respectively. Typical uncertainties on the 
merger remnant profiles are $15$\% at $4$ \Reff.}
\label{fig:Vsigma7457}
\end{figure}

In this context, a particularly interesting case study is offered by NGC~$7457$ that we have discussed in \S \ref{sec:NGC7457}. This nearby lenticular galaxy contains in fact both 
signatures of a recent merger \citep{Chomiuk} and a disky GC kinematics with a possible counter-rotating core in the innermost regions \citep{Sil'chenko}. 
In Figure~\ref{fig:Vsigma7457} we compare the $V_{\rm rot}  / \sigma$ profile of the stars \citep{Simien} and of our GC measurements (including the dataset of Chomiuk) 
with N-body simulations of merging with different mass-ratios of 1:3 \citep{Cretton} and 1:10 \citep{Bournaud}, respectively. To do this, we select all the GCs 
within $1$ arcmin from the photometric major axis and we construct two bins of $8$ objects each. 

Results in Figure~\ref{fig:Vsigma7457} show that our data are in good agreement with the $1$:$10$ remnant, even though
the discrepancies occurring at small and large radii seem to suggest that the mass-ratio involved in the formation of this 
galaxy was probably smaller than this value. These results, along with the counter-rotating core in the centre and the intermediate-age GCs, seem to favour the minor merger scenario. 
We also note that the S$0$ galaxies simulated in B$+05$ are characterised by a strong rotation of the blue GC subpopulation if formed via merging, but we do not see any significant
change of the GC kinematics with the GC colour in NGC~$7457$. 
This effect could indeed be real, but it could also be driven by the low number statistics or by the large uncertainties ($ \pm 0.2$ mag) due to the conversion from \textit{HST} to WIYN magnitudes.

\section{Summary and Conclusions}
\label{sec:Summary}

We have examined the GC systems in twelve early-type galaxies (from lenticular to large ellipticals over a range of galaxy mass), 
nine of which published for the first time, with particular emphasis on the kinematics of their blue and red GC subpopulations. 

For the new presented data, we have used wide-field ground-based imaging (mainly from Subaru/Suprime-cam) and \textit{HST} observations, 
finding that eight out of nine galaxies have a significant bimodal GC colour distribution. The study of the spatial distribution of these two subpopulations
has revealed that red GCs are generally more centrally concentrated and that they have a slope similar to the surface brightness of the underlying galaxy. 
This is in agreement with the idea that red and blue GC subpopulations trace the spheroidal component and the halo component of the host galaxies, respectively.

Multi-object spectroscopy was performed for a bright sample of GCs using Keck/DEIMOS that provided an average velocity resolution of $\sim 15\kms$
(a factor of 3 better than most previous studies). We find a variety of kinematic profiles for both the blue and the red GC subpopulations. The salient results are:

\begin{enumerate}
\item The GC kinematics (rotation amplitude, velocity dispersion and rotation axis) varies with GC mean colour. In particular, we find GC systems which have a sharp 
kinematic transition at the blue-red dividing colour, implying that the GC colour bimodality is real and extends to kinematic bimodality.

\item The rotation velocity and velocity dispersion of the red GC subpopulation mimics the host galaxy stellar kinematics (including those of planetary nebulae). 
This property supports the scenario in which the red GCs form together with the bulk of stars of the host galaxy, for instance during the `turbulent disk phase' at high redshift.
The rotation of the blue GCs is typically consistent with zero, but their velocity dispersion is always higher or equal to that of the red GCs, especially in the outer regions.

\item The GC kinematics combined with other kinematic studies can reveal interesting features. We have found that 
the blue GCs in NGC~$821$ mimic the minor axis rotation of the PNe, but not that of the stars, suggesting that there might be a blue GC-PNe connection that 
trace a recent merging event. Also, our data seem to be consistent with the idea that the S$0$ galaxy NGC~$7457$ was formed via a minor merger with a $1$:$10$
mass ratio.
\end{enumerate}

We have supplemented our dataset with ten additional GC systems from the literature and compared their kinematics to host galaxy properties.  
We have compared our results with numerical simulations, finding no strong evidence that GC systems have formed via major-mergers.
We find that the correlation between the $V_{\rm rms}$ of the GC systems and the mass of the host galaxy 
(from lenticular to massive ellipticals) holds for less massive galaxies and that it is tighter for the red GC subpopulation. 
The blue GCs in more massive galaxies have increasing \Vrms\ profiles, whereas the red GCs have always shallower \Vrms\ profiles. 
A study of the GC velocity kurtosis, suggests that blue GCs generally appear to be isotropic or tangentially-biased in the outer regions, unlike the red GCs which instead have more
radially-biased orbits.

\section{Acknowledgements} 
We thank the anonymous referee for the constructive feedback. 
AJB acknowledges the support of the Gordon \& Betty Moore Foundation. 
CF acknowledges co-funding under the Marie Curie Actions of the European Commission (FP7-COFUND).
KLR and JRH acknowledge support from NSF Career award AST-$0847109$ (PI: Rhode).
This work was supported by the National Science Foundation through grants AST-$0808099$, AST-$0909237$ and AST-$1211995$.
Some of the data presented herein were obtained at the W. M. Keck Observatory, operated as a scientific partnership among the California Institute of Technology, 
the University of California and the National Aeronautics and Space Administration, and made possible by the generous financial support of the W. M. Keck Foundation. 
The authors wish to recognise and acknowledge the very significant cultural role and reverence that the summit of Mauna Kea has always had within the indigenous
Hawaiian community. The analysis pipeline used to reduce the DEIMOS data was developed at UC Berkeley with support from NSF grant AST-0071048. 
Based in part on data collected at Subaru Telescope and obtained from the SMOKA  (which is operated by the Astronomy Data Centre, National Astronomical Observatory of Japan),
via a Gemini Observatory time exchange.
The authors acknowledge the data analysis facilities provided by IRAF, which is distributed by the National
Optical Astronomy Observatories and operated by AURA, Inc., under cooperative agreement with the National Science Foundation.
We have used the data products from the 2MASS, which is a joint project of the University of Massachusetts and the Infrared Processing and Analysis Centre/California Institute 
of Technology, funded by the National Aeronautics and Space Administration and the National Science Foundation.
This research has made use of the NASA/IPAC Extragalactic Database (NED) which is operated by the Jet Propulsion Laboratory, California Institute of Technology, 
under contract with the National Aeronautics and Space Administration. 
The figures for this article have been created using the excellent LevelScheme scientific figure preparation system \citep{Caprio}.

\bibliographystyle{mn2e}
\bibliography{Pota}

\appendix
\section{Velocity bias correction}
\label{appendix}
A feature of the kinematic modelling used in this work (equation~\ref{eq:Vobs2}) is that the rotation amplitude is biased to higher values when the kinematic position angle is unconstrained. 
This effect becomes important in systems with $V_{\rm rot} \ll \sigma$, because the rotation (and its direction) is fully embedded in the dispersion.
There are two ways to approach this issue: numerically, via a Monte Carlo simulation, or analytically, using the Box's bias measure in non-linear model theory \citep{Box}. 
The first method is the most straightforward, whereas the latter is computationally more complicated. Here we adopt a similar approach to \citet{Strader11} and correct 
for the bias via a Monte Carlo simulation. 

In practice, for each data bin containing $N$ GCs (see Figure \ref{fig:kinematics}), we compute the best fit rotation by minimizing equation \ref{eq:GC_LR}. 
We then generate $1000$ artificial datasets, of the same size $N$, drawn from
the best-fit model. We repeat the fit for each generated dataset using the original best-fit model as starting point. We define the velocity bias as the median difference between the computed and the
simulated (known) rotation amplitude. This approach is adopted to correct for the bias also in Figure \ref{fig:CV}, Table \ref{tab:KinematicsTable} and Table \ref{tab: survey_literature}.

We find that the magnitude of the bias correction depends both on the bin size and on the $(V_{\rm rot} / \sigma)$. 
In the case of $(V_{\rm rot} / \sigma) \gtrsim 1$, the bias is negligible regardless the bin size. 
For bins with $(V_{\rm rot} / \sigma) \sim 0.4 \, (0.6)$ and  $N\le20$, the bias can be up to $\sim 50 \, (40) \kms$. 
Within the same $(V_{\rm rot} / \sigma)$ values, the bias can decrease down to $\sim 30 (20)\kms$ for $N\approx50$. 
This agrees with the rule of the thumb of \citet{Strader11} to identify the severity of the bias, i.e. the bias is small if $(V_{\rm rot} / \sigma) \ge 0.55 \times \sqrt{20/N}$. 
If this condition is satisfied, we find that the bias is of the order of (or less then) our nominal median uncertainty of 15 \kms.

\end{document}